\documentclass[aps,pra,twocolumn,showpacs,superscriptaddress,floatfix]{revtex4-1}
\usepackage{array}
\usepackage{graphicx,amsmath,amssymb}
\usepackage[usenames]{color}
\usepackage[dvipsnames]{xcolor}
\usepackage[colorlinks=true,linkcolor=blue,anchorcolor=blue,citecolor=blue,urlcolor=blue]{hyperref}
\begin{document}

\title{Geometric perspective on quantum Rabi model: Super adiabatic
universality and curvature non-adiabatic effects}
\author{Zu-Jian Ying}
\email{yingzj@lzu.edu.cn}

\affiliation{School of Physical Science and Technology, Lanzhou University,
Lanzhou 730000, China} \affiliation{Key Laboratory for Quantum Theory and
Applications of MoE, Lanzhou Center for Theoretical Physics, Lanzhou
University, Lanzhou 730000, China}

\begin{abstract}
Quantum Rabi model (QRM) is a fundamental model in light-matter
interactions (LMIs), whereas this finite-component system manifests infinite
physics to be explored and a full understanding has not yet been reached
after a hundred years since it was proposed. In the present work we create a
geometric perspective for the QRM by mapping to a curved nanowire system, with curvature-induced spin-orbit coupling (SOC) manipulatable by LMI strength and frequency ratio, which is capable of revealing and perceiving the property universality and diversity from the ground state to the whole spectrum in a unified picture.
Indeed, we unveil a super university in the adiabatic regime which manifests a universal scaling of spin texture ubiquitously valid for different couplings, different frequencies and different energy levels. The curve frame also facilitates our finding that a direction changeover of tangential spin swing emerges in the non-adiabatic spin texture in association with the quantum phase transition (QPT) in the QRM, which endows a geometric non-adiabaticity interpretation and connotation to the QPT.
We also clarify that, among the curvature-induced effects, the SOC accounts for the spectral bifurcation and oscillations.
Our method and analysis, readily generalized
to other light-matter models, opens a dialogue between geometry and physics, thus paving a novel way to
gain deeper insight into LMIs.
\end{abstract}

\pacs{ }
\maketitle



\section{Introduction}

\label{Sect-Intro}

Light-matter interaction (LMI)~\cite{Eckle-Book-Models,Diaz2019RevModPhy,Kockum2019NRP,Qin2024PhysRep} plays a ubiquitous role in our physical
world, with a broad relevance to quantum information and quantum computation~\cite{Diaz2019RevModPhy,Romero2012,Stassi2020QuComput,Stassi2018,Macri2018},
quantum metrology~\cite{Garbe2020,Montenegro2021-Metrology,Chu2021-Metrology,Garbe2021-Metrology,Ilias2022-Metrology, Ying2022-Metrology,Gietka2023PRL-Squeezing,YangZheng2023SciChina,Hotter2024-Metrology,Alushi2024PRL,Mukhopadhyay2024PRL,Mihailescuy2024, Ying-Topo-JC-nonHermitian-Fisher,*Ying-Topo-JC-nonHermitian-Fisher-Cover,Ying-g2hz-QFI-2024,*Ying-g2hz-QFI-2024-Cover, Ying-g1g2hz-QFI-2025,Ying2025g2A4,*Ying2025g2A4-Cover,Ying-g2Stark-QFI-2025,Gietka2025PRL100802,Mihailescu2025CQMtutorial,QiaoFeng2026SpinSqueeze,QiuYi2025gA2},
condensed matter~\cite{Kockum2019NRP}, and cold atoms~\cite{LinRashbaBECExp2013Review,LinRashbaBECExp2011,LiuYing02025exoticSOC2Ring,*LiuYing02025exoticSOC2Ring-Cover,LiuYing02025KaleidoscopeDDI}.
With the extraordinary experimental
progresses in the past two decades~\cite{Diaz2019RevModPhy,Wallraff2004,Gunter2009,Niemczyk2010,Peropadre2010,FornDiaz2017, Ciuti2005EarlyUSC,Aji2009EarlyUSC,Forn-Diaz2010,Scalari2012,Xiang2013,Yoshihara2017NatPhys,Kockum2017,Qin-ExpLightMatter-2018,LiPengBo-Magnon-PRL-2024}, LMIs have entered a
contemporary era of ultra-strong~\cite{Ciuti2005EarlyUSC,Aji2009EarlyUSC,Diaz2019RevModPhy,Kockum2019NRP,Wallraff2004,Gunter2009,Niemczyk2010, Peropadre2010,FornDiaz2017,Forn-Diaz2010,Scalari2012,Xiang2013,Yoshihara2017NatPhys,Kockum2017,Bayer2017DeepStrong,Qin2024PhysRep,Ulstrong-JC-2}
and deep-strong~\cite{WangYouJQ2023DeepStrong,Yoshihara2017NatPhys,Bayer2017DeepStrong,DeepStrong-JC-Huang-2024}
couplings. With the advantages of high controllability and tunability~\cite{Qin-ExpLightMatter-2018,LiPengBo-Magnon-PRL-2024}, LMIs
not only provide an ideal platform for developments of advanced quantum technologies but also
open a test ground for novel quantum
theories~\cite{Braak2011,ChenQH2012,Batchelor2015,
Ashhab2013,Ying2015,LiuM2017PRL,Liu2021AQT,Hwang2015PRL,Hwang2016PRL,Irish2017,
Ying-g2hz-QFI-2024,*Ying-g2hz-QFI-2024-Cover,Ying-g1g2hz-QFI-2025,Ying2025g2A4,*Ying2025g2A4-Cover,Ying-g2Stark-QFI-2025,
Ying-2018-arxiv,Ying2020-nonlinear-bias,Ying-2021-AQT,*Ying-2021-AQT-Cover, Ying-gapped-top, Ying-Stark-top,*Ying-Stark-top-Cover,
Ying-Spin-Winding,*Ying-Spin-Winding-Cover,
Ying-JCwinding,Ying-Topo-JC-nonHermitian,*Ying-Topo-JC-nonHermitian-Cover,Ying-Topo-JC-nonHermitian-Fisher,*Ying-Topo-JC-nonHermitian-Fisher-Cover,Ying-gC-by-QFI-2024,
Grimaudo2022q2QPT,Grimaudo2023-Entropy,Grimaudo2024PRR,Zhu2024PRL,DeepStrong-JC-Huang-2024,PengJie2019,PengJ2021PRL,Padilla2022,Gao2022Rabi-dimer,GaoXL2025SPT,
Braak2019Symmetry,HiddenSymMangazeev2021,HiddenSymLi2021,HiddenSymBustos2021,
Irish2017,Irish-class-quan-corresp,
Felicetti2015-TwoPhotonProcess,e-collpase-Garbe-2017,e-collpase-Duan-2016,CongLei2019,Rico2020,
PengXHPRL2024RabiNMR,
Li2020conical,
WangYouJQ2023DeepStrong,
KuangLM2024AQT,Ulstrong-JC-2,
Yan2023-AQT,ZhengHang2017,
Garbe2020,Montenegro2021-Metrology,Chu2021-Metrology,Garbe2021-Metrology,Ilias2022-Metrology, Ying2022-Metrology,YangZheng2023SciChina,Gietka2023PRL-Squeezing,Hotter2024-Metrology,Alushi2024PRL,Mukhopadhyay2024PRL,Mihailescuy2024,
Gietka2025PRL100802,Mihailescu2025CQMtutorial,QiuYi2025gA2,
Boite2016-Photon-Blockade,Ridolfo2012-Photon-Blockade,LiaoJQ2020BlockadeJC,Ma2026PRL-Strong-2PB,Felicetti2026PRXQuantumBlockade, Carmichael1985,Birnbaum2005,Shamailov2010,Liew2010,Hamsen2017,Garziano2017,Flayac2017,Han2026WeakTPB,QiaoFeng2026SpinSqueeze,QiaoFeng2016AsymPolaron}.

A most fundamental model of LMIs is the quantum Rabi model (QRM)\cite{rabi1936,Rabi-Braak,Eckle-Book-Models}. The indispensable role of the
counter-rotating terms discovered in enhanced couplings~\cite{PRX-Xie-Anistropy} also renders the QRM to be more important. Surprisingly, this finite-component system manifests infinite
physics to be explored. As a matter of fact, despite of the seemingly simple form of the QRM, its exact solution had remained unavailable until the recent
milestone work by D. Braak who finally revealed its integrability~\cite{Braak2011}. This achievement has
triggered a massive dialogue~\cite{Solano2011} between mathematics and
physics~\cite{Braak2011,Solano2011,Boite2020,Liu2021AQT,
Ashhab2013,Ying2015,Liu2021AQT,Hwang2015PRL,Hwang2016PRL,Irish2017,
Ying-g2hz-QFI-2024,*Ying-g2hz-QFI-2024-Cover,Ying-g1g2hz-QFI-2025,Ying2025g2A4,*Ying2025g2A4-Cover,Ying-g2Stark-QFI-2025, LiuM2017PRL,Ying-2018-arxiv,Ying2020-nonlinear-bias,Ying-2021-AQT,*Ying-2021-AQT-Cover, Ying-gapped-top, Ying-Stark-top,*Ying-Stark-top-Cover, Ying-Spin-Winding,*Ying-Spin-Winding-Cover, Ying-JCwinding,Ying-Topo-JC-nonHermitian,*Ying-Topo-JC-nonHermitian-Cover,Ying-Topo-JC-nonHermitian-Fisher,*Ying-Topo-JC-nonHermitian-Fisher-Cover,Ying-gC-by-QFI-2024, Grimaudo2022q2QPT,Grimaudo2023-Entropy,Grimaudo2024PRR,Zhu2024PRL,DeepStrong-JC-Huang-2024,PengJie2019,Padilla2022,Gao2022Rabi-dimer,GaoXL2025SPT,QiuYi2025gA2, Garbe2020,Montenegro2021-Metrology,Chu2021-Metrology,Garbe2021-Metrology,Ilias2022-Metrology, Ying2022-Metrology,YangZheng2023SciChina,Gietka2023PRL-Squeezing,Hotter2024-Metrology,Alushi2024PRL,Mukhopadhyay2024PRL,Mihailescuy2024, Ying-Topo-JC-nonHermitian-Fisher,*Ying-Topo-JC-nonHermitian-Fisher-Cover,Ying-g2hz-QFI-2024,*Ying-g2hz-QFI-2024-Cover, Ying-g1g2hz-QFI-2025,Ying2025g2A4,*Ying2025g2A4-Cover,Ying-g2Stark-QFI-2025,Gietka2025PRL100802,Mihailescu2025CQMtutorial,QiuYi2025gA2, Bera2014Polaron, CongLei2017,CongLei2019,ChenQH2012, Braak2019Symmetry, Wolf2012,FelicettiPRL2020,Felicetti2018-mixed-TPP-SPP,Felicetti2015-TwoPhotonProcess,Simone2018,Alushi2023PRX, Irish2014,Irish-class-quan-corresp, PRX-Xie-Anistropy,Batchelor2015,XieQ-2017JPA, e-collpase-Garbe-2017,e-collpase-Duan-2016,Rico2020, Boite2016-Photon-Blockade,Ridolfo2012-Photon-Blockade,Li2020conical,Ma2020Nonlinear, ZhangYY2016,ZhengHang2017,Zheng2017,Yan2023-AQT,Chen-2021-NC,Liu2015, ChenGang2011-GVM,ChenGang2012,FengMang2013,HiddenSymMangazeev2021,HiddenSymLi2021,HiddenSymBustos2021,Casanova2018npj,JC-Larson2021,Ulstrong-JC-2,Eckle-2017JPA,*Eckle-2017JPA-b, Stark-Grimsmo2013,Stark-Grimsmo2014,Lu-2018-1,Xie2019-Stark,Stark-Cong2020,Cong2022Peter,Qin-ExpLightMatter-2018,
LiPengBo-Magnon-PRL-2024,PengJ2021PRL,Gao2022Rabi-aniso,QiaoFeng2026SpinSqueeze,KuangLM2024AQT,
WangYouJQ2023DeepStrong,PengXHPRL2024RabiNMR,
QiaoFeng2016AsymPolaron,
Boite2016-Photon-Blockade,Ridolfo2012-Photon-Blockade,LiaoJQ2020BlockadeJC,Ma2026PRL-Strong-2PB,Felicetti2026PRXQuantumBlockade,
Hamsen2017,Garziano2017,Flayac2017,FengLJ2021TPB,Kudlaszyk2019PRATPB,QianBin2018praTPB,Miranowicz2013praTPB,Han2026WeakTPB,Ying2026StarkTPB}.
In such a trend, a rich phenomenology has been explored in the QRM and its extensions. Indeed, aside from the integrability~\cite{Braak2011,ChenQH2012,Batchelor2015}, numerous findings have been yielded, including
finite-component QPTs~\cite{Ashhab2013,Ying2015,Liu2021AQT,Hwang2015PRL,Hwang2016PRL,Irish2017,
Ying-g2hz-QFI-2024,*Ying-g2hz-QFI-2024-Cover,Ying-g1g2hz-QFI-2025,Ying2025g2A4,*Ying2025g2A4-Cover,Ying-g2Stark-QFI-2025,
LiuM2017PRL,Ying-2018-arxiv,Ying2020-nonlinear-bias,Ying-2021-AQT,*Ying-2021-AQT-Cover,
Ying-gapped-top,
Ying-Stark-top,*Ying-Stark-top-Cover,
Ying-Spin-Winding,*Ying-Spin-Winding-Cover,
Ying-JCwinding,Ying-Topo-JC-nonHermitian,*Ying-Topo-JC-nonHermitian-Cover,Ying-Topo-JC-nonHermitian-Fisher,*Ying-Topo-JC-nonHermitian-Fisher-Cover,Ying-gC-by-QFI-2024, Grimaudo2022q2QPT,Grimaudo2023-Entropy,Grimaudo2024PRR,Zhu2024PRL,DeepStrong-JC-Huang-2024,PengJie2019,Padilla2022,Gao2022Rabi-dimer,GaoXL2025SPT},
multicriticalities and multiple points~\cite{Ying2020-nonlinear-bias,Ying-2021-AQT,Ying-gapped-top,Ying-Stark-top}, hidden
symmetry~\cite{Braak2019Symmetry,HiddenSymMangazeev2021,HiddenSymLi2021,HiddenSymBustos2021},
various patterns of symmetry breaking~\cite{Ying2020-nonlinear-bias,Ying-2018-arxiv,Ying-2021-AQT},
universality classification~\cite{Hwang2015PRL,LiuM2017PRL,Irish2017,Ying-2021-AQT,Ying-Stark-top,Ying-Topo-JC-nonHermitian-Fisher,*Ying-Topo-JC-nonHermitian-Fisher-Cover},
spectral collapse~\cite{Felicetti2015-TwoPhotonProcess,e-collpase-Garbe-2017,e-collpase-Duan-2016,CongLei2019,Rico2020} and stabilization~\cite{Ying2025g2A4,*Ying2025g2A4-Cover},
spectral conical intersections~\cite{Li2020conical},
classical-quantum correspondence~\cite{Irish-class-quan-corresp}, no-node theorem~\cite{Ying-gapped-top},
single-qubit conventional and unconventional topological phase transitions~\cite{Ying-2021-AQT,Ying-gapped-top,Ying-Stark-top,Ying-Spin-Winding,Ying-JCwinding},
coexistence and simultaneous occurrence of Landau-class and topological-class phase transitions~\cite{Ying-2021-AQT,Ying-Stark-top,Ying-JCwinding,Ying-Topo-JC-nonHermitian-Fisher,*Ying-Topo-JC-nonHermitian-Fisher-Cover}, robust topological feature against nonhermiticity~\cite{Ying-Topo-JC-nonHermitian,*Ying-Topo-JC-nonHermitian-Cover,Ying-Topo-JC-nonHermitian-Fisher,*Ying-Topo-JC-nonHermitian-Fisher-Cover},
squeezing and critical resources for quantum metrology~\cite{Garbe2020,Montenegro2021-Metrology,Chu2021-Metrology,Garbe2021-Metrology,Ilias2022-Metrology, Ying2022-Metrology,YangZheng2023SciChina,Gietka2023PRL-Squeezing,Hotter2024-Metrology,Alushi2024PRL,Mukhopadhyay2024PRL,Mihailescuy2024,
Ying-Topo-JC-nonHermitian-Fisher,*Ying-Topo-JC-nonHermitian-Fisher-Cover,Ying-g2hz-QFI-2024,*Ying-g2hz-QFI-2024-Cover,
Ying-g1g2hz-QFI-2025,Ying2025g2A4,*Ying2025g2A4-Cover,Ying-g2Stark-QFI-2025,Gietka2025PRL100802,Mihailescu2025CQMtutorial,QiuYi2025gA2},
photon blockade effect~\cite{Boite2016-Photon-Blockade,Ridolfo2012-Photon-Blockade,LiaoJQ2020BlockadeJC,Ma2026PRL-Strong-2PB,Felicetti2026PRXQuantumBlockade,
Carmichael1985,Birnbaum2005,Shamailov2010,Liew2010,Hamsen2017,Garziano2017,Flayac2017,
FengLJ2021TPB,Kudlaszyk2019PRATPB,QianBin2018praTPB,Miranowicz2013praTPB,Han2026WeakTPB,Ying2026StarkTPB},
and so on. Even undergoing these intensive studies, more physics can be still extracted and it is recognized that the
physics we can learn or get inspired from the QRM might be infinite~\cite{QiaoFeng2016AsymPolaron} and a full understanding has not yet been reached
after a hundred years since it was proposed~\cite{rabi1936,Rabi-Braak}.

Among the rich phenomenology, a particularly intriguing property that the QRM possesses is the finite-component QPT~\cite{Ashhab2013,Ying2015,Liu2021AQT,Hwang2015PRL,Hwang2016PRL,Irish2017,
Ying-g2hz-QFI-2024,*Ying-g2hz-QFI-2024-Cover,Ying-g1g2hz-QFI-2025,Ying2025g2A4,*Ying2025g2A4-Cover,Ying-g2Stark-QFI-2025,
LiuM2017PRL,Ying-2018-arxiv,Ying2020-nonlinear-bias,Ying-2021-AQT,*Ying-2021-AQT-Cover,
Ying-gapped-top,
Ying-Stark-top,*Ying-Stark-top-Cover,
Ying-Spin-Winding,*Ying-Spin-Winding-Cover,
Ying-JCwinding,Ying-Topo-JC-nonHermitian,*Ying-Topo-JC-nonHermitian-Cover,Ying-Topo-JC-nonHermitian-Fisher,*Ying-Topo-JC-nonHermitian-Fisher-Cover,Ying-gC-by-QFI-2024, Grimaudo2022q2QPT,Grimaudo2023-Entropy,Grimaudo2024PRR,Zhu2024PRL,DeepStrong-JC-Huang-2024,PengJie2019,Padilla2022,Gao2022Rabi-dimer,GaoXL2025SPT} which may be applied for quantum metrology~\cite{Garbe2020,Montenegro2021-Metrology,Chu2021-Metrology,Garbe2021-Metrology,Ilias2022-Metrology, Ying2022-Metrology,Gietka2023PRL-Squeezing,YangZheng2023SciChina,Hotter2024-Metrology,Alushi2024PRL,Mukhopadhyay2024PRL,Mihailescuy2024, Ying-Topo-JC-nonHermitian-Fisher,*Ying-Topo-JC-nonHermitian-Fisher-Cover,Ying-g2hz-QFI-2024,*Ying-g2hz-QFI-2024-Cover, Ying-g1g2hz-QFI-2025,Ying2025g2A4,*Ying2025g2A4-Cover,Ying-g2Stark-QFI-2025,Gietka2025PRL100802,Mihailescu2025CQMtutorial,QiaoFeng2026SpinSqueeze,QiuYi2025gA2,QiaoFeng2016AsymPolaron}. With anisotropy the QRM even opens a mini-world of QPTS~\cite{Ying-gapped-top}. A special character of the QPT is that it exhibits the critical universality via scaling relation and critical exponent with respect to the anisotropy in the low frequency limit~\cite{LiuM2017PRL,Ying-Stark-top}. However, at finite frequencies such a critical universality breaks down and the physical properties are diversified~\cite{Ying-Stark-top,*Ying-Stark-top-Cover}. Surprisingly, among the diversity new class of universality arises from the unnoticed nature of topological phase transitions~\cite{Ying-2021-AQT,*Ying-2021-AQT-Cover,Ying-gapped-top,Ying-Stark-top,*Ying-Stark-top-Cover,Ying-Spin-Winding,*Ying-Spin-Winding-Cover,Ying-JCwinding,Ying-Topo-JC-nonHermitian}. With the hint and inspiration of this universality-diversity-university finding progress, one may wonder if there is any other universality still hidden and whether the breakdown of hidden universality can bring us novel physics or insight.

For such kinds of expectations or aspirations of going beyond the conventional pictures and current knowledge on the QRM, one may need novel method and innovative perspective.
So far, in the literature a bunch of approaches and methods~\cite{Boite2020} have been raised and
developed for investigations on the QRM and its extensions, such as exact
solution in Bargmann-space representation~\cite{Braak2011}, Bogoliubov
transformation~\cite{ChenQH2012}, exact diagonalization~\cite%
{Ying2020-nonlinear-bias,Ying-Spin-Winding}, the rotating-wave approximation
(RWA)~\cite{JC-model}, the generalized RWA~\cite{Irish2007GRWA}, symmetry
analysis~\cite{Braak2019Symmetry,HiddenSymMangazeev2021,HiddenSymLi2021,HiddenSymBustos2021, Ying-2021-AQT,*Ying-2021-AQT-Cover,Ying-JCwinding,Ying-Topo-JC-nonHermitian-Fisher,*Ying-Topo-JC-nonHermitian-Fisher-Cover,Ying-g2hz-QFI-2024}, the adiabatic approximation~\cite{Irish-2005-AA}, the generalized
variational method~\cite{ChenGang2011-GVM,ChenGang2012}, Schrieffer-Wolff
transformation~\cite{Hwang2015PRL,LiuM2017PRL}, the
mean-photon-number-dependent variational method~\cite{Liu2015}, mean-field
method~\cite{Hwang2016PRL,PengJie2019}, variational displaced coherent state
method~\cite{Irish2014,ZhangYY2016,Hwang2010,Bera2014Polaron}, the symmetric polaron
picture~\cite{Ying2015,CongLei2019,Ying2020-nonlinear-bias,Ying-gapped-top} and asymmetric-polaron picture~\cite{QiaoFeng2016AsymPolaron},
and so forth.  These methods have played significant roles in the afore-mentioned findings. However, generally speaking, apart from those designed for a special aspect such as integrability, these methods either are limited to the ground state, or aim to provide numerical or analytical approximation for excited states in a special regime. A unified picture individually and globally analyzing the total spectrum from the ground state, excited states to the entire spectral structure in all regimes, especially with a transparent physics, is still lacking.

In the present work we create a
geometric perspective for the QRM by mapping to a curved nanowire system~\cite{Ying2016Ellipse,Ying2017curvedSC,Ying2020PRR,Gentile2022NatElec,Nagasawa2013Rings}. The curvature effectively induces spin-orbit coupling (SOC), imaginary out-of-plane field and quadratic curvature potential, with the curvature amplitude manipulatable by LMI strength and frequency ratio. This perspective is capable of revealing and perceiving the property universality and diversity from the ground state to the whole energy spectrum in a unified picture.
Indeed, we unveil a super university in the adiabatic regime which manifests a universal scaling of spin texture ubiquitously valid for different couplings, different frequencies and different energy levels. The super university becomes violated in large couplings, high frequencies and excited states, and diversity arises as displayed by non-adiabatic spin texture which is more directly observed in the curve frame. Still, in this diversified situation we again notice a new universality as the tangential spin shares the same swing directions. This seemingly simple new universality is actually not trivial, as we find a changeover of spin swing directions which turns out to be associated with the QPT of the QRM.  This finding endows a geometric non-adiabaticity interpretation and connotation to the QPT.
Furthermore, we find that finite curvature leads to a spectral bifurcation in the distribution of level spacings. We further clarify that, among the three curvature-induced terms, it is the SOC that is responsible for both the spectral bifurcation and spectral oscillation.
Our method and analysis paves a geometric way to gain deeper insight into LMIs.

The paper is organized as follows. Section~\ref{Sect-Model} introduces the
QRM. Section~\ref{Sect-local-basis} establishes the local basis which is
capable of capturing the QPT of the QRM. Section~\ref{Sect-Curve-Rabi}
creates the geometric perspective by mapping the QRM to a curved nanowire
system. Section~\ref{Sect-Univ-div} reveals the super universality in the adiabatic regime. In the curve
frame the changeover of non-adiabatic spin swing direction is noticed. The association with the QPT is unveiled, with the
analytical critical coupling and frequency obtained. Section~\ref{Sect-bifurcation}
demonstrates that the curvature leads to the spectral
bifurcation in the distribution of level spacings. The
curvature-induced SOC is clarified to be finally responsible for the spectral bifurcation
and oscillations.
Section~\ref{Sect-Conclusions} summarizes the conclusions and gives some
prospects from the geometric perspective.

\section{Model}

\label{Sect-Model}

The QRM~\cite{rabi1936,Rabi-Braak,Eckle-Book-Models} is a most fundamental
model for light-matter interactions, describing the coupling of a two-level
system (qubit) with a single-mode light field. Under the rotating-wave
approximation it is equivalent to the Jaynes-Cummings model~\cite%
{JC-model,JC-Larson2021}. The QRM and its extensions also have a broad
relevance, being a fundamental building block for quantum information and
quantum computation~\cite{Diaz2019RevModPhy,Romero2012,Stassi2020QuComput,Stassi2018,Macri2018},
applied in quantum metrology~\cite{Garbe2020,Montenegro2021-Metrology,Chu2021-Metrology,Garbe2021-Metrology,Ilias2022-Metrology, Ying2022-Metrology,YangZheng2023SciChina,Gietka2023PRL-Squeezing,Hotter2024-Metrology,Alushi2024PRL,Mukhopadhyay2024PRL,Mihailescuy2024, Ying-Topo-JC-nonHermitian-Fisher,*Ying-Topo-JC-nonHermitian-Fisher-Cover,Ying-g2hz-QFI-2024,*Ying-g2hz-QFI-2024-Cover, Ying-g1g2hz-QFI-2025,Ying2025g2A4,*Ying2025g2A4-Cover,Ying-g2Stark-QFI-2025,
Gietka2025PRL100802,Mihailescu2025CQMtutorial,QiuYi2025gA2,QiaoFeng2016AsymPolaron},
bridged to many-body systems~\cite{LiuM2017PRL,Irish2017}, connected to
models in condense matter~\cite{Kockum2019NRP} and couplings in cold atoms~%
\cite%
{LinRashbaBECExp2013Review,LinRashbaBECExp2011,LiuYing02025exoticSOC2Ring,*LiuYing02025exoticSOC2Ring-Cover,LiuYing02025KaleidoscopeDDI}%
. The Hamiltonian of the standard QRM reads
\begin{equation}
H_{R}=\omega {\hat{a}^{\dagger }\hat{a}}+\frac{\Omega }{2}\sigma
_{x}+g\sigma _{z}(\hat{a}^{\dagger }+\hat{a})  \label{eq:rabi}
\end{equation}%
where $\hat{a}^{\dagger }(\hat{a})$ is the bosonic creation (annihilation)
operator for the light field with frequency $\omega $. The two-level system
is represented by the Pauli matrices $\sigma _{x,y,z}$ with $\Omega $ being
the level splitting. The coupling strength is denoted by $g$. In recent
years, ultra-strong~\cite%
{Ciuti2005EarlyUSC,Aji2009EarlyUSC,Diaz2019RevModPhy,Kockum2019NRP,Wallraff2004,Gunter2009,Niemczyk2010, Peropadre2010,FornDiaz2017,Forn-Diaz2010,Scalari2012,Xiang2013,Yoshihara2017NatPhys,Kockum2017,Bayer2017DeepStrong,Qin2024PhysRep}
and deep-strong~\cite%
{WangYouJQ2023DeepStrong,Yoshihara2017NatPhys,Bayer2017DeepStrong} couplings
have been realized in superconducting circuit systems. Here we have adopted the spin
notation \cite{Irish2014,Ying2015} which is more convenient to represents
the two flux states in the superconducting flux-qubit circuit systems~\cite%
{flux-qubit-Mooij-1999} by $\sigma _{z}=\pm $. One can exchange $\sigma _{x}$
and $\sigma _{z}$ to retrieve conventional form of the QRM by a spin
rotation around the axis $\vec{x}+\vec{z}$.

\begin{figure*}[t]
\includegraphics[width=1.0\linewidth]{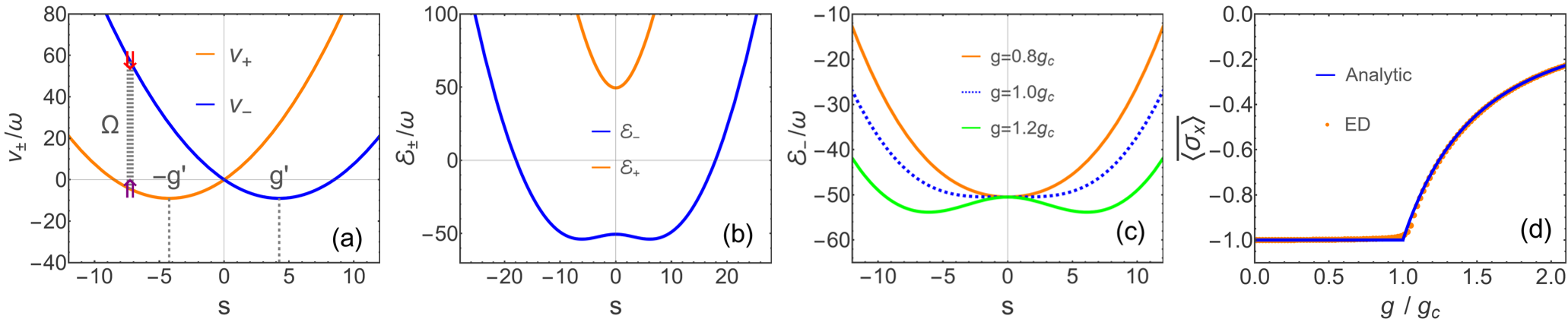}
\caption{Reliability of the adiabatic basis: Capturing the quantum phase transition (QPT).
(a) Description of QRM in effective $s$ space.
(b) Local energy ${\cal E}_{\pm}$.
(c) Dependence of lower-branch energy ${\cal E}_{-}$ around $g_c$.
(d) Comparison of the spin expectation $\overline{\langle \sigma _x\rangle}$ [Eqs.~\eqref{Sx-below-gc} and \eqref{Sx-above-gc}] with exact diagonalization (ED). Here $\omega=0.01 /\Omega$ in all panels and $g=1.2 g_c$ in (a) and (b). }
\label{Fig-V-rotated}
\end{figure*}
\begin{figure*}[t]
\includegraphics[width=1.0\linewidth]{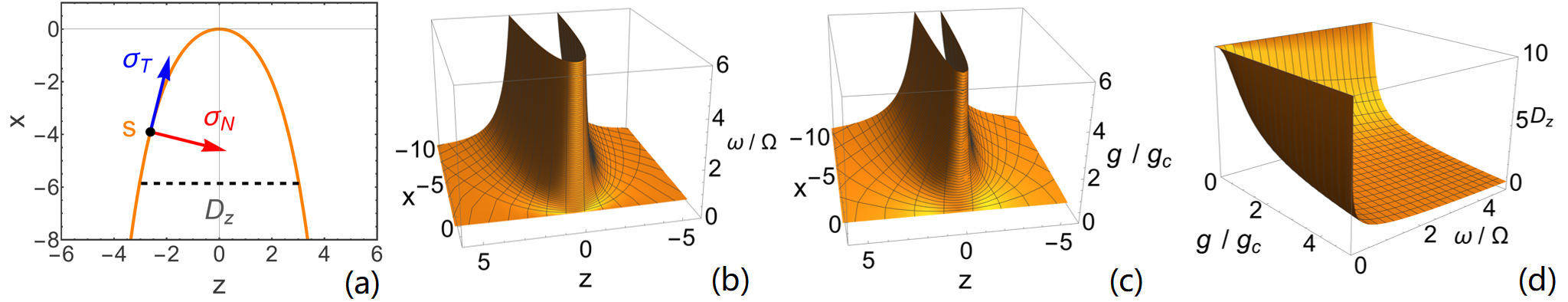}
\caption{Mapping the QRM to curved nanowire system with curvature controlled by the coupling or frequency.
(a) Effective geometric path of the QRM in the $z$--$x$ plane at $\omega=0.5\Omega$ and $g=0.8g_c$. (b) 3D plot for the
geometric path depending on coupling $g$ at $\omega=0.5\Omega$. (c) 3D plot for the geometric path
depending on frequency $\omega$ at $g=0.8g_c$. (d) 3D plot for the path width $D_z$
depending on $g$ and $\omega$ at $|s|=5$.}
\label{Fig-wire-2D3D}
\end{figure*}

\section{Locally diagonalized basis and the QPT of the QRM}
\label{Sect-local-basis}

\subsection{Locally diagonalized basis}

By the transformation $\hat{a}^{\dagger }=(\hat{s}-i\hat{p})/\sqrt{2}$, $%
\hat{a}=(\hat{s}+i\hat{p})/\sqrt{2}$, where $\hat{s}=s$ and $\hat{p}=-i\frac{%
\partial }{\partial s},$ the Hamiltonian can be rewritten in the position
space as%
\begin{equation}
H=\frac{\omega }{2}\hat{p}^{2}+H_{v}+H_{\Omega }+H_{g}+\varepsilon _{0}
\end{equation}%
where the first term is the effective kinetic energy, $H_{v}=\frac{1}{2}%
\omega s^{2}$ denotes the harmonic potential energy from the bosonic field, $%
H_{\Omega }=\frac{\Omega }{2}\sigma _{x}$ is the atomic level splitting
energy, $H_{g}=\sqrt{2}gs\sigma _{z}$ describes the coupling in spatial form
and $\varepsilon _{0}=-\frac{1}{2}\omega $ is a constant energy.
Experimentally in superconducting circuit systems the spin $\sigma _{z}=\pm $
can be represented by two flux states in a flux qubit system, while the
position $s$ and momentum $p$ can be simulated by the flux and charge of
Josephson junctions in another superconducting circuit coupled to the flux
qubit~\cite{flux-qubit-Mooij-1999,Bertet2005mixedModel,you024532}.

We collect the energy contributions apart from the kinetic part, $H_{\Omega }
$, $H_{v}$ and $H_{g}$, in matrix form in the $\sigma _{z}$ basis
\begin{eqnarray}
H_{v\Omega g} &=&H_{v}+H_{\Omega }+H_{g}  \nonumber \\
&=&\left(
\begin{array}{cc}
\frac{1}{2}\omega s^{2}+\sqrt{2}gs+\varepsilon _{0} & \frac{1}{2}\Omega  \\
\frac{1}{2}\Omega  & \frac{1}{2}\omega s^{2}-\sqrt{2}gs+\varepsilon _{0}%
\end{array}%
\right)   \nonumber \\
&=&\left(
\begin{array}{cc}
v_{+}+\varepsilon _{0} & \frac{1}{2}\Omega  \\
\frac{1}{2}\Omega  & v_{-}+\varepsilon _{0}
\end{array}%
\right) ,
\end{eqnarray}%
where $v_{\pm }=\frac{1}{2}\omega \left( s\pm g^{\prime }\right) ^{2}-\frac{1}{2}\omega g^{\prime 2}$ is
the spin-dependent potential and $g^{\prime }=\sqrt{2}g/\omega $. Figure \ref%
{Fig-V-rotated}(a) {}pictorializes $H_{v\Omega g}$ which has a clear
interpretation: The coupling separates the harmonic potentials to be $v_{\pm
}$ for the spin-up ($+$) and spin-down ($-$) components in the opposite
directions by displacements $-g^{\prime }$ and $+g^{\prime }$, while the
spin is flipping with the strength $\Omega $.

We can locally diagonalize $H_{v\Omega g}$~\cite{Ying-Stark-top,Ying2020-nonlinear-bias,Ying-2018-arxiv} as
\begin{equation}
H_{v\Omega g}\longrightarrow \left(
\begin{array}{cc}
{\cal E}_{+} & 0 \\
0 & {\cal E}_{-}
\end{array}
\right)
\end{equation}
with $\eta =\pm $ labels the two local energy branches
\begin{equation}
{\cal E}_{\eta }=\frac{1}{2}\omega s^{2}+\eta \sqrt{8^{2}g^{2}s^{2}+\Omega
^{2}}-\frac{1}{2}\omega .
\end{equation}%
The corresponding local basis takes the form
\begin{equation}
\left\vert \eta \right\rangle =\left(
\begin{array}{c}
\phi _{\eta +} \\
\phi _{\eta -}
\end{array}
\right) =\left(
\begin{array}{c}
\left( 2\sqrt{2}gs+\eta \sqrt{8g^{2}s^{2}+\Omega ^{2}}\right) /N_{\eta } \\
\Omega /N_{\eta }
\end{array}
\right)   \label{local-basis}
\end{equation}
with the normalization factor
\begin{equation}
N_{\eta }=\sqrt{16g^{2}s^{2}+2\Omega ^{2}+4\eta gs\sqrt{16g^{2}s^{2}+2\Omega
^{2}}}.
\end{equation}

\subsection{Reliability of the local basis: Capturing the quantum phase
transition (QPT)}

In the low-frequency limit, $\omega /\Omega \rightarrow 0,$ the QRM
manifests a finite-component QPT in the ground state. In the locally
diagonalized basis, the critical point of the QPT can be readily extracted.
Indeed, in such a limit, the kinetic energy is of an order $\omega $, while
around the transition the potential energy is of an order $\Omega $; the
ratio between the wave-packet size and the potential size is of order
$\sqrt{\omega /\Omega }$~\cite{Ying2020-nonlinear-bias}. In such a situation, one
can apply the semiclassical approximation by regarding the particle as a
classical mass point in position space while keeping the quantum part in the
spin space. Thus, the ground state has a vanishing kinetic energy so that
the energy reduces to ${\cal E}_{\eta }$. We plot ${\cal E}_{\pm }$ in Fig.~\ref{Fig-V-rotated}(b)
which shows that ${\cal E}_{+}$ is much higher than ${\cal E}_{-}$ at low frequencies,
separated by an energy gap of an order $\Omega $.
The ground state finally lies in the ${\cal E}_{-}$ branch. Before
the transition ${\cal E}_{-}$ is in a single-well structure [orange (gray)
line in Fig.\ref{Fig-V-rotated}(c)], while after the transition ${\cal E}_{-}$
exhibits a double-well profile [green (light gray) line in Fig.~\ref{Fig-V-rotated}(c)].
The critical coupling at the transition [blue dashed
line in Fig.\ref{Fig-V-rotated}(c)] can be directly extracted by the
vanishing point of the second derivative of ${\cal E}_{-}$
\begin{equation}
\frac{\partial ^{2}{\cal E}_{-}}{\partial s^{2}}=\omega -\frac{4g^{2}\Omega
^{2}}{\left( 8g^{2}s^{2}+\Omega ^{2}\right) ^{3/2}}=0
\end{equation}%
at the origin $s=0$, which leads to the transition point at \cite{Ashhab2013},
\begin{equation}
g=g_{c}=\frac{\sqrt{\omega \Omega }}{2}.  \label{gC0}
\end{equation}%
This indicates that the locally diagonalized basis can capture the leading
physics of the QPT in the low-frequency limit.

Furthermore, concerning the critical properties around the transition, in
the semiclassical approximation the lowest energy can be extracted by
minimization of the ${\cal E}_{-}$%
\begin{equation}
\frac{\partial {\cal E}_{-}}{\partial s}=\omega s-\frac{4g^{2}s}{\sqrt{8g^{2}s^{2}+\Omega ^{2}}}=0,
\end{equation}
which yields the optimal position
\begin{eqnarray}
s &=&s_{\min }=0,\quad \text{for }g<g_{c}; \\
s &=&s_{\min }=\pm \frac{\sqrt{16g^{4}-\omega ^{2}\Omega ^{2}}}{2\sqrt{2}
g\omega },\quad \text{for }g>g_{c}
\end{eqnarray}
and the spin expectations%
\begin{equation}
\overline{\langle \sigma _{z}\rangle }=\overline{\langle \sigma _{y}\rangle }
=0
\end{equation}
and
\begin{eqnarray}
\overline{\langle \sigma _{x}\rangle } &=&-1,\quad \text{for }g<g_{c}; \label{Sx-below-gc}\\
\overline{\langle \sigma _{x}\rangle } &=&-\frac{g_{c}^{2}}{g^{2}},\quad
\text{for }g>g_{c}.   \label{Sx-above-gc}
\end{eqnarray}
We plot $\overline{\langle \sigma _{x}\rangle }$ in Fig.~\ref{Fig-V-rotated}
(d) which shows the above analytical result (blue solid line) in good
agreement with the exact diagonalization (ED) results (orange dots) in all coupling regimes.

\section{Geometric perspective on the QRM}

\label{Sect-Curve-Rabi}

\subsection{Picking up the kinetic energy at finite frequencies}

At finite frequencies, the semiclassical approximation will be violated, in
such a situation we need to pick up the kinetic energy. Actually the local
energy ${\cal E}_{\eta }$ can be regarded as an effective potential. A
direct way to include the kinetic energy is to add the kinetic term to the
effective potential in each energy branch so that the eigen equation reads
\begin{equation}
\left( \frac{\omega }{2}\hat{p}^{2}+{\cal E}_{\eta }\right) \psi _{\eta
}\left( s\right) =E_{\eta }\psi _{\eta }\left( s\right) .
\label{EigenEq-T+V}
\end{equation}%
Note that, apart from the quadratic term, ${\cal E}_{\eta }$ contains a
square root function $\sqrt{8^{2}s^{2}+\Omega ^{2}}$ which is highly
nonlinear and not exactly tractable in the basis of the photon-number Fock
states. Nevertheless, rather than expansion on Fock states, we can treat the
eigen equation Eq.~\eqref{EigenEq-T+V} in spatial dimension by finite
difference method which is capable of keeping the full form of the highly
nonlinear effective potential. We will see that Eq.~\eqref{EigenEq-T+V} works
only in the low-frequency limit and is actually an incomplete inclusion of
the kinetic energy.

\subsection{Non-adiabatic curvature contributions and mapping to nanowire
system}

As a matter of fact, full contributions of the kinetic energy should include
not only the kinetic energy in position space but also the rotation of the
local basis. The exact form of wave function should be%
\begin{equation}
\left\vert \psi \right\rangle =\sum_{\eta =\pm }\psi _{\eta }\left( s\right)
\left\vert \eta \right\rangle
\end{equation}%
in the locally diagonalized basis $\left\vert \eta \right\rangle $ in Eq.~%
\eqref{local-basis}. Note that $\left\vert \eta \right\rangle $ is also
position-dependent so that the derivative of the wave function should acting
not only on the spatial part $\psi _{\eta }\left( s\right) $ but also on
spin part (basis $\left\vert \eta \right\rangle $)
\begin{equation}
\partial _{s}\left\vert \psi \right\rangle =\sum_{\eta =\pm }\psi _{\eta
}^{\prime }\left( s\right) \left\vert \eta \right\rangle +\sum_{\eta =\pm
}\psi _{\eta }\left( s\right) \partial _{s}\left\vert \eta \right\rangle
\end{equation}%
where the prime represents the derivative with respect to $s$. Therefore,
the Laplacian is actually composed of three terms
\begin{equation}
\partial _{s}^{2}\left\vert \psi \right\rangle =\sum_{\eta =\pm }\left( \psi
_{\eta }^{\prime \prime }\left( s\right) \left\vert \eta \right\rangle
+2\psi _{\eta }^{\prime }\left( s\right) \partial _{s}\left\vert \eta
\right\rangle +\psi _{\eta }\left( s\right) \partial _{s}^{2}\left\vert \eta
\right\rangle \right) .
\end{equation}%
Explicitly we find
\begin{eqnarray}
\partial _{s}\left\vert \eta \right\rangle  &=&K_{\eta }\left( s\right)
\left\vert -\eta \right\rangle ,  \label{dbasis-ds-1} \\
\partial _{s}^{2}\left\vert \eta \right\rangle  &=&K_{\eta }^{\prime }\left(
s\right) \left\vert -\eta \right\rangle +K_{\eta }\left( s\right)
^{2}\left\vert \eta \right\rangle ,  \label{dbasis-ds-2}
\end{eqnarray}%
where
\begin{equation}
K_{\eta }\left( s\right) =-\eta \frac{1}{2}K\left( s\right)
\end{equation}%
and
\begin{equation}
K\left( s\right) =\frac{2\sqrt{2}g\Omega }{8g^{2}s^{2}+\Omega ^{2}},\quad
K^{\prime }\left( s\right) =\frac{32\sqrt{2}g^{3}s\Omega }{\left(
8g^{2}s^{2}+\Omega ^{2}\right) ^{2}}.
\end{equation}

The arising of the $\partial _{s}\left\vert \eta \right\rangle $ and $%
\partial _{s}^{2}\left\vert \eta \right\rangle $ terms means that the state
does not adiabatically follow the local basis, thus being non-adiabatic
effects. From Eqs. (\ref{dbasis-ds-1}) and (\ref{dbasis-ds-1}) we see that
these non-adiabatic effects will couple the two wave-function components $%
\psi _{\eta }$ in the local basis, in contrast to the decoupled form in (\ref%
{EigenEq-T+V}). Then, we get the full eigen equation in the local basis
\begin{eqnarray}
E\sum_{\eta } &&\psi _{\eta }\left( s\right) \left\vert \eta \right\rangle =-%
\frac{\omega }{2}\sum_{\eta =\pm }\psi _{\eta }^{\prime \prime }\left(
s\right) \left\vert \eta \right\rangle +2K_{\eta }\left( s\right) \psi
_{\eta }^{\prime }\left( s\right) \left\vert -\eta \right\rangle   \nonumber
\\
&&+K_{\eta }^{\prime }\left( s\right) \psi _{\eta }\left( s\right)
\left\vert -\eta \right\rangle +\left[ K_{\eta }\left( s\right) ^{2}+{\cal E}%
_{\eta }\right] \psi _{\eta }\left( s\right) \left\vert \eta \right\rangle .
\label{EigenEq-new}
\end{eqnarray}%
Eq.(\ref{EigenEq-new}) can be described by the new Hamiltonian
\begin{eqnarray}
\widetilde{H} &=&\frac{\omega }{2}\widetilde{p}^{2}+\alpha _{{\rm SOC}
}\left( s\right) \widetilde{\sigma }_{y}\widetilde{p} - \frac{1}{2}\widetilde{
\Omega }_{y}\left( s\right) i\widetilde{\sigma }_{y}+v_{K}\left( s\right)
\nonumber \\
&&+\frac{\left[ {\cal E}_{+}\left( s\right) -{\cal E}_{-}\left( s\right)
\right] }{2}\widetilde{\sigma }_{z}+\frac{\left[ {\cal E}_{+}\left( s\right)
+{\cal E}_{-}\left( s\right) \right] }{2},  \label{H-curve}
\end{eqnarray}%
where%
\begin{eqnarray}
\alpha _{{\rm SOC}}\left( s\right)  &=&\frac{\omega }{2}K\left( s\right) , \\
\widetilde{\Omega }_{y}\left( s\right)  &=&\frac{\omega }{2}K^{\prime
}\left( s\right) , \\
v_{K}\left( s\right)  &=&\frac{\omega }{2}\frac{1}{4}K\left( s\right) ^{2}.
\end{eqnarray}%
As will be more clearly demonstrated in next section, this new Hamiltonian
in Eq.~(\ref{H-curve}) actually describes a curved nanowire system \cite%
{Ying2016Ellipse,Ying2017curvedSC,Ying2020PRR,Gentile2022NatElec,Nagasawa2013Rings}%
, with an inhomogeneous Rashba-type SOC in the curve frame, and $K\left(
x\right) $ is the curvature along the curved nanowire. Differently from the
conventional in-plane Rashba SOC in nanowires, the Rashba SOC here is
out-of-plane.

In the new Hamiltonian $\widetilde{H}$, the tilde over the spin operator
indicates action on the rotated basis $\left\vert \eta \right\rangle $ in
the curve frame, still represented by Pauli matrices as the unrotated ones $%
\sigma _{x,y,z}$ in the lab frame
\begin{equation}
\widetilde{\sigma }_{z}=\left(
\begin{array}{cc}
1 & 0 \\
0 & -1%
\end{array}%
\right) ,\ \widetilde{\sigma }_{z}=\left(
\begin{array}{cc}
0 & 1 \\
1 & 0%
\end{array}%
\right) ,\ \widetilde{\sigma }_{y}=\left(
\begin{array}{cc}
0 & -i \\
i & 0%
\end{array}%
\right) .
\end{equation}%
Therefore, the tilded momentum operator
\begin{equation}
\widetilde{p}=-i\partial _{s}
\end{equation}%
now acts only on the basis coefficient $\psi _{\eta }\left( s\right) $, not
any more on the basis $\left\vert \eta \right\rangle $ itself which has
become position-independent in the curve frame.

In such a formalism, we see in Eq. (\ref{H-curve}) that the curvature
induces three terms: an effective Rashba SOC (the $\alpha _{{\rm SOC}}$
term), an imaginary out-of-plane field (the $\widetilde{\Omega }_{y}$ term),
and a quadratic curvature potential $v_{K}\left( x\right) $. These
curvature-induced effects, especially the SOC, play some key roles in the
non-adiabatic properties and the variation of spectral structure.

\begin{figure*}[t]
\includegraphics[width=1.0\linewidth]{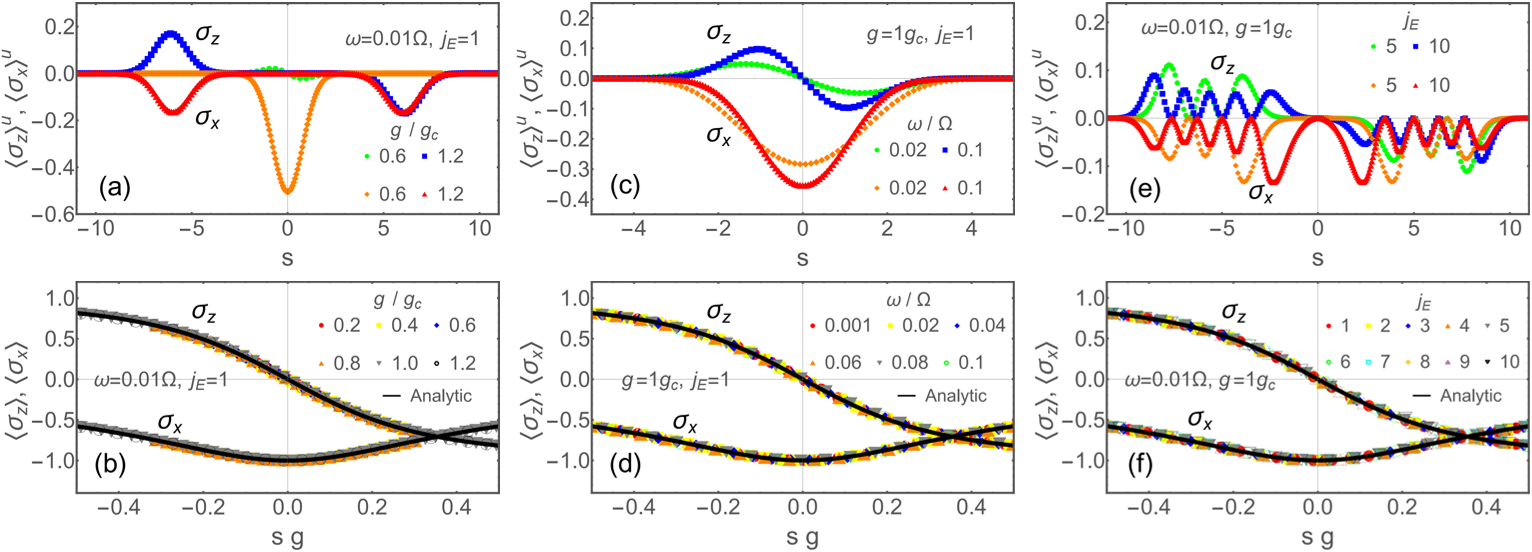}
\caption{Super adiabatic universality: Universal scaling relation of the spin texture $\langle \sigma _{z,x}\rangle$ [(b), (d) and (f)] 
ubiquitously valid for different couplings $g$ [(a) and (b)], different frequencies $\omega$ [(c) and (d)], different energy levels $j_E$ [(e) and (f)].
(a),(b) Different $g/g_c$ at $\omega=0.01$ for $j_E=1$ (ground state). 
(c),(d) Different $\omega/\Omega$ at $g=1.0g_c$ for $j_E=1$. 
(e),(f) Different excited states $j_E$ at $\omega=0.01$ and $g=1.0g_c$. 
In (a),(c) and (e) the diversified spin textures are unscaled (with superscript $u$), while in (b),(d) and (f) they are normalized by the density $\rho$ and position scaled with $sg$. All the spin textures (symbols) collapse into a same line [(b),(d) and (f)] captured by the analytic one (solid line) [Eqs.~\eqref{Analytic-Sz} and \eqref{Analytic-Sx}].
}
\label{Fig-SuperUniv}
\end{figure*}

\subsection{Geometric structure and manipulation the curvature by coupling
and frequency}

The geometric structure of the new Hamiltonian $\widetilde{H}$ can be more
clearly seen as $ds$ is the infinitesimal distance along a nanowire in the $z
$--$x$ plane
\begin{equation}
dz=ds\cos \varphi ,\quad dx=ds\sin \varphi ,
\end{equation}%
while the tangential and normal directions of the wire are represented by
the direction vectors
\begin{eqnarray}
\vec{T} &=&\vec{z}\cos \varphi +\vec{x}\sin \varphi , \\
\vec{N} &=&\vec{z}\sin \varphi -\vec{x}\cos \varphi .
\end{eqnarray}%
We illustrate the nanowire in Fig.\ref{Fig-wire-2D3D}(a). The curvature $%
K\left( s\right) $ connects the tangential and normal directions and their
variations by the relations
\begin{eqnarray}
\frac{\partial \vec{T}}{\partial s} &=&K\left( s\right) \vec{N}, \\
\frac{\partial \vec{N}}{\partial s} &=&-K\left( s\right) \vec{T}.
\end{eqnarray}%
In fact, the curvature is the variation speed of the rotation, as indicated
by the explicitly relation
\begin{equation}
K\left( s\right) =-\varphi ^{\prime }\left( s\right) .  \label{K=dphi}
\end{equation}%
Correspondingly, the spin component in the tangential and normal directions
are given by
\begin{eqnarray}
\widetilde{\sigma }_{x} &\longrightarrow &\sigma _{T}=\sigma _{z}\cos
\varphi +\sigma _{x}\sin \varphi , \\
\widetilde{\sigma }_{z} &\longrightarrow &\sigma _{N}=\sigma _{z}\sin
\varphi -\sigma _{x}\cos \varphi
\end{eqnarray}%
where $\widetilde{\sigma }_{x},\widetilde{\sigma }_{z}$ are represented in
the rotation frame while $\sigma _{T},\sigma _{N}$ in the lab frame. The
counterparts for the variations of the spin read
\begin{eqnarray}
\partial _{s}\sigma _{T} &=&K\left( s\right) \sigma _{N},  \label{dSpinT} \\
\partial _{s}\sigma _{N} &=&-K\left( s\right) \sigma _{T}.
\end{eqnarray}%
Explicitly we obtain the local spin expectation
\begin{eqnarray}
\langle \sigma _{z}\rangle _{\eta } &=&\frac{\langle \sigma _{z}\rangle
_{\eta }^{u}}{\rho _{\eta }}=\frac{\eta 2\sqrt{2}gs}{\sqrt{%
8g^{2}s^{2}+\Omega ^{2}}},  \label{Analytic-Sz} \\
\langle \sigma _{x}\rangle _{\eta } &=&\frac{\langle \sigma _{x}\rangle
_{\eta }^{u}}{\rho _{\eta }}=\frac{\eta \Omega }{\sqrt{8g^{2}s^{2}+\Omega
^{2}}},  \label{Analytic-Sx}
\end{eqnarray}%
where the superscript $u$ denotes the expectation unscaled by the local
density $\rho _{\eta }$
\begin{eqnarray}
\langle \sigma _{z}\rangle _{\eta }^{u} &=&\phi _{\eta +}\phi _{\eta +}-\phi
_{\eta -}\phi _{\eta -}, \\
\langle \sigma _{x}\rangle _{\eta }^{u} &=&\phi _{\eta +}\phi _{\eta -}+\phi
_{\eta -}\phi _{\eta +}, \\
\rho _{\eta } &=&\phi _{\eta +}\phi _{\eta +}+\phi _{\eta -}\phi _{\eta -}.
\end{eqnarray}%
As the adiabatic basis, we set
\begin{equation}
\langle \sigma _{z}\rangle _{\eta }=\sin \varphi ,\quad \langle \sigma
_{x}\rangle _{\eta }=-\cos \varphi   \label{SxzCosSin}
\end{equation}%
for $\eta =-$, so that the basis spin is oriented in the normal
direction. Combining Eq.(\ref{dSpinT}) and the basis exchange action $%
\widetilde{\sigma }_{x}\left\vert \eta \right\rangle =\left\vert -\eta
\right\rangle $ in the lab frame, we can find the relation of the wire
curvature and the basis rotation rate%
\begin{equation}
K\left( s\right) =K_{-\eta }\left( s\right) -K_{\eta }\left( s\right)
=2K_{-\eta }\left( s\right)   \label{K=2K}
\end{equation}%
where we set $\eta =+$.

Either from relation (\ref{SxzCosSin}) or Eq.(\ref{K=dphi}) we can extract the explicit rotation angle%
\begin{equation}
\varphi \left( s\right) =-\arctan \frac{2\sqrt{2}gs}{\Omega }.
\end{equation}%
The geometric path are then analytically available as
\begin{eqnarray}
x &=&\frac{\Omega -\sqrt{8g^{2}s^{2}+\Omega ^{2}}}{2\sqrt{2}g}, \\
z &=&\frac{\Omega }{2\sqrt{2}g}\tanh ^{-1}\left( \frac{2\sqrt{2}gs}{\sqrt{%
8g^{2}s^{2}+\Omega ^{2}}}\right) ,
\end{eqnarray}%
where $\tanh ^{-1}\left( x\right) =\frac{1}{2}\ln \frac{1+x}{1-x}$ is the
inverse hyperbolic tangent function.

Besides the example of the geometric path of the effective curved nanowire
in Fig.\ref{Fig-wire-2D3D}(a), we also illustrate the 3D plots for
dependence of the geometric path on the frequency ratio $\omega /\Omega $
[panel (b)], the coupling $g/g_{c}$ [panel (c)], as well as the path width
[panel (d)]%
\begin{equation}
D_{z}=\left\vert z\left( s\right) -z\left( -s\right) \right\vert =\frac{%
\Omega }{\sqrt{2}g}\tanh ^{-1}\left( \frac{2\sqrt{2}g\left\vert s\right\vert
}{\sqrt{8g^{2}s^{2}+\Omega ^{2}}}\right)
\end{equation}%
as a function of $\omega /\Omega $ and $g/g_{c}$ at a fixed $s$. From the
narrower profile in Figs.\ref{Fig-wire-2D3D}(b) and \ref{Fig-wire-2D3D}(c)
or from the smaller $D_{z}$ in Fig.\ref{Fig-wire-2D3D}(d) with a larger
value of $\omega /\Omega $ or $g/g_{c}$, we see that the nanowire is more
bent with a higher frequency or a stronger coupling. So one can manipulate
the geometric structure and curvature, and thus the curvature induced
properties, of the QRM by the variation of frequency or coupling strength.

\section{Universality and diversity in the geometric perspective}

\label{Sect-Univ-div}

\subsection{Spin texture in the curve frame and classification of adiabatic
and non-adiabatic states}

The spin texture of a state in the curve frame can be obtained by%
\begin{eqnarray}
\langle \sigma _{T}\rangle  &=&\langle \sigma _{z}\rangle \cos \varphi
+\langle \sigma _{x}\rangle \sin \varphi   \nonumber \\
&=&-\langle \sigma _{z}\rangle \langle \sigma _{x}\rangle _{\eta }+\langle
\sigma _{x}\rangle \langle \sigma _{z}\rangle _{\eta },  \label{SpinT-by-Sxz}
\\
\langle \sigma _{N}\rangle  &=&\langle \sigma _{z}\rangle \sin \varphi
-\langle \sigma _{x}\rangle \cos \varphi   \nonumber \\
&=&\langle \sigma _{z}\rangle \langle \sigma _{z}\rangle _{\eta }+\langle
\sigma _{x}\rangle \langle \sigma _{x}\rangle _{\eta },  \label{SpinN-by-Sxz}
\end{eqnarray}%
where $\langle \sigma _{z}\rangle $ and $\langle \sigma _{x}\rangle $ are
the spin texture in the lab frame. When the amplitude of $\langle \sigma
_{N}\rangle $ is close to the unity the spin is basically aligned along the
normal direction, the state can be regarded as in the adiabatic regime;
Otherwise, when the amplitude of $\langle \sigma _{N}\rangle $ is
considerably deviated from the unity, the state is in the non-adiabatic
regime. The above formulations in the curve frame facilitate the
understanding and classification of our two following findings: At low
frequencies we find a super universality of spin texture which belongs to
the adiabatic cases with unity $\langle \sigma _{N}\rangle $, while at
finite frequencies amidst the breaking down of the super universality we
notice a new non-adiabatic universality in the spin swing direction of $%
\langle \sigma _{T}\rangle $ and its association with the QPT of the QRM, as
addressed in the following sections.

\subsection{Super universality of spin texture in the adiabatic regime}

In low frequencies and for low-energy levels, we find a super university in
the sense that the universal scaling relation of spin texture is
ubiquitously valid for different couplings, different frequencies and
different energy levels. In Fig.~\ref{Fig-SuperUniv}(a) we illustrate some
examples of unscaled spin texture $\langle \sigma _{x}\rangle ^{u}$ and $%
\langle \sigma _{z}\rangle ^{u}$ at different couplings for the ground state
($j_{E}=1$) at a fixed frequency $\omega =0.01\Omega $, as calculated by
ED~\cite{Ying2020-nonlinear-bias,Ying-Stark-top,*Ying-Stark-top-Cover,Ying-g1g2hz-QFI-2025}.
Before the QPT ($g<g_{c}$) the profile of $\langle \sigma _{x}\rangle ^{u}$
is single-peaked, while beyond the QPT ($g>g_{c}$) $\langle \sigma
_{x}\rangle ^{u}$ manifests a double-peaked structure. In contrast to the
symmetric $\langle \sigma _{x}\rangle ^{u}$, the shape $\langle \sigma
_{z}\rangle ^{u}$ is antisymmetric with respect to $s$ and always has two
peaks with opposite signs which are adjacent before the transition and
become separated beyond the transition. However, despite their diversity,
when we normalize all the spin textures by the local density $\rho \left(
s\right) $ and scale the position by the coupling as $sg$, all the coupling cases
collapse into a same line, as in Fig.~\ref{Fig-SuperUniv}(b). In contrast to
the conventional critical universality~\cite{LiuM2017PRL,Irish2017,Ying-2021-AQT,Ying-Stark-top}
in which different values of parameter such as anisotropy share the same critical variations but never collapse with respect to
the couplings, the universality for different couplings revealed here is a
much stronger one.

The same scaling behavior happens for different frequencies, with unscaled
spin textures illustrated in Fig.~\ref{Fig-SuperUniv}(c) and collapsed ones
in Fig.~\ref{Fig-SuperUniv}(c). Such universality is well preserved for the
ground state from the low-frequency limit up to $\omega =0.1\Omega $. More
surprising is the validity of the universality over different excited states
which have more oscillations in the unscaled spin texture as in Fig.~\ref{Fig-SuperUniv}(e)
but still collapse to the same scaled ones in Fig.~\ref{Fig-SuperUniv}(f).

All the collapsed lines are well captured by Eqs.~(\ref{Analytic-Sz}) and (\ref{Analytic-Sx}) (solid lines)
so the spin textures of the eigen states are
basically equal to that of the local basis
\begin{equation}
\langle \sigma _{z}\rangle \doteq \langle \sigma _{z}\rangle _{\eta },\quad
\langle \sigma _{x}\rangle \doteq \langle \sigma _{x}\rangle _{\eta }.
\end{equation}%
Therefore, from Eq.~(\ref{SpinN-by-Sxz}) we have normal spin component equal
to unity
\begin{equation}
\langle \sigma _{N}\rangle \doteq \langle \sigma _{z}\rangle _{\eta
}^{2}+\langle \sigma _{x}\rangle _{\eta }^{2}=1,
\end{equation}
which belongs to the adiabatic case.

\begin{figure*}[t]
\includegraphics[width=1\linewidth]{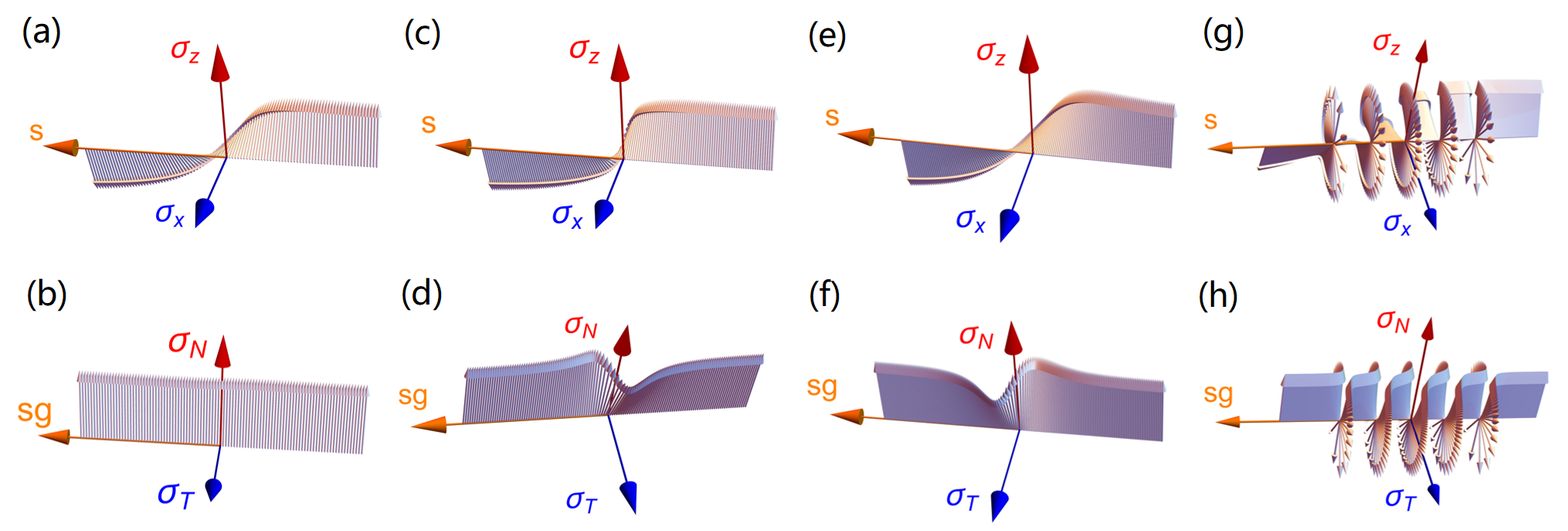}
\caption{Non-adiabatic 3D spin texture [(c)--(h)] compared with adiabatic one [(a) and (b)] in the lab frame [(a),(c),(e) and (g)] and curve frame [(b),(d),(f) and (h)].
(a),(b) $\omega=0.1\Omega$, $g=1.0 g_c$ and $j_E=1$.
(c),(d) $\omega=0.1\Omega$ and $g=3.0 g_c$ and $j_E=1$.
(e),(f) $\omega=2.0\Omega$ and $g=1.0 g_c$ and $j_E=1$.
(g),(h) $\omega=0.1\Omega$ and $g=1.0 g_c$ and $j_E=6$.
In the curve frame [(b),(d),(f) and (h)] it is direct to see the non-adiabaticity arising when the spin trajectory deviates from the normal direction (labeled by $\sigma_N$), as in a large coupling (d), high frequency (f) or excited state (h).
}
\label{Fig-non-adiabatic}
\end{figure*}
\begin{figure*}[t]
\includegraphics[width=1\linewidth]{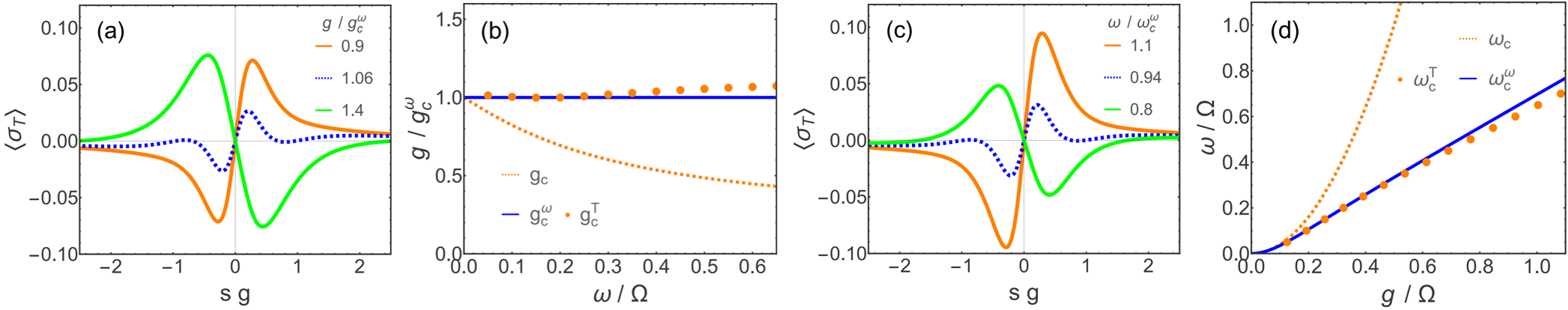}
\caption{Changeover of non-adiabatic spin-swing direction in association with the QPT of the QRM.
(a) Spin swing in the tangential component $\langle \sigma _T \rangle$ around the QPT: $g/g_c^{\omega}=0.9$ [orange (gray) solid], $1.06$ (blue dotted]) and $1.4$ [green (light gray) solid] at $\omega=0.5\Omega$.
(b) Frequency dependence of critical coupling $g_c^T$ (ED, dots) of spin swing changeover in agreement with the analytical critical coupling $g_c^{\omega}$ (blue solid line) of the QPT.
(c) Spin swing in $\langle \sigma _T \rangle$ around the critical frequency: $\omega/\omega_c^{\omega}=1.1$ [orange (gray) solid], $0.94$ (blue dotted) and $0.8$ [green (light gray) solid] at $g=0.8g_c^{\omega}$.
(d) Coupling dependence of critical frequency $\omega_c^{T}$ (ED, dots) of the tangential spin-swing changeover in agreement with the analytical critical coupling $\omega_c^{\omega}$ of the QPT. Here, $g_c^{\omega}$ and $\omega_c^{\omega}$ at finite frequencies are analytically expressed in Eqs.~\eqref{gCw} and \eqref{wCw}, while the ones in the low-frequency limit $g_c$ and $\omega_c$ [dotted lines in (b) and (d)] are given in Eqs.~\eqref{gC0} and \eqref{wC0}.
}
\label{Fig-sT-transition}
\end{figure*}

\subsection{Non-adiabaticity arising in strong coupling, high frequency and
excited states}

The super adiabatic university holds for low frequencies and low-energy
levels, while non-adiabaticity arises in very strong couplings, high
frequencies and excited states. In Fig.~\ref{Fig-non-adiabatic} we illustrate
the non-adiabaticity in these situations. To see the non-adiabaticity more
clearly we show the local spin texture perpendicular to the wire distance $s$
or scaled one $sg$, with panels (a), (c), (e) and (g) in the lab frame while
panels (b), (d), (f) and (h) in the curve frame. As an adiabatic reference,
Figures~\ref{Fig-non-adiabatic}(a) and \ref{Fig-non-adiabatic}(b) give the
case in the ground state $j_{E}=1$ at $g=1.0g_{c}$ and $\omega =0.1\Omega $
which is at the edge of adiabatic regime. We see that the spin texture is
indeed aligned along the normal direction in Fig.~\ref{Fig-non-adiabatic}(b) at
all wire distances. However, deviations from the normal direction show up
and the spin trajectory is swinging when a very large coupling is present [$g=3.0g_{c}$ in
Fig.~\ref{Fig-non-adiabatic}(d)] or a high frequency is set [$\omega
=2.0\Omega $ in Fig.~\ref{Fig-non-adiabatic}(f)]. And now in an excited
state the spin trajectory is even strongly winding as in Fig.~\ref{Fig-non-adiabatic}(h).

\begin{figure*}[t]
\includegraphics[width=1.0\linewidth]{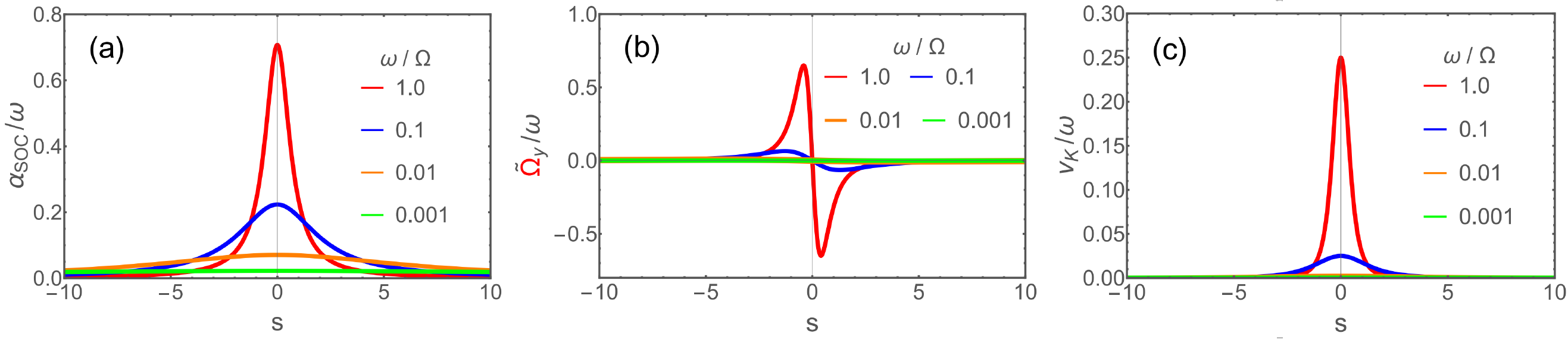}
\caption{Curvature-induced three effective terms at different frequencies: (a) spin-orbit coupling $%
\protect\alpha _{\mathrm{SOC}}$, (b) out-of-plane field $\widetilde{\Omega }%
_{y}$, (c) quadratic curvature potential $v_{K}\left( x\right) $. Here $g=1.0g_c$.}
\label{Fig-K-dK-K2}
\end{figure*}

The non-adiabaticity can be readily understood from the new Hamiltonian $\widetilde{H}$ in \eqref{H-curve} in the curve frame,
as the curvature-induced SOC,
$\alpha _{{\rm SOC}}\left( s\right) \widetilde{\sigma }_{y}\widetilde{p}$,
involves the spin $\widetilde{%
\sigma }_{y}$ which plays the role of rotating the spin in $\langle \sigma
_{N}\rangle $--$\langle \sigma _{T}\rangle $ plane. Note that the rotation
strength $\alpha _{{\rm SOC}}\left( s\right) $ is proportional to the
curvature $K\left( s\right) $ which is small at low frequencies but becomes
considerable at finite frequencies, as shown in Fig.~\ref{Fig-K-dK-K2}(a).
Therefore, the spin swing at low frequencies is negligible thus appearing to
be adiabatic, while at high frequencies the spin swing becomes obvious. The
coupling also strengthens the curvature, as shown in Fig.~\ref{Fig-wire-2D3D}%
(b), thus playing a similarly stronger rotation role when the coupling $g$ is large.
The SOC is also proportional to the momentum $\widetilde{p}$ whose amplitude
grows in the excited states, consequently the rotation effect is enhanced in
the excited states even if the coupling $g$ and frequency remains untuned.
Of course, in the low-frequency limit for this momentum-enhanced SOC to take
effect it needs higher excited states to compensate the effect weakening due
to smaller curvature, which in turn accounts for the validity of the super
adiabatic universality in excited states at $\omega =0.01\Omega $ in Fig.~\ref{Fig-SuperUniv}(f).

\begin{figure*}[t]
\includegraphics[width=1.0%
\linewidth]{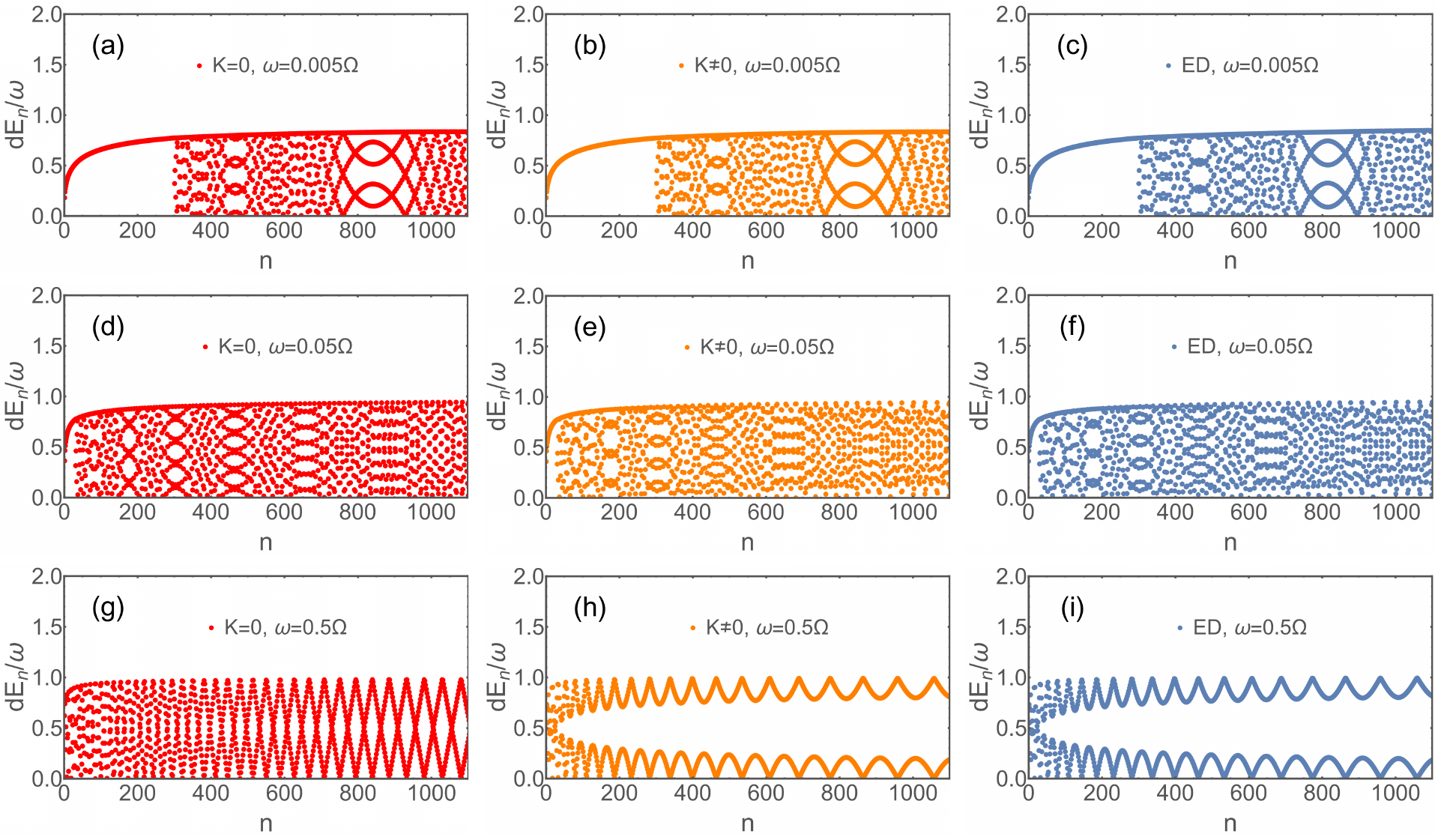}
\caption{Curvature-induced variation of the spectral structure. Energy level spacing $dE_{n}=E_{n+1}-E_{n}$ in the absence of
curvature [$K=0$, (a),(d),(g)] and in the presence of curvature [$K \neq 0$,
(b),(e),(h)], compared to exact diagonalization (ED) [(c),(f),(i)] at finite
frequencies. (a)--(c) $\protect\omega=0.005 \Omega$, (d)--(f) $\protect\omega%
=0.05 \Omega$, (g)--(i) $\protect\omega=0.5 \Omega$. Here the coupling is set to be $g=1.0g_c$.}
\label{Fig-dEn-vs-ED}
\end{figure*}

\begin{figure*}[t]
\includegraphics[width=1.0\linewidth]{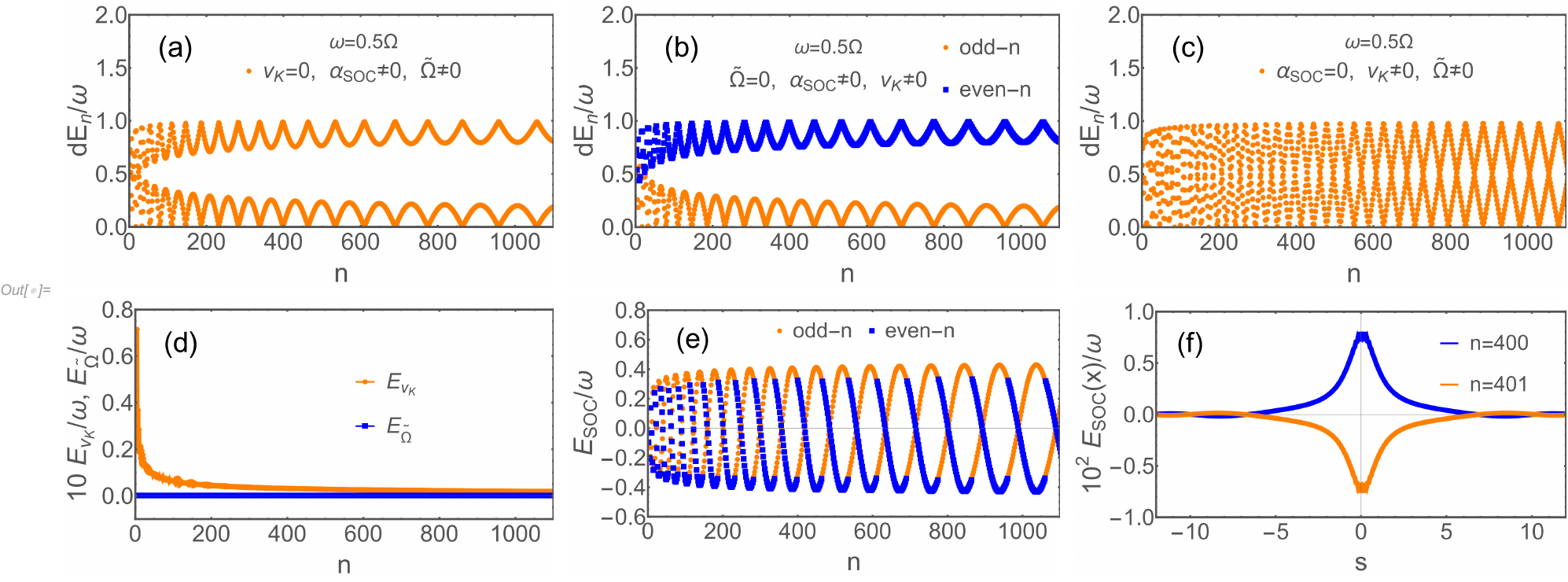}
\caption{Origin of the spectral bifurcation and oscillations at finite frequencies: The curvature-induced SOC is responsible for both. Level spacing $dE_{n}$ by setting (a) $%
v_K(x)=0$, (b) $\tilde{\Omega}(x)=0$, and (c) $\protect\alpha_{\mathrm{SOC}%
}=0$, at $\protect\omega =0.5\Omega$. The spectral bifurcation distribution
maintains in (a) and (b) as in Figs.\ref{Fig-dEn-vs-ED}(h)(i) but
breaks down in (c). The SOC energy is oscillating (e), while the other energies are not (d). }
\label{Fig-SOC-dEn}
\end{figure*}

\subsection{Changeover of non-adiabatic spin swing direction in association
with the QPT}

Although in Figs.~\ref{Fig-non-adiabatic}(d) and \ref{Fig-non-adiabatic}(f)
the modes of the spin swing seem similar, we notice the difference in the
swing direction. In fact, in the case of Fig.~\ref{Fig-non-adiabatic}(d) the
spin trajectory in the negative (positive) axis of $sg$ swings to the
negative (positive) axis of tangential spin $\langle \sigma _{T}\rangle $,
while in Fig.~\ref{Fig-non-adiabatic}(d) the swing direction is reversed. It
turns out that this changeover of spin swing direction is associated with the
QPT of the QRM. In the low-frequency limit the critical coupling is given by
$g_{c}$ in Eq.~(\ref{gC0}), while at a finite frequency the critical coupling
shifts from $g_{c}$ to
\begin{equation}
g_{c}^{\omega }=\sqrt{\omega ^{2}+\sqrt{\omega ^{4}+g_{c}^{4}}}  \label{gCw}
\end{equation}%
due to wave-packet broadening~\cite{Ying2015}. We illustrate some line plots
of the tangential spin $\langle \sigma _{T}\rangle $ in Fig.~\ref{Fig-sT-transition}(a)
which display the reversal of the swing direction
more clearly. Here in the figure we see that the swings in $\langle \sigma
_{T}\rangle $ indeed have reversed directions before the QPT at $%
g=0.9g_{c}^{\omega }$ [orange (gray) solid line] and beyond the QPT at $%
g=1.4g_{c}^{\omega }$ [green (light gray) solid line]. In Fig.~\ref{Fig-sT-transition}(a)
we also plot the case at $g=1.06g_{c}^{\omega }$
(blue dotted) in which the spin trajectory hunches up and touches the $sg$
axis, about to cross the axis, which is the critical moment to start the
reversal of swing direction. We denote this critical coupling by $g_{c}^{T}$%
. Figure~\ref{Fig-sT-transition}(b) compares $g_{c}^{T}$ (red dots) and $%
g_{c}^{\omega }$ (blue solid line) which coincide basically at all
frequencies.

On the other hand, one can fix the coupling and find the changeover of
spin-swing direction at $\omega _{c}^{T}$ in variation of the frequency.
From the expressions of $g_{c}$ and $g_{c}^{\omega }$ we obtain
the critical frequency of the QPT as functions of the coupling
\begin{eqnarray}
\omega _{c} &=&\frac{4g^{2}}{\Omega },  \label{wC0} \\
\omega _{c}^{\omega } &=&\frac{4g^{2}}{\sqrt{32g^{2}+\Omega ^{2}}},
\label{wCw}
\end{eqnarray}%
which are valid in the low-frequency limit and at finite frequencies,
respectively. In Fig.~\ref{Fig-sT-transition}(c) we also illustrate three
examples of tangential spin trajectories above [orange (gray) solid
line], around (blue dotted line) and below [green (light gray) solid line] the
critical frequency. Figure~\ref{Fig-sT-transition}(d) shows $\omega _{c}^{T}$
(red dots) and $g_{c}^{\omega }$ (blue solid line) in good agreements.

The coincidences between $g_{c}^{T}$ ($\omega _{c}^{T}$) and $g_{c}^{\omega}$ ($\omega _{c}^{\omega}$) indicate
that the changeover of non-adiabatic spin swing direction is associated with
the QPT of the QRM. This in turn endows a non-adiabaticity interpretation
and connotation to the QPT from the geometric physics point of view.
Moreover, tracking of $\omega _{c}^{T}$ may also provide an alternative way
for detection of the QPT.

\section{Insight for variation of spectral structure in the geometric
perspective}

\label{Sect-bifurcation}

\subsection{Curvature-induced variation of spectral structure}

Aside from the geometric analysis for the properties of eigen states, we can
also gain novel insight into the variation of spectral structure in the
geometric perspective. Actually from the curve Hamiltonian (\ref{H-curve})
we can track the geometric contribution to the spectral structure by the
turning off and turning on the curvature. For this purpose we solve the
eigen system by finite difference method with discretization in the $s$
dimension, in comparison with the ED results in the Fock space. The spectral
structures at different frequencies are displayed by the energy level spacing%
\begin{equation}
dE_{n}=E_{n+1}-E_{n}
\end{equation}%
as shown in Fig.~\ref{Fig-dEn-vs-ED}.

At a very low frequency $\omega =0.005\Omega $, turning off [$K=0$, Fig.~\ref{Fig-dEn-vs-ED}(a)]
or turning on [$K\neq 0$, Fig.~\ref{Fig-dEn-vs-ED}(b)]
the curvature the spectral structure show little difference, both cases are
producing basically the same spectral structure as in ED [Fig.~\ref{Fig-dEn-vs-ED}(c)].
Here in the low energy level below $n=n_c\sim 300$,
the $dE_{n}$ manifests an anharmonicity (non-uniform spacing) with single-value
curve like increasing, which is coming from the lower energy branch ${\cal E}_{-}$ with the nonliear $\sqrt{8^{2}g^{2}s^{2}+%
\Omega ^{2}}$ term. Above $n_c$,
the spectral distribution turns to be a wide texture, which
occurs when the two energy branches ${\cal E}_{-}$ and ${\cal E}_{+}$ meet. Actually the energy gap between
the bottoms of the two energy branches can be estimated by
\begin{eqnarray}
\Delta _{{\cal E}} &=&{\cal E}_{+}^{\min }-{\cal E}_{-}^{\min }  \nonumber \\
&=&\left[ \theta \left( g_{c}-g\right) +\theta \left( g-g_{c}\right) \frac{%
\left( g^{2}+g_{c}^{2}\right) ^{2}}{4g^{2}g_{c}^{2}}\right] \Omega
\end{eqnarray}%
where $\theta \left( x\right) $ is the heaviside function, as ${\cal E}%
_{+}^{\min }=\left( \Omega -\omega \right) /2$ while ${\cal E}_{-}^{\min
}=\left( -\Omega -\omega \right) /2$ for $g<g_{c}$ and ${\cal E}_{-}^{\min
}=-\omega /2-\left( g^{2}+g_{c}^{2}\right) ^{2}/\left(
4g^{2}g_{c}^{2}\right) $ for $g>g_{c}$. In the low-frequency limit, the
level spacing is of order $\omega $, so that the level number within the gap
is of order $n_c\sim \Delta _{{\cal E}}/\omega \sim \Omega /\omega =\mathcal{O}\left(
10^{2}\right) .$

At a low frequency $\omega =0.05\Omega $, the wide texture distribution
extends to low levels, as $n_c\sim \Omega /\omega =\mathcal{O}\left( 10^{1}\right) $
now. In this situation, turning off the curvature [Fig.~\ref{Fig-dEn-vs-ED}(d)]
cannot capture the exact spectral texture of the ED
result [Fig.~\ref{Fig-dEn-vs-ED}(f)], while turning on the curvature [Fig.~\ref{Fig-dEn-vs-ED}(d)]
reproduces the exact spectral texture.

At an intermediate frequency $\omega =0.5\Omega $, turning off the curvature
[Fig.~\ref{Fig-dEn-vs-ED}(g)] still leads to a wide spectral texture, but
qualitatively different from the bifurcated structure in the ED spectrum
[Fig.~\ref{Fig-dEn-vs-ED}(i)]. In contrast, turning on the curvature [Fig.~\ref{Fig-dEn-vs-ED}(h)]
yields the same bifurcated structure as the ED result.

These comparisons show that the\ spectral bifurcation is arising from the
curvature effect.

\subsection{Deeper origin: Curvature-induced SOC responsible for both
spectral bifurcation and spectral oscillations}

We further clarify that it is the curvature-induced SOC that is responsible
for the spectral bifurcation. As decomposed in the curve Hamiltonian (\ref{H-curve})
the curvature induces three terms, including the SOC ($\alpha _{{\rm SOC}}$),
imaginary out-of-plane field ($\widetilde{\Omega }_{y}$), and
curvature potential ($v_{K}$). We show the corresponding strengths in Fig.~\ref{Fig-SOC-dEn}
which shows that they are relatively comparable and the
weights of roles cannot be simply judged by their amplitudes. Nevertheless,
we can turn off $\alpha _{{\rm SOC}}$, $\widetilde{\Omega }_{y}$, and $v_{K}$
one by one to track their contributions.

Figures~\ref{Fig-SOC-dEn}(a) and \ref{Fig-SOC-dEn}(b) show the spectral
outcomes by setting $v_{K}=0$ and $\widetilde{\Omega }_{y}=0$, respectively.
We see that the spectral bifurcation is retained in both cases. These two
cases commonly keep the SOC, which already gives the hint that the SOC is
crucial for the spectral bifurcation. When we only turn off the SOC in Fig.~\ref{Fig-SOC-dEn}(c)
by setting $\alpha _{{\rm SOC}}=0$, the spectral
bifurcation disappears and the spectrum is completely different from the
exact one. Thus it is confirmed that the SOC is finally responsible for the
spectral bifurcation.

In fact, as shown in Fig.~\ref{Fig-SOC-dEn}(e), the SOC energies in odd
[orange (gray) squares] and even [blue (dark gray) squares] levels are
separated, alternatively having negative and positive signs except a sign
exchange in level crossings. We illustrate an example of neighboring levels
in Fig.~\ref{Fig-SOC-dEn}(f), where one can see that the SOC energies in the
even and odd cases have opposite signs even locally at all the positions. As
a result, the total energies in odd and even levels are also separated to
form the bifurcation as shown in Fig.~\ref{Fig-SOC-dEn}(b).

Furthermore, we find that the oscillations in each bifurcated energy branch
are also arising from the SOC. Indeed, as demonstrated in Fig.~\ref{Fig-SOC-dEn}(d),
the $\widetilde{\Omega }_{y}$ term [blue (dark gray)] has
a vanishing contribution in all levels, which actually guarantees the real energy spectrum, and neither does the $v_{K}$ [orange (gray)]
energy oscillate noticeably. Only the SOC energy manifests strong
oscillations. This means that the SOC is also responsible for the spectral
oscillations.

\section{Conclusions and discussions}

\label{Sect-Conclusions}

In the present work we have established a geometric perspective for the QRM
which is a fundamental model for light-matter interactions. In the
locally-diagonalized basis we mapped the QRM to a curved nanowire system.
The curvature induces three terms including the SOC, imaginary out-of-plane
field, and quadratic curvature potential. The curvature can be controlled
either by the strength of light-matter coupling or the frequency ratio. Such
a geometric description enables us to reveal and perceive the property
universality and diversity from the ground state to the whole spectrum in a
unified picture.

First, we unveiled a super university in the adiabatic regime which
manifests a universal scaling of spin texture ubiquitously valid for
different couplings, different frequencies and different energy levels. Such
a ubiquitousness of universality provides a sharp contrast to the conventional critical
universality~\cite{LiuM2017PRL,Irish2017,Ying-Stark-top},
in which different values of parameters such as anisotropy~\cite{LiuM2017PRL,Ying-Stark-top} share the same critical variation usually only in the vicinity
of a QPT~\cite{Sachdev-QPT,Irish2017}, but never behave as invariant for all the
couplings. Also, unlike the super universality, the conventional critical
universality of the QPT holds only for the ground state, but invalid for the
excite states.

Second, we studied the non-adiabaticity of the spin texture which arises in
very large couplings, high frequencies or excited states. In such a
situation, the super universality become violated and properties are
diversified. Nevertheless, in the ground states we found a kind of new
universality amidst the diversity: the tangential spin swing has a same
direction before the critical coupling $g_{c}^{T}$ at a given frequency or
above the critical coupling $\omega _{c}^{T}$ at fixed coupling, while it
shares a reversed spin-swing direction beyond the critical values. We find
such a direction changeover in the tangential spin swing correspond to the
variation boundary $g_{c}^{\omega }$~\cite{Ying2015} and $\omega
_{c}^{\omega }$ of the QPT at finite frequency in the QRM. Thus the
spin-swing direction changeover is associated with the QPT~\cite{Ashhab2013,Ying2015,Liu2021AQT,Hwang2015PRL,Hwang2016PRL,Irish2017,
Ying-g2hz-QFI-2024,*Ying-g2hz-QFI-2024-Cover,Ying-g1g2hz-QFI-2025,Ying2025g2A4,*Ying2025g2A4-Cover,Ying-g2Stark-QFI-2025, LiuM2017PRL,Ying-2018-arxiv,Ying2020-nonlinear-bias,Ying-2021-AQT,*Ying-2021-AQT-Cover,Ying-gapped-top, Ying-Stark-top,*Ying-Stark-top-Cover, Ying-Spin-Winding,*Ying-Spin-Winding-Cover, Ying-JCwinding,Ying-Topo-JC-nonHermitian,*Ying-Topo-JC-nonHermitian-Cover,Ying-Topo-JC-nonHermitian-Fisher,*Ying-Topo-JC-nonHermitian-Fisher-Cover,Ying-gC-by-QFI-2024, Grimaudo2022q2QPT,Grimaudo2023-Entropy,Grimaudo2024PRR,Zhu2024PRL,DeepStrong-JC-Huang-2024,PengJie2019,Padilla2022,Gao2022Rabi-dimer,GaoXL2025SPT}
in the QRM. Such a finding endows a geometric non-adiabaticity
interpretation and connotation to the QPT of the QRM.

Third, we demonstrated that the curvature plays a crucial role to form the
spectral bifurcation in the distribution of level spacings at finite
frequencies. We further clarified that, among the three curvature-induced
effects, it is the SOC that is responsible for both the spectral bifurcation and oscillations.

As a final remark, it is worthwhile to mention that, although the QRM has no explicit
SOC~\cite{Ying-2021-AQT,*Ying-2021-AQT-Cover,Ying-gapped-top,Ying-Spin-Winding,*Ying-Spin-Winding-Cover},
our analysis reveals that curvature-induced SOC can still arise from the
geometric effect. Our method and analysis can be readily generalized to
other light-matter models, such as
anisotropy~\cite{Liu2021AQT,Ying-2021-AQT,*Ying-2021-AQT-Cover,Ying-gapped-top,Ying-Spin-Winding,*Ying-Spin-Winding-Cover,PengJie2019}, asymmetry~\cite{Ying2020-nonlinear-bias,Ying-g1g2hz-QFI-2025}, nonlinear
coupling~\cite{Ying2020-nonlinear-bias,Ying-2018-arxiv,Ying-g1g2hz-QFI-2025,Ying2022-Metrology,Ying-g2hz-QFI-2024,*Ying-g2hz-QFI-2024-Cover, Ying2025g2A4,*Ying2025g2A4-Cover,Felicetti2015-TwoPhotonProcess,e-collpase-Garbe-2017,e-collpase-Duan-2016,CongLei2019,Rico2020}
and Stark coupling~\cite{Eckle-2017JPA,*Eckle-2017JPA-b,Stark-Cong2020,Ying-Stark-top,*Ying-Stark-top-Cover,Ying-g2Stark-QFI-2025,
Stark-Grimsmo2013,Stark-Grimsmo2014,Stark-Cong2020,Ying2026StarkTPB}. We speculate that our treatment may also provide more understanding in
dynamics~\cite{Wolf2012} and the level statics in the integrability~\cite{Braak2011} to quantum chaos~\cite{EmaryPRE2003DickeChaos} variation in
increasing qubits. The formalism we established may also facilitate extracting the non-adiabatic
Aharonov-Anandan~\cite{AAphase1987} geometric phase in light-matter interactions and is inspiring for bridging more general non-Hermitian and Hermitian systems by geometry~\cite{Ying2026GeomNonhermitian}.
We leave these discussions for future works. On the other
hand, experimentally in superconducting circuit systems the spin can be
represented by two flux states in a flux qubit system~\cite{flux-qubit-Mooij-1999,Bertet2005mixedModel,you024532}. The position $s$ and
momentum $p$ can be simulated by the flux and the charge of Josephson
junctions~\cite{flux-qubit-Mooij-1999,Bertet2005mixedModel,you024532} in
another superconducting circuit system coupled to the qubit system with
ultra-strong~\cite{Ciuti2005EarlyUSC,Aji2009EarlyUSC,Diaz2019RevModPhy,Kockum2019NRP,Wallraff2004,Gunter2009,Niemczyk2010, Peropadre2010,FornDiaz2017,Forn-Diaz2010,Scalari2012,Xiang2013,Yoshihara2017NatPhys,Kockum2017,Bayer2017DeepStrong,Qin2024PhysRep,Ulstrong-JC-2}
or deep-strong~\cite{WangYouJQ2023DeepStrong,Yoshihara2017NatPhys,Bayer2017DeepStrong,DeepStrong-JC-Huang-2024}
couplings. These parameters may be tuned and detected by interference
devices and magnetometer~\cite{you024532}. Thus, the properties revealed in
our perspective may also be realistically relevant and experimentally
measurable. In such prospects, it can be expected that the geometric
perspective will open a dialogue between geometry and physics, aside from
the one between mathematics and physics~\cite{Braak2011,Solano2011}, thus
paving a novel way to gain deeper insight into light-matter interactions.

\section*{Acknowledgment}

This work was supported by the National Natural Science Foundation of China
(Grants No. 12474358 and No. 12247101).

\bibliography{Refs-2026-9-Photon-Blockade-Curvature}

\begin{thebibliography}{170}%
\makeatletter
\providecommand \@ifxundefined [1]{%
 \@ifx{#1\undefined}
}%
\providecommand \@ifnum [1]{%
 \ifnum #1\expandafter \@firstoftwo
 \else \expandafter \@secondoftwo
 \fi
}%
\providecommand \@ifx [1]{%
 \ifx #1\expandafter \@firstoftwo
 \else \expandafter \@secondoftwo
 \fi
}%
\providecommand \natexlab [1]{#1}%
\providecommand \enquote  [1]{``#1''}%
\providecommand \bibnamefont  [1]{#1}%
\providecommand \bibfnamefont [1]{#1}%
\providecommand \citenamefont [1]{#1}%
\providecommand \href@noop [0]{\@secondoftwo}%
\providecommand \href [0]{\begingroup \@sanitize@url \@href}%
\providecommand \@href[1]{\@@startlink{#1}\@@href}%
\providecommand \@@href[1]{\endgroup#1\@@endlink}%
\providecommand \@sanitize@url [0]{\catcode `\\12\catcode `\$12\catcode
  `\&12\catcode `\#12\catcode `\^12\catcode `\_12\catcode `\%12\relax}%
\providecommand \@@startlink[1]{}%
\providecommand \@@endlink[0]{}%
\providecommand \url  [0]{\begingroup\@sanitize@url \@url }%
\providecommand \@url [1]{\endgroup\@href {#1}{\urlprefix }}%
\providecommand \urlprefix  [0]{URL }%
\providecommand \Eprint [0]{\href }%
\providecommand \doibase [0]{http://dx.doi.org/}%
\providecommand \selectlanguage [0]{\@gobble}%
\providecommand \bibinfo  [0]{\@secondoftwo}%
\providecommand \bibfield  [0]{\@secondoftwo}%
\providecommand \translation [1]{[#1]}%
\providecommand \BibitemOpen [0]{}%
\providecommand \bibitemStop [0]{}%
\providecommand \bibitemNoStop [0]{.\EOS\space}%
\providecommand \EOS [0]{\spacefactor3000\relax}%
\providecommand \BibitemShut  [1]{\csname bibitem#1\endcsname}%
\let\auto@bib@innerbib\@empty
\bibitem [{\citenamefont {Eckle}(2019)}]{Eckle-Book-Models}%
  \BibitemOpen
  \bibfield  {author} {\bibinfo {author} {\bibfnamefont {H.-P.}\ \bibnamefont
  {Eckle}},\ }\href@noop {} {\emph {\bibinfo {title} {Models of Quantum
  Matter}}}\ (\bibinfo  {publisher} {Oxford University, Oxford},\ \bibinfo
  {year} {2019})\BibitemShut {NoStop}%
\bibitem [{\citenamefont {Forn-D\'{\i}az}\ \emph {et~al.}(2019)\citenamefont
  {Forn-D\'{\i}az}, \citenamefont {Lamata}, \citenamefont {Rico}, \citenamefont
  {Kono},\ and\ \citenamefont {Solano}}]{Diaz2019RevModPhy}%
  \BibitemOpen
  \bibfield  {author} {\bibinfo {author} {\bibfnamefont {P.}~\bibnamefont
  {Forn-D\'{\i}az}}, \bibinfo {author} {\bibfnamefont {L.}~\bibnamefont
  {Lamata}}, \bibinfo {author} {\bibfnamefont {E.}~\bibnamefont {Rico}},
  \bibinfo {author} {\bibfnamefont {J.}~\bibnamefont {Kono}}, \ and\ \bibinfo
  {author} {\bibfnamefont {E.}~\bibnamefont {Solano}},\ }\href {\doibase
  10.1103/RevModPhys.91.025005} {\bibfield  {journal} {\bibinfo  {journal}
  {Rev. Mod. Phys.}\ }\textbf {\bibinfo {volume} {91}},\ \bibinfo {pages}
  {025005} (\bibinfo {year} {2019})}\BibitemShut {NoStop}%
\bibitem [{\citenamefont {Kockum}\ \emph {et~al.}(2019)\citenamefont {Kockum},
  \citenamefont {Miranowicz}, \citenamefont {De~Liberato}, \citenamefont
  {Savasta},\ and\ \citenamefont {Nori}}]{Kockum2019NRP}%
  \BibitemOpen
  \bibfield  {author} {\bibinfo {author} {\bibfnamefont {A.~F.}\ \bibnamefont
  {Kockum}}, \bibinfo {author} {\bibfnamefont {A.}~\bibnamefont {Miranowicz}},
  \bibinfo {author} {\bibfnamefont {S.}~\bibnamefont {De~Liberato}}, \bibinfo
  {author} {\bibfnamefont {S.}~\bibnamefont {Savasta}}, \ and\ \bibinfo
  {author} {\bibfnamefont {F.}~\bibnamefont {Nori}},\ }\href {\doibase
  10.1038/s42254-018-0006-2} {\bibfield  {journal} {\bibinfo  {journal} {Nat.
  Rev. Phys.}\ }\textbf {\bibinfo {volume} {1}},\ \bibinfo {pages} {19}
  (\bibinfo {year} {2019})}\BibitemShut {NoStop}%
\bibitem [{\citenamefont {Qin}\ \emph {et~al.}(2024)\citenamefont {Qin},
  \citenamefont {Kockum}, \citenamefont {Mu{\~n}oz}, \citenamefont
  {Miranowicz},\ and\ \citenamefont {Nori}}]{Qin2024PhysRep}%
  \BibitemOpen
  \bibfield  {author} {\bibinfo {author} {\bibfnamefont {W.}~\bibnamefont
  {Qin}}, \bibinfo {author} {\bibfnamefont {A.~F.}\ \bibnamefont {Kockum}},
  \bibinfo {author} {\bibfnamefont {C.~S.}\ \bibnamefont {Mu{\~n}oz}}, \bibinfo
  {author} {\bibfnamefont {A.}~\bibnamefont {Miranowicz}}, \ and\ \bibinfo
  {author} {\bibfnamefont {F.}~\bibnamefont {Nori}},\ }\href {\doibase
  https://doi.org/10.1016/j.physrep.2024.05.003} {\bibfield  {journal}
  {\bibinfo  {journal} {Phys. Rep.}\ }\textbf {\bibinfo {volume} {1078}},\
  \bibinfo {pages} {1} (\bibinfo {year} {2024})}\BibitemShut {NoStop}%
\bibitem [{\citenamefont {Romero}\ \emph {et~al.}(2012)\citenamefont {Romero},
  \citenamefont {Ballester}, \citenamefont {Wang}, \citenamefont {Scarani},\
  and\ \citenamefont {Solano}}]{Romero2012}%
  \BibitemOpen
  \bibfield  {author} {\bibinfo {author} {\bibfnamefont {G.}~\bibnamefont
  {Romero}}, \bibinfo {author} {\bibfnamefont {D.}~\bibnamefont {Ballester}},
  \bibinfo {author} {\bibfnamefont {Y.~M.}\ \bibnamefont {Wang}}, \bibinfo
  {author} {\bibfnamefont {V.}~\bibnamefont {Scarani}}, \ and\ \bibinfo
  {author} {\bibfnamefont {E.}~\bibnamefont {Solano}},\ }\href {\doibase
  10.1103/PhysRevLett.108.120501} {\bibfield  {journal} {\bibinfo  {journal}
  {Phys. Rev. Lett.}\ }\textbf {\bibinfo {volume} {108}},\ \bibinfo {pages}
  {120501} (\bibinfo {year} {2012})}\BibitemShut {NoStop}%
\bibitem [{\citenamefont {Stassi}\ \emph {et~al.}(2020)\citenamefont {Stassi},
  \citenamefont {Cirio},\ and\ \citenamefont {Nori}}]{Stassi2020QuComput}%
  \BibitemOpen
  \bibfield  {author} {\bibinfo {author} {\bibfnamefont {R.}~\bibnamefont
  {Stassi}}, \bibinfo {author} {\bibfnamefont {M.}~\bibnamefont {Cirio}}, \
  and\ \bibinfo {author} {\bibfnamefont {F.}~\bibnamefont {Nori}},\ }\href
  {\doibase 10.1038/s41534-020-00294} {\bibfield  {journal} {\bibinfo
  {journal} {npj Quantum Information}\ }\textbf {\bibinfo {volume} {6}},\
  \bibinfo {pages} {67} (\bibinfo {year} {2020})}\BibitemShut {NoStop}%
\bibitem [{\citenamefont {Stassi}\ and\ \citenamefont
  {Nori}(2018)}]{Stassi2018}%
  \BibitemOpen
  \bibfield  {author} {\bibinfo {author} {\bibfnamefont {R.}~\bibnamefont
  {Stassi}}\ and\ \bibinfo {author} {\bibfnamefont {F.}~\bibnamefont {Nori}},\
  }\href {\doibase 10.1103/PhysRevA.97.033823} {\bibfield  {journal} {\bibinfo
  {journal} {Phys. Rev. A}\ }\textbf {\bibinfo {volume} {97}},\ \bibinfo
  {pages} {033823} (\bibinfo {year} {2018})}\BibitemShut {NoStop}%
\bibitem [{\citenamefont {Macr\`{\i}}\ \emph {et~al.}(2018)\citenamefont
  {Macr\`{\i}}, \citenamefont {Nori},\ and\ \citenamefont
  {Kockum}}]{Macri2018}%
  \BibitemOpen
  \bibfield  {author} {\bibinfo {author} {\bibfnamefont {V.}~\bibnamefont
  {Macr\`{\i}}}, \bibinfo {author} {\bibfnamefont {F.}~\bibnamefont {Nori}}, \
  and\ \bibinfo {author} {\bibfnamefont {A.~F.}\ \bibnamefont {Kockum}},\
  }\href {\doibase 10.1103/PhysRevA.98.062327} {\bibfield  {journal} {\bibinfo
  {journal} {Phys. Rev. A}\ }\textbf {\bibinfo {volume} {98}},\ \bibinfo
  {pages} {062327} (\bibinfo {year} {2018})}\BibitemShut {NoStop}%
\bibitem [{\citenamefont {Garbe}\ \emph {et~al.}(2020)\citenamefont {Garbe},
  \citenamefont {Bina}, \citenamefont {Keller}, \citenamefont {Paris},\ and\
  \citenamefont {Felicetti}}]{Garbe2020}%
  \BibitemOpen
  \bibfield  {author} {\bibinfo {author} {\bibfnamefont {L.}~\bibnamefont
  {Garbe}}, \bibinfo {author} {\bibfnamefont {M.}~\bibnamefont {Bina}},
  \bibinfo {author} {\bibfnamefont {A.}~\bibnamefont {Keller}}, \bibinfo
  {author} {\bibfnamefont {M.~G.~A.}\ \bibnamefont {Paris}}, \ and\ \bibinfo
  {author} {\bibfnamefont {S.}~\bibnamefont {Felicetti}},\ }\href {\doibase
  10.1103/PhysRevLett.124.120504} {\bibfield  {journal} {\bibinfo  {journal}
  {Phys. Rev. Lett.}\ }\textbf {\bibinfo {volume} {124}},\ \bibinfo {pages}
  {120504} (\bibinfo {year} {2020})}\BibitemShut {NoStop}%
\bibitem [{\citenamefont {Montenegro}\ \emph {et~al.}(2021)\citenamefont
  {Montenegro}, \citenamefont {Mishra},\ and\ \citenamefont
  {Bayat}}]{Montenegro2021-Metrology}%
  \BibitemOpen
  \bibfield  {author} {\bibinfo {author} {\bibfnamefont {V.}~\bibnamefont
  {Montenegro}}, \bibinfo {author} {\bibfnamefont {U.}~\bibnamefont {Mishra}},
  \ and\ \bibinfo {author} {\bibfnamefont {A.}~\bibnamefont {Bayat}},\ }\href
  {\doibase 10.1103/PhysRevLett.126.200501} {\bibfield  {journal} {\bibinfo
  {journal} {Phys. Rev. Lett.}\ }\textbf {\bibinfo {volume} {126}},\ \bibinfo
  {pages} {200501} (\bibinfo {year} {2021})}\BibitemShut {NoStop}%
\bibitem [{\citenamefont {Chu}\ \emph {et~al.}(2021)\citenamefont {Chu},
  \citenamefont {Zhang}, \citenamefont {Yu},\ and\ \citenamefont
  {Cai}}]{Chu2021-Metrology}%
  \BibitemOpen
  \bibfield  {author} {\bibinfo {author} {\bibfnamefont {Y.}~\bibnamefont
  {Chu}}, \bibinfo {author} {\bibfnamefont {S.}~\bibnamefont {Zhang}}, \bibinfo
  {author} {\bibfnamefont {B.}~\bibnamefont {Yu}}, \ and\ \bibinfo {author}
  {\bibfnamefont {J.}~\bibnamefont {Cai}},\ }\href {\doibase
  10.1103/PhysRevLett.126.010502} {\bibfield  {journal} {\bibinfo  {journal}
  {Phys. Rev. Lett.}\ }\textbf {\bibinfo {volume} {126}},\ \bibinfo {pages}
  {010502} (\bibinfo {year} {2021})}\BibitemShut {NoStop}%
\bibitem [{\citenamefont {Garbe}\ \emph {et~al.}(2022)\citenamefont {Garbe},
  \citenamefont {Abah}, \citenamefont {Felicetti},\ and\ \citenamefont
  {Puebla}}]{Garbe2021-Metrology}%
  \BibitemOpen
  \bibfield  {author} {\bibinfo {author} {\bibfnamefont {L.}~\bibnamefont
  {Garbe}}, \bibinfo {author} {\bibfnamefont {O.}~\bibnamefont {Abah}},
  \bibinfo {author} {\bibfnamefont {S.}~\bibnamefont {Felicetti}}, \ and\
  \bibinfo {author} {\bibfnamefont {R.}~\bibnamefont {Puebla}},\ }\href
  {\doibase 10.1103/PhysRevResearch.4.043061} {\bibfield  {journal} {\bibinfo
  {journal} {Phys. Rev. Res.}\ }\textbf {\bibinfo {volume} {4}},\ \bibinfo
  {pages} {043061} (\bibinfo {year} {2022})}\BibitemShut {NoStop}%
\bibitem [{\citenamefont {Ilias}\ \emph {et~al.}(2022)\citenamefont {Ilias},
  \citenamefont {Yang}, \citenamefont {Huelga},\ and\ \citenamefont
  {Plenio}}]{Ilias2022-Metrology}%
  \BibitemOpen
  \bibfield  {author} {\bibinfo {author} {\bibfnamefont {T.}~\bibnamefont
  {Ilias}}, \bibinfo {author} {\bibfnamefont {D.}~\bibnamefont {Yang}},
  \bibinfo {author} {\bibfnamefont {S.~F.}\ \bibnamefont {Huelga}}, \ and\
  \bibinfo {author} {\bibfnamefont {M.~B.}\ \bibnamefont {Plenio}},\ }\href
  {\doibase 10.1103/PRXQuantum.3.010354} {\bibfield  {journal} {\bibinfo
  {journal} {PRX Quantum}\ }\textbf {\bibinfo {volume} {3}},\ \bibinfo {pages}
  {010354} (\bibinfo {year} {2022})}\BibitemShut {NoStop}%
\bibitem [{\citenamefont {Ying}\ \emph {et~al.}(2022)\citenamefont {Ying},
  \citenamefont {Felicetti}, \citenamefont {Liu},\ and\ \citenamefont
  {Braak}}]{Ying2022-Metrology}%
  \BibitemOpen
  \bibfield  {author} {\bibinfo {author} {\bibfnamefont {Z.-J.}\ \bibnamefont
  {Ying}}, \bibinfo {author} {\bibfnamefont {S.}~\bibnamefont {Felicetti}},
  \bibinfo {author} {\bibfnamefont {G.}~\bibnamefont {Liu}}, \ and\ \bibinfo
  {author} {\bibfnamefont {D.}~\bibnamefont {Braak}},\ }\href {\doibase
  10.3390/e24081015} {\bibfield  {journal} {\bibinfo  {journal} {Entropy}\
  }\textbf {\bibinfo {volume} {24}},\ \bibinfo {pages} {1015} (\bibinfo {year}
  {2022})}\BibitemShut {NoStop}%
\bibitem [{\citenamefont {Gietka}\ \emph {et~al.}(2023)\citenamefont {Gietka},
  \citenamefont {Hotter},\ and\ \citenamefont
  {Ritsch}}]{Gietka2023PRL-Squeezing}%
  \BibitemOpen
  \bibfield  {author} {\bibinfo {author} {\bibfnamefont {K.}~\bibnamefont
  {Gietka}}, \bibinfo {author} {\bibfnamefont {C.}~\bibnamefont {Hotter}}, \
  and\ \bibinfo {author} {\bibfnamefont {H.}~\bibnamefont {Ritsch}},\ }\href
  {\doibase 10.1103/PhysRevLett.131.223604} {\bibfield  {journal} {\bibinfo
  {journal} {Phys. Rev. Lett.}\ }\textbf {\bibinfo {volume} {131}},\ \bibinfo
  {pages} {223604} (\bibinfo {year} {2023})}\BibitemShut {NoStop}%
\bibitem [{\citenamefont {Zhu}\ \emph {et~al.}(2023)\citenamefont {Zhu},
  \citenamefont {L\"u}, \citenamefont {Ning}, \citenamefont {Wu}, \citenamefont
  {Shen}, \citenamefont {Yang},\ and\ \citenamefont
  {Zheng}}]{YangZheng2023SciChina}%
  \BibitemOpen
  \bibfield  {author} {\bibinfo {author} {\bibfnamefont {X.}~\bibnamefont
  {Zhu}}, \bibinfo {author} {\bibfnamefont {J.-H.}\ \bibnamefont {L\"u}},
  \bibinfo {author} {\bibfnamefont {W.}~\bibnamefont {Ning}}, \bibinfo {author}
  {\bibfnamefont {F.}~\bibnamefont {Wu}}, \bibinfo {author} {\bibfnamefont
  {L.-T.}\ \bibnamefont {Shen}}, \bibinfo {author} {\bibfnamefont {Z.-B.}\
  \bibnamefont {Yang}}, \ and\ \bibinfo {author} {\bibfnamefont {S.-B.}\
  \bibnamefont {Zheng}},\ }\href {\doibase 10.1007/s11433-022-2073-9}
  {\bibfield  {journal} {\bibinfo  {journal} {Sci. China Phys. Mech. Astron.}\
  }\textbf {\bibinfo {volume} {66}},\ \bibinfo {pages} {250313} (\bibinfo
  {year} {2023})}\BibitemShut {NoStop}%
\bibitem [{\citenamefont {Hotter}\ \emph {et~al.}(2024)\citenamefont {Hotter},
  \citenamefont {Ritsch},\ and\ \citenamefont {Gietka}}]{Hotter2024-Metrology}%
  \BibitemOpen
  \bibfield  {author} {\bibinfo {author} {\bibfnamefont {C.}~\bibnamefont
  {Hotter}}, \bibinfo {author} {\bibfnamefont {H.}~\bibnamefont {Ritsch}}, \
  and\ \bibinfo {author} {\bibfnamefont {K.}~\bibnamefont {Gietka}},\ }\href
  {\doibase 10.1103/PhysRevLett.132.060801} {\bibfield  {journal} {\bibinfo
  {journal} {Phys. Rev. Lett.}\ }\textbf {\bibinfo {volume} {132}},\ \bibinfo
  {pages} {060801} (\bibinfo {year} {2024})}\BibitemShut {NoStop}%
\bibitem [{\citenamefont {Alushi}\ \emph {et~al.}(2024)\citenamefont {Alushi},
  \citenamefont {G\'orecki}, \citenamefont {Felicetti},\ and\ \citenamefont
  {Di~Candia}}]{Alushi2024PRL}%
  \BibitemOpen
  \bibfield  {author} {\bibinfo {author} {\bibfnamefont {U.}~\bibnamefont
  {Alushi}}, \bibinfo {author} {\bibfnamefont {W.}~\bibnamefont {G\'orecki}},
  \bibinfo {author} {\bibfnamefont {S.}~\bibnamefont {Felicetti}}, \ and\
  \bibinfo {author} {\bibfnamefont {R.}~\bibnamefont {Di~Candia}},\ }\href
  {\doibase 10.1103/PhysRevLett.133.040801} {\bibfield  {journal} {\bibinfo
  {journal} {Phys. Rev. Lett.}\ }\textbf {\bibinfo {volume} {133}},\ \bibinfo
  {pages} {040801} (\bibinfo {year} {2024})}\BibitemShut {NoStop}%
\bibitem [{\citenamefont {Mukhopadhyay}\ and\ \citenamefont
  {Bayat}(2024)}]{Mukhopadhyay2024PRL}%
  \BibitemOpen
  \bibfield  {author} {\bibinfo {author} {\bibfnamefont {C.}~\bibnamefont
  {Mukhopadhyay}}\ and\ \bibinfo {author} {\bibfnamefont {A.}~\bibnamefont
  {Bayat}},\ }\href {\doibase 10.1103/PhysRevLett.133.120601} {\bibfield
  {journal} {\bibinfo  {journal} {Phys. Rev. Lett.}\ }\textbf {\bibinfo
  {volume} {133}},\ \bibinfo {pages} {120601} (\bibinfo {year}
  {2024})}\BibitemShut {NoStop}%
\bibitem [{\citenamefont {Mihailescu}\ \emph {et~al.}(2024)\citenamefont
  {Mihailescu}, \citenamefont {Bayat}, \citenamefont {Campbell},\ and\
  \citenamefont {Mitchell}}]{Mihailescuy2024}%
  \BibitemOpen
  \bibfield  {author} {\bibinfo {author} {\bibfnamefont {G.}~\bibnamefont
  {Mihailescu}}, \bibinfo {author} {\bibfnamefont {A.}~\bibnamefont {Bayat}},
  \bibinfo {author} {\bibfnamefont {S.}~\bibnamefont {Campbell}}, \ and\
  \bibinfo {author} {\bibfnamefont {A.~K.}\ \bibnamefont {Mitchell}},\ }\href
  {\doibase 10.1088/2058-9565/ad438d} {\bibfield  {journal} {\bibinfo
  {journal} {Quantum Sci. Technol.}\ }\textbf {\bibinfo {volume} {9}},\
  \bibinfo {pages} {035033} (\bibinfo {year} {2024})}\BibitemShut {NoStop}%
\bibitem [{\citenamefont
  {Ying}(2024{\natexlab{a}})}]{Ying-Topo-JC-nonHermitian-Fisher}%
  \BibitemOpen
  \bibfield  {author} {\bibinfo {author} {\bibfnamefont {Z.-J.}\ \bibnamefont
  {Ying}},\ }\href {\doibase 10.1002/qute.202400288} {\bibfield  {journal}
  {\bibinfo  {journal} {Adv. Quantum Technol.}\ }\textbf {\bibinfo {volume}
  {7}},\ \bibinfo {pages} {2400288} (\bibinfo {year}
  {2024}{\natexlab{a}})}\BibitemShut {NoStop}%
\bibitem [{\citenamefont
  {Ying}(2024{\natexlab{b}})}]{Ying-Topo-JC-nonHermitian-Fisher-Cover}%
  \BibitemOpen
  \bibfield  {author} {\bibinfo {author} {\bibfnamefont {Z.-J.}\ \bibnamefont
  {Ying}},\ }\href {\doibase 10.1002/qute.202470029} {\bibfield  {journal}
  {\bibinfo  {journal} {Adv. Quantum Technol.}\ }\textbf {\bibinfo {volume}
  {7}},\ \bibinfo {pages} {2470029} (\bibinfo {year} {2024}{\natexlab{b}})},\
  \bibinfo {note} {[Back Cover: (Adv. Quantum Technol. 10/2024)]}\BibitemShut
  {NoStop}%
\bibitem [{\citenamefont {Ying}(2025{\natexlab{a}})}]{Ying-g2hz-QFI-2024}%
  \BibitemOpen
  \bibfield  {author} {\bibinfo {author} {\bibfnamefont {Z.-J.}\ \bibnamefont
  {Ying}},\ }\href {\doibase 10.1002/qute.202400630} {\bibfield  {journal}
  {\bibinfo  {journal} {Adv. Quantum Technol.}\ }\textbf {\bibinfo {volume}
  {8}},\ \bibinfo {pages} {2400630} (\bibinfo {year}
  {2025}{\natexlab{a}})}\BibitemShut {NoStop}%
\bibitem [{\citenamefont
  {Ying}(2025{\natexlab{b}})}]{Ying-g2hz-QFI-2024-Cover}%
  \BibitemOpen
  \bibfield  {author} {\bibinfo {author} {\bibfnamefont {Z.-J.}\ \bibnamefont
  {Ying}},\ }\href {\doibase 10.1002/qute.202570015} {\bibfield  {journal}
  {\bibinfo  {journal} {Adv. Quantum Technol.}\ }\textbf {\bibinfo {volume}
  {8}},\ \bibinfo {pages} {2570015} (\bibinfo {year} {2025}{\natexlab{b}})},\
  \bibinfo {note} {[Back Cover: (Adv. Quantum Technol. 7/2025)]}\BibitemShut
  {NoStop}%
\bibitem [{\citenamefont {Ying}(2025{\natexlab{c}})}]{Ying-g1g2hz-QFI-2025}%
  \BibitemOpen
  \bibfield  {author} {\bibinfo {author} {\bibfnamefont {Z.-J.}\ \bibnamefont
  {Ying}},\ }\href {\doibase 10.1103/rtd9-z7cl} {\bibfield  {journal} {\bibinfo
   {journal} {Phys. Rev. A}\ }\textbf {\bibinfo {volume} {112}},\ \bibinfo
  {pages} {032626} (\bibinfo {year} {2025}{\natexlab{c}})}\BibitemShut
  {NoStop}%
\bibitem [{\citenamefont {Ying}\ \emph
  {et~al.}(2025{\natexlab{a}})\citenamefont {Ying}, \citenamefont {Han},
  \citenamefont {Li}, \citenamefont {Felicetti},\ and\ \citenamefont
  {Braak}}]{Ying2025g2A4}%
  \BibitemOpen
  \bibfield  {author} {\bibinfo {author} {\bibfnamefont {Z.-J.}\ \bibnamefont
  {Ying}}, \bibinfo {author} {\bibfnamefont {H.-H.}\ \bibnamefont {Han}},
  \bibinfo {author} {\bibfnamefont {B.-J.}\ \bibnamefont {Li}}, \bibinfo
  {author} {\bibfnamefont {S.}~\bibnamefont {Felicetti}}, \ and\ \bibinfo
  {author} {\bibfnamefont {D.}~\bibnamefont {Braak}},\ }\href {\doibase
  10.1002/qute.202500263} {\bibfield  {journal} {\bibinfo  {journal} {Adv.
  Quantum Technol.}\ }\textbf {\bibinfo {volume} {8}},\ \bibinfo {pages}
  {e00263} (\bibinfo {year} {2025}{\natexlab{a}})}\BibitemShut {NoStop}%
\bibitem [{\citenamefont {Ying}\ \emph
  {et~al.}(2025{\natexlab{b}})\citenamefont {Ying}, \citenamefont {Han},
  \citenamefont {Li}, \citenamefont {Felicetti},\ and\ \citenamefont
  {Braak}}]{Ying2025g2A4-Cover}%
  \BibitemOpen
  \bibfield  {author} {\bibinfo {author} {\bibfnamefont {Z.-J.}\ \bibnamefont
  {Ying}}, \bibinfo {author} {\bibfnamefont {H.-H.}\ \bibnamefont {Han}},
  \bibinfo {author} {\bibfnamefont {B.-J.}\ \bibnamefont {Li}}, \bibinfo
  {author} {\bibfnamefont {S.}~\bibnamefont {Felicetti}}, \ and\ \bibinfo
  {author} {\bibfnamefont {D.}~\bibnamefont {Braak}},\ }\href {\doibase
  https://doi.org/10.1002/qute.70061} {\bibfield  {journal} {\bibinfo
  {journal} {Adv. Quantum Technol.}\ }\textbf {\bibinfo {volume} {8}},\
  \bibinfo {pages} {e70061} (\bibinfo {year} {2025}{\natexlab{b}})},\ \bibinfo
  {note} {[Back Cover: (Adv. Quantum Technol. 11/2025)]}\BibitemShut {NoStop}%
\bibitem [{\citenamefont {Ying}(2025{\natexlab{d}})}]{Ying-g2Stark-QFI-2025}%
  \BibitemOpen
  \bibfield  {author} {\bibinfo {author} {\bibfnamefont {Z.-J.}\ \bibnamefont
  {Ying}},\ }\href {\doibase 10.1103/wnvk-lq8x} {\bibfield  {journal} {\bibinfo
   {journal} {Phys. Rev. A}\ }\textbf {\bibinfo {volume} {112}},\ \bibinfo
  {pages} {052617} (\bibinfo {year} {2025}{\natexlab{d}})}\BibitemShut
  {NoStop}%
\bibitem [{\citenamefont {Hotter}\ \emph {et~al.}(2025)\citenamefont {Hotter},
  \citenamefont {Miranowicz},\ and\ \citenamefont
  {Gietka}}]{Gietka2025PRL100802}%
  \BibitemOpen
  \bibfield  {author} {\bibinfo {author} {\bibfnamefont {C.}~\bibnamefont
  {Hotter}}, \bibinfo {author} {\bibfnamefont {A.}~\bibnamefont {Miranowicz}},
  \ and\ \bibinfo {author} {\bibfnamefont {K.}~\bibnamefont {Gietka}},\ }\href
  {\doibase 10.1103/cxvs-5pb1} {\bibfield  {journal} {\bibinfo  {journal}
  {Phys. Rev. Lett.}\ }\textbf {\bibinfo {volume} {135}},\ \bibinfo {pages}
  {100802} (\bibinfo {year} {2025})}\BibitemShut {NoStop}%
\bibitem [{\citenamefont {Mihailescu}\ \emph {et~al.}(2026)\citenamefont
  {Mihailescu}, \citenamefont {Alushi}, \citenamefont {Di~Candia},
  \citenamefont {Felicetti},\ and\ \citenamefont
  {Gietka}}]{Mihailescu2025CQMtutorial}%
  \BibitemOpen
  \bibfield  {author} {\bibinfo {author} {\bibfnamefont {G.}~\bibnamefont
  {Mihailescu}}, \bibinfo {author} {\bibfnamefont {U.}~\bibnamefont {Alushi}},
  \bibinfo {author} {\bibfnamefont {R.}~\bibnamefont {Di~Candia}}, \bibinfo
  {author} {\bibfnamefont {S.}~\bibnamefont {Felicetti}}, \ and\ \bibinfo
  {author} {\bibfnamefont {K.}~\bibnamefont {Gietka}},\ }\href {\doibase
  10.1103/v7mf-yh8n} {\bibfield  {journal} {\bibinfo  {journal} {PRX Quantum}\
  }\textbf {\bibinfo {volume} {7}},\ \bibinfo {pages} {020201} (\bibinfo {year}
  {2026})}\BibitemShut {NoStop}%
\bibitem [{\citenamefont {Qiao}\ and\ \citenamefont
  {Ying}(2026)}]{QiaoFeng2026SpinSqueeze}%
  \BibitemOpen
  \bibfield  {author} {\bibinfo {author} {\bibfnamefont {F.}~\bibnamefont
  {Qiao}}\ and\ \bibinfo {author} {\bibfnamefont {Z.-J.}\ \bibnamefont
  {Ying}},\ }\href {\doibase 10.1103/cmb2-sktt} {\bibfield  {journal} {\bibinfo
   {journal} {Phys. Rev. A}\ }\textbf {\bibinfo {volume} {113}},\ \bibinfo
  {pages} {043503} (\bibinfo {year} {2026})}\BibitemShut {NoStop}%
\bibitem [{\citenamefont {Chen}\ \emph {et~al.}(2026)\citenamefont {Chen},
  \citenamefont {Qiao},\ and\ \citenamefont {Ying}}]{QiuYi2025gA2}%
  \BibitemOpen
  \bibfield  {author} {\bibinfo {author} {\bibfnamefont {Q.-Y.}\ \bibnamefont
  {Chen}}, \bibinfo {author} {\bibfnamefont {F.}~\bibnamefont {Qiao}}, \ and\
  \bibinfo {author} {\bibfnamefont {Z.-J.}\ \bibnamefont {Ying}},\ }\href
  {\doibase 10.1103/gbcp-2crx} {\bibfield  {journal} {\bibinfo  {journal}
  {Phys. Rev. A}\ }\textbf {\bibinfo {volume} {113}},\ \bibinfo {pages}
  {062423} (\bibinfo {year} {2026})}\BibitemShut {NoStop}%
\bibitem [{\citenamefont {Galitski}\ and\ \citenamefont
  {Spielman}(2013)}]{LinRashbaBECExp2013Review}%
  \BibitemOpen
  \bibfield  {author} {\bibinfo {author} {\bibfnamefont {V.}~\bibnamefont
  {Galitski}}\ and\ \bibinfo {author} {\bibfnamefont {I.~B.}\ \bibnamefont
  {Spielman}},\ }\href {\doibase 10.1038/nature11841} {\bibfield  {journal}
  {\bibinfo  {journal} {Nature}\ }\textbf {\bibinfo {volume} {494}},\ \bibinfo
  {pages} {49} (\bibinfo {year} {2013})}\BibitemShut {NoStop}%
\bibitem [{\citenamefont {Lin}\ \emph {et~al.}(2011)\citenamefont {Lin},
  \citenamefont {Jim\'{e}nez-Garc\'{\i}a},\ and\ \citenamefont
  {Spielman}}]{LinRashbaBECExp2011}%
  \BibitemOpen
  \bibfield  {author} {\bibinfo {author} {\bibfnamefont {Y.-J.}\ \bibnamefont
  {Lin}}, \bibinfo {author} {\bibfnamefont {K.}~\bibnamefont
  {Jim\'{e}nez-Garc\'{\i}a}}, \ and\ \bibinfo {author} {\bibfnamefont {I.~B.}\
  \bibnamefont {Spielman}},\ }\href {\doibase 10.1038/nature09887} {\bibfield
  {journal} {\bibinfo  {journal} {Nature}\ }\textbf {\bibinfo {volume} {471}},\
  \bibinfo {pages} {83} (\bibinfo {year} {2011})}\BibitemShut {NoStop}%
\bibitem [{\citenamefont {Liu}\ and\ \citenamefont
  {Ying}(2025{\natexlab{a}})}]{LiuYing02025exoticSOC2Ring}%
  \BibitemOpen
  \bibfield  {author} {\bibinfo {author} {\bibfnamefont {Y.}~\bibnamefont
  {Liu}}\ and\ \bibinfo {author} {\bibfnamefont {Z.-J.}\ \bibnamefont {Ying}},\
  }\href {\doibase https://doi.org/10.1002/qute.202500431} {\bibfield
  {journal} {\bibinfo  {journal} {Adv. Quantum Technol.}\ }\textbf {\bibinfo
  {volume} {8}},\ \bibinfo {pages} {e00431} (\bibinfo {year}
  {2025}{\natexlab{a}})}\BibitemShut {NoStop}%
\bibitem [{\citenamefont {Liu}\ and\ \citenamefont
  {Ying}(2025{\natexlab{b}})}]{LiuYing02025exoticSOC2Ring-Cover}%
  \BibitemOpen
  \bibfield  {author} {\bibinfo {author} {\bibfnamefont {Y.}~\bibnamefont
  {Liu}}\ and\ \bibinfo {author} {\bibfnamefont {Z.-J.}\ \bibnamefont {Ying}},\
  }\href {\doibase https://doi.org/10.1002/qute.70062} {\bibfield  {journal}
  {\bibinfo  {journal} {Adv. Quantum Technol.}\ }\textbf {\bibinfo {volume}
  {8}},\ \bibinfo {pages} {e70062} (\bibinfo {year} {2025}{\natexlab{b}})},\
  \bibinfo {note} {[Front Cover: (Adv. Quantum Technol. 11/2025)]}\BibitemShut
  {NoStop}%
\bibitem [{\citenamefont {Liu}\ and\ \citenamefont
  {Ying}(2025{\natexlab{c}})}]{LiuYing02025KaleidoscopeDDI}%
  \BibitemOpen
  \bibfield  {author} {\bibinfo {author} {\bibfnamefont {Y.}~\bibnamefont
  {Liu}}\ and\ \bibinfo {author} {\bibfnamefont {Z.-J.}\ \bibnamefont {Ying}},\
  }\href {\doibase https://doi.org/10.1002/qute.202500475} {\bibfield
  {journal} {\bibinfo  {journal} {Adv. Quantum Technol.}\ }\textbf {\bibinfo
  {volume} {8}},\ \bibinfo {pages} {e00475} (\bibinfo {year}
  {2025}{\natexlab{c}})}\BibitemShut {NoStop}%
\bibitem [{\citenamefont {Wallraff}\ \emph {et~al.}(2004)\citenamefont
  {Wallraff}, \citenamefont {Schuster}, \citenamefont {Blais}, \citenamefont
  {Frunzio}, \citenamefont {Huang}, \citenamefont {Majer}, \citenamefont
  {Kumar}, \citenamefont {Girvin},\ and\ \citenamefont
  {Schoelkopf}}]{Wallraff2004}%
  \BibitemOpen
  \bibfield  {author} {\bibinfo {author} {\bibfnamefont {A.}~\bibnamefont
  {Wallraff}}, \bibinfo {author} {\bibfnamefont {D.~I.}\ \bibnamefont
  {Schuster}}, \bibinfo {author} {\bibfnamefont {A.}~\bibnamefont {Blais}},
  \bibinfo {author} {\bibfnamefont {L.}~\bibnamefont {Frunzio}}, \bibinfo
  {author} {\bibfnamefont {R.-S.}\ \bibnamefont {Huang}}, \bibinfo {author}
  {\bibfnamefont {J.}~\bibnamefont {Majer}}, \bibinfo {author} {\bibfnamefont
  {S.}~\bibnamefont {Kumar}}, \bibinfo {author} {\bibfnamefont {S.~M.}\
  \bibnamefont {Girvin}}, \ and\ \bibinfo {author} {\bibfnamefont {R.~J.}\
  \bibnamefont {Schoelkopf}},\ }\href {\doibase 10.1038/nature02851} {\bibfield
   {journal} {\bibinfo  {journal} {Nature}\ }\textbf {\bibinfo {volume}
  {431}},\ \bibinfo {pages} {162} (\bibinfo {year} {2004})}\BibitemShut
  {NoStop}%
\bibitem [{\citenamefont {G\"{u}nter}\ \emph {et~al.}(2009)\citenamefont
  {G\"{u}nter}, \citenamefont {Anappara}, \citenamefont {Hees}, \citenamefont
  {Sell}, \citenamefont {Biasiol}, \citenamefont {Sorba}, \citenamefont
  {De~Liberato}, \citenamefont {Ciuti}, \citenamefont {Tredicucci},
  \citenamefont {Leitenstorfer},\ and\ \citenamefont {Huber}}]{Gunter2009}%
  \BibitemOpen
  \bibfield  {author} {\bibinfo {author} {\bibfnamefont {G.}~\bibnamefont
  {G\"{u}nter}}, \bibinfo {author} {\bibfnamefont {A.~A.}\ \bibnamefont
  {Anappara}}, \bibinfo {author} {\bibfnamefont {J.}~\bibnamefont {Hees}},
  \bibinfo {author} {\bibfnamefont {A.}~\bibnamefont {Sell}}, \bibinfo {author}
  {\bibfnamefont {G.}~\bibnamefont {Biasiol}}, \bibinfo {author} {\bibfnamefont
  {L.}~\bibnamefont {Sorba}}, \bibinfo {author} {\bibfnamefont
  {S.}~\bibnamefont {De~Liberato}}, \bibinfo {author} {\bibfnamefont
  {C.}~\bibnamefont {Ciuti}}, \bibinfo {author} {\bibfnamefont
  {A.}~\bibnamefont {Tredicucci}}, \bibinfo {author} {\bibfnamefont
  {A.}~\bibnamefont {Leitenstorfer}}, \ and\ \bibinfo {author} {\bibfnamefont
  {R.}~\bibnamefont {Huber}},\ }\href {\doibase 10.1038/nature07838} {\bibfield
   {journal} {\bibinfo  {journal} {Nature}\ }\textbf {\bibinfo {volume}
  {458}},\ \bibinfo {pages} {178} (\bibinfo {year} {2009})}\BibitemShut
  {NoStop}%
\bibitem [{\citenamefont {Niemczyk}\ \emph {et~al.}(2010)\citenamefont
  {Niemczyk}, \citenamefont {Deppe}, \citenamefont {Huebl}, \citenamefont
  {Menzel}, \citenamefont {Hocke}, \citenamefont {Schwarz}, \citenamefont
  {Garcia-Ripoll}, \citenamefont {Zueco}, \citenamefont {H\"{u}mmer},
  \citenamefont {Solano}, \citenamefont {Marx},\ and\ \citenamefont
  {Gross}}]{Niemczyk2010}%
  \BibitemOpen
  \bibfield  {author} {\bibinfo {author} {\bibfnamefont {T.}~\bibnamefont
  {Niemczyk}}, \bibinfo {author} {\bibfnamefont {F.}~\bibnamefont {Deppe}},
  \bibinfo {author} {\bibfnamefont {H.}~\bibnamefont {Huebl}}, \bibinfo
  {author} {\bibfnamefont {E.~P.}\ \bibnamefont {Menzel}}, \bibinfo {author}
  {\bibfnamefont {F.}~\bibnamefont {Hocke}}, \bibinfo {author} {\bibfnamefont
  {M.~J.}\ \bibnamefont {Schwarz}}, \bibinfo {author} {\bibfnamefont {J.~J.}\
  \bibnamefont {Garcia-Ripoll}}, \bibinfo {author} {\bibfnamefont
  {D.}~\bibnamefont {Zueco}}, \bibinfo {author} {\bibfnamefont
  {T.}~\bibnamefont {H\"{u}mmer}}, \bibinfo {author} {\bibfnamefont
  {E.}~\bibnamefont {Solano}}, \bibinfo {author} {\bibfnamefont
  {A.}~\bibnamefont {Marx}}, \ and\ \bibinfo {author} {\bibfnamefont
  {R.}~\bibnamefont {Gross}},\ }\href {\doibase 10.1038/nphys1730} {\bibfield
  {journal} {\bibinfo  {journal} {Nat. Phys.}\ }\textbf {\bibinfo {volume}
  {6}},\ \bibinfo {pages} {772} (\bibinfo {year} {2010})}\BibitemShut {NoStop}%
\bibitem [{\citenamefont {Peropadre}\ \emph {et~al.}(2010)\citenamefont
  {Peropadre}, \citenamefont {Forn-D\'{\i}az}, \citenamefont {Solano},\ and\
  \citenamefont {Garc\'{\i}a-Ripoll}}]{Peropadre2010}%
  \BibitemOpen
  \bibfield  {author} {\bibinfo {author} {\bibfnamefont {B.}~\bibnamefont
  {Peropadre}}, \bibinfo {author} {\bibfnamefont {P.}~\bibnamefont
  {Forn-D\'{\i}az}}, \bibinfo {author} {\bibfnamefont {E.}~\bibnamefont
  {Solano}}, \ and\ \bibinfo {author} {\bibfnamefont {J.~J.}\ \bibnamefont
  {Garc\'{\i}a-Ripoll}},\ }\href {\doibase 10.1103/PhysRevLett.105.023601}
  {\bibfield  {journal} {\bibinfo  {journal} {Phys. Rev. Lett.}\ }\textbf
  {\bibinfo {volume} {105}},\ \bibinfo {pages} {023601} (\bibinfo {year}
  {2010})}\BibitemShut {NoStop}%
\bibitem [{\citenamefont {Forn-D\'{\i}az}\ \emph {et~al.}(2017)\citenamefont
  {Forn-D\'{\i}az}, \citenamefont {Garc\'{\i}a-Ripoll}, \citenamefont
  {Peropadre}, \citenamefont {Orgiazzi}, \citenamefont {Yurtalan},
  \citenamefont {Belyansky}, \citenamefont {Wilson},\ and\ \citenamefont
  {Lupascu}}]{FornDiaz2017}%
  \BibitemOpen
  \bibfield  {author} {\bibinfo {author} {\bibfnamefont {P.}~\bibnamefont
  {Forn-D\'{\i}az}}, \bibinfo {author} {\bibfnamefont {J.~J.}\ \bibnamefont
  {Garc\'{\i}a-Ripoll}}, \bibinfo {author} {\bibfnamefont {B.}~\bibnamefont
  {Peropadre}}, \bibinfo {author} {\bibfnamefont {J.-L.}\ \bibnamefont
  {Orgiazzi}}, \bibinfo {author} {\bibfnamefont {M.~A.}\ \bibnamefont
  {Yurtalan}}, \bibinfo {author} {\bibfnamefont {R.}~\bibnamefont {Belyansky}},
  \bibinfo {author} {\bibfnamefont {C.~M.}\ \bibnamefont {Wilson}}, \ and\
  \bibinfo {author} {\bibfnamefont {A.}~\bibnamefont {Lupascu}},\ }\href
  {\doibase 10.1038/nphys3905} {\bibfield  {journal} {\bibinfo  {journal} {Nat.
  Phys.}\ }\textbf {\bibinfo {volume} {13}},\ \bibinfo {pages} {39} (\bibinfo
  {year} {2017})}\BibitemShut {NoStop}%
\bibitem [{\citenamefont {Ciuti}\ \emph {et~al.}(2005)\citenamefont {Ciuti},
  \citenamefont {Bastard},\ and\ \citenamefont
  {Carusotto}}]{Ciuti2005EarlyUSC}%
  \BibitemOpen
  \bibfield  {author} {\bibinfo {author} {\bibfnamefont {C.}~\bibnamefont
  {Ciuti}}, \bibinfo {author} {\bibfnamefont {G.}~\bibnamefont {Bastard}}, \
  and\ \bibinfo {author} {\bibfnamefont {I.}~\bibnamefont {Carusotto}},\ }\href
  {\doibase 10.1103/PhysRevB.72.115303} {\bibfield  {journal} {\bibinfo
  {journal} {Phys. Rev. B}\ }\textbf {\bibinfo {volume} {72}},\ \bibinfo
  {pages} {115303} (\bibinfo {year} {2005})}\BibitemShut {NoStop}%
\bibitem [{\citenamefont {Anappara}\ \emph {et~al.}(2009)\citenamefont
  {Anappara}, \citenamefont {De~Liberato}, \citenamefont {Tredicucci},
  \citenamefont {Ciuti}, \citenamefont {Biasiol}, \citenamefont {Sorba},\ and\
  \citenamefont {Beltram}}]{Aji2009EarlyUSC}%
  \BibitemOpen
  \bibfield  {author} {\bibinfo {author} {\bibfnamefont {A.~A.}\ \bibnamefont
  {Anappara}}, \bibinfo {author} {\bibfnamefont {S.}~\bibnamefont
  {De~Liberato}}, \bibinfo {author} {\bibfnamefont {A.}~\bibnamefont
  {Tredicucci}}, \bibinfo {author} {\bibfnamefont {C.}~\bibnamefont {Ciuti}},
  \bibinfo {author} {\bibfnamefont {G.}~\bibnamefont {Biasiol}}, \bibinfo
  {author} {\bibfnamefont {L.}~\bibnamefont {Sorba}}, \ and\ \bibinfo {author}
  {\bibfnamefont {F.}~\bibnamefont {Beltram}},\ }\href {\doibase
  10.1103/PhysRevB.79.201303} {\bibfield  {journal} {\bibinfo  {journal} {Phys.
  Rev. B}\ }\textbf {\bibinfo {volume} {79}},\ \bibinfo {pages} {201303(R)}
  (\bibinfo {year} {2009})}\BibitemShut {NoStop}%
\bibitem [{\citenamefont {Forn-D\'{\i}az}\ \emph {et~al.}(2010)\citenamefont
  {Forn-D\'{\i}az}, \citenamefont {Lisenfeld}, \citenamefont {Marcos},
  \citenamefont {Garc\'{\i}a-Ripoll}, \citenamefont {Solano}, \citenamefont
  {Harmans},\ and\ \citenamefont {Mooij}}]{Forn-Diaz2010}%
  \BibitemOpen
  \bibfield  {author} {\bibinfo {author} {\bibfnamefont {P.}~\bibnamefont
  {Forn-D\'{\i}az}}, \bibinfo {author} {\bibfnamefont {J.}~\bibnamefont
  {Lisenfeld}}, \bibinfo {author} {\bibfnamefont {D.}~\bibnamefont {Marcos}},
  \bibinfo {author} {\bibfnamefont {J.~J.}\ \bibnamefont {Garc\'{\i}a-Ripoll}},
  \bibinfo {author} {\bibfnamefont {E.}~\bibnamefont {Solano}}, \bibinfo
  {author} {\bibfnamefont {C.~J. P.~M.}\ \bibnamefont {Harmans}}, \ and\
  \bibinfo {author} {\bibfnamefont {J.~E.}\ \bibnamefont {Mooij}},\ }\href
  {\doibase 10.1103/PhysRevLett.105.237001} {\bibfield  {journal} {\bibinfo
  {journal} {Phys. Rev. Lett.}\ }\textbf {\bibinfo {volume} {105}},\ \bibinfo
  {pages} {237001} (\bibinfo {year} {2010})}\BibitemShut {NoStop}%
\bibitem [{\citenamefont {Scalari}\ \emph {et~al.}(2012)\citenamefont
  {Scalari}, \citenamefont {Maissen}, \citenamefont {Tur\v{c}inkov\'{a}},
  \citenamefont {Hagenm\"{u}ller}, \citenamefont {De~Liberato}, \citenamefont
  {Ciuti}, \citenamefont {Reichl}, \citenamefont {Schuh}, \citenamefont
  {Wegscheider}, \citenamefont {Beck},\ and\ \citenamefont
  {Faist}}]{Scalari2012}%
  \BibitemOpen
  \bibfield  {author} {\bibinfo {author} {\bibfnamefont {G.}~\bibnamefont
  {Scalari}}, \bibinfo {author} {\bibfnamefont {C.}~\bibnamefont {Maissen}},
  \bibinfo {author} {\bibfnamefont {D.}~\bibnamefont {Tur\v{c}inkov\'{a}}},
  \bibinfo {author} {\bibfnamefont {D.}~\bibnamefont {Hagenm\"{u}ller}},
  \bibinfo {author} {\bibfnamefont {S.}~\bibnamefont {De~Liberato}}, \bibinfo
  {author} {\bibfnamefont {C.}~\bibnamefont {Ciuti}}, \bibinfo {author}
  {\bibfnamefont {C.}~\bibnamefont {Reichl}}, \bibinfo {author} {\bibfnamefont
  {D.}~\bibnamefont {Schuh}}, \bibinfo {author} {\bibfnamefont
  {W.}~\bibnamefont {Wegscheider}}, \bibinfo {author} {\bibfnamefont
  {M.}~\bibnamefont {Beck}}, \ and\ \bibinfo {author} {\bibfnamefont
  {J.}~\bibnamefont {Faist}},\ }\href {\doibase 10.1126/science.1216022}
  {\bibfield  {journal} {\bibinfo  {journal} {Science}\ }\textbf {\bibinfo
  {volume} {335}},\ \bibinfo {pages} {1323} (\bibinfo {year}
  {2012})}\BibitemShut {NoStop}%
\bibitem [{\citenamefont {Xiang}\ \emph {et~al.}(2013)\citenamefont {Xiang},
  \citenamefont {Ashhab}, \citenamefont {You},\ and\ \citenamefont
  {Nori}}]{Xiang2013}%
  \BibitemOpen
  \bibfield  {author} {\bibinfo {author} {\bibfnamefont {Z.-L.}\ \bibnamefont
  {Xiang}}, \bibinfo {author} {\bibfnamefont {S.}~\bibnamefont {Ashhab}},
  \bibinfo {author} {\bibfnamefont {J.~Q.}\ \bibnamefont {You}}, \ and\
  \bibinfo {author} {\bibfnamefont {F.}~\bibnamefont {Nori}},\ }\href {\doibase
  10.1103/RevModPhys.85.623} {\bibfield  {journal} {\bibinfo  {journal} {Rev.
  Mod. Phys.}\ }\textbf {\bibinfo {volume} {85}},\ \bibinfo {pages} {623}
  (\bibinfo {year} {2013})}\BibitemShut {NoStop}%
\bibitem [{\citenamefont {Yoshihara}\ \emph {et~al.}(2017)\citenamefont
  {Yoshihara}, \citenamefont {Fuse}, \citenamefont {Ashhab}, \citenamefont
  {Kakuyanagi}, \citenamefont {Saito},\ and\ \citenamefont
  {Semba}}]{Yoshihara2017NatPhys}%
  \BibitemOpen
  \bibfield  {author} {\bibinfo {author} {\bibfnamefont {F.}~\bibnamefont
  {Yoshihara}}, \bibinfo {author} {\bibfnamefont {T.}~\bibnamefont {Fuse}},
  \bibinfo {author} {\bibfnamefont {S.}~\bibnamefont {Ashhab}}, \bibinfo
  {author} {\bibfnamefont {K.}~\bibnamefont {Kakuyanagi}}, \bibinfo {author}
  {\bibfnamefont {S.}~\bibnamefont {Saito}}, \ and\ \bibinfo {author}
  {\bibfnamefont {K.}~\bibnamefont {Semba}},\ }\href {\doibase
  10.1038/nphys3906} {\bibfield  {journal} {\bibinfo  {journal} {Nat. Phys.}\
  }\textbf {\bibinfo {volume} {13}},\ \bibinfo {pages} {44} (\bibinfo {year}
  {2017})}\BibitemShut {NoStop}%
\bibitem [{\citenamefont {Gu}\ \emph {et~al.}(2017)\citenamefont {Gu},
  \citenamefont {Kockum}, \citenamefont {Miranowicz}, \citenamefont {Liu},\
  and\ \citenamefont {Nori}}]{Kockum2017}%
  \BibitemOpen
  \bibfield  {author} {\bibinfo {author} {\bibfnamefont {X.}~\bibnamefont
  {Gu}}, \bibinfo {author} {\bibfnamefont {A.~F.}\ \bibnamefont {Kockum}},
  \bibinfo {author} {\bibfnamefont {A.}~\bibnamefont {Miranowicz}}, \bibinfo
  {author} {\bibfnamefont {Y.-X.}\ \bibnamefont {Liu}}, \ and\ \bibinfo
  {author} {\bibfnamefont {F.}~\bibnamefont {Nori}},\ }\href {\doibase
  https://doi.org/10.1016/j.physrep.2017.10.002} {\bibfield  {journal}
  {\bibinfo  {journal} {Physics Reports}\ }\textbf {\bibinfo {volume}
  {718-719}},\ \bibinfo {pages} {1 } (\bibinfo {year} {2017})}\BibitemShut
  {NoStop}%
\bibitem [{\citenamefont {Qin}\ \emph {et~al.}(2018)\citenamefont {Qin},
  \citenamefont {Miranowicz}, \citenamefont {Li}, \citenamefont {L\"u},
  \citenamefont {You},\ and\ \citenamefont {Nori}}]{Qin-ExpLightMatter-2018}%
  \BibitemOpen
  \bibfield  {author} {\bibinfo {author} {\bibfnamefont {W.}~\bibnamefont
  {Qin}}, \bibinfo {author} {\bibfnamefont {A.}~\bibnamefont {Miranowicz}},
  \bibinfo {author} {\bibfnamefont {P.-B.}\ \bibnamefont {Li}}, \bibinfo
  {author} {\bibfnamefont {X.-Y.}\ \bibnamefont {L\"u}}, \bibinfo {author}
  {\bibfnamefont {J.~Q.}\ \bibnamefont {You}}, \ and\ \bibinfo {author}
  {\bibfnamefont {F.}~\bibnamefont {Nori}},\ }\href {\doibase
  10.1103/PhysRevLett.120.093601} {\bibfield  {journal} {\bibinfo  {journal}
  {Phys. Rev. Lett.}\ }\textbf {\bibinfo {volume} {120}},\ \bibinfo {pages}
  {093601} (\bibinfo {year} {2018})}\BibitemShut {NoStop}%
\bibitem [{\citenamefont {Pan}\ \emph {et~al.}(2024)\citenamefont {Pan},
  \citenamefont {Li}, \citenamefont {Hei}, \citenamefont {Zhang}, \citenamefont
  {Mochizuki}, \citenamefont {Li},\ and\ \citenamefont
  {Nori}}]{LiPengBo-Magnon-PRL-2024}%
  \BibitemOpen
  \bibfield  {author} {\bibinfo {author} {\bibfnamefont {X.-F.}\ \bibnamefont
  {Pan}}, \bibinfo {author} {\bibfnamefont {P.-B.}\ \bibnamefont {Li}},
  \bibinfo {author} {\bibfnamefont {X.-L.}\ \bibnamefont {Hei}}, \bibinfo
  {author} {\bibfnamefont {X.}~\bibnamefont {Zhang}}, \bibinfo {author}
  {\bibfnamefont {M.}~\bibnamefont {Mochizuki}}, \bibinfo {author}
  {\bibfnamefont {F.-L.}\ \bibnamefont {Li}}, \ and\ \bibinfo {author}
  {\bibfnamefont {F.}~\bibnamefont {Nori}},\ }\href {\doibase
  10.1103/PhysRevLett.132.193601} {\bibfield  {journal} {\bibinfo  {journal}
  {Phys. Rev. Lett.}\ }\textbf {\bibinfo {volume} {132}},\ \bibinfo {pages}
  {193601} (\bibinfo {year} {2024})}\BibitemShut {NoStop}%
\bibitem [{\citenamefont {Bayer}\ \emph {et~al.}(2017)\citenamefont {Bayer},
  \citenamefont {Pozimski}, \citenamefont {Schambeck}, \citenamefont {Schuh},
  \citenamefont {Huber}, \citenamefont {Bougeard},\ and\ \citenamefont
  {Lange}}]{Bayer2017DeepStrong}%
  \BibitemOpen
  \bibfield  {author} {\bibinfo {author} {\bibfnamefont {A.}~\bibnamefont
  {Bayer}}, \bibinfo {author} {\bibfnamefont {M.}~\bibnamefont {Pozimski}},
  \bibinfo {author} {\bibfnamefont {S.}~\bibnamefont {Schambeck}}, \bibinfo
  {author} {\bibfnamefont {D.}~\bibnamefont {Schuh}}, \bibinfo {author}
  {\bibfnamefont {R.}~\bibnamefont {Huber}}, \bibinfo {author} {\bibfnamefont
  {D.}~\bibnamefont {Bougeard}}, \ and\ \bibinfo {author} {\bibfnamefont
  {C.}~\bibnamefont {Lange}},\ }\href {\doibase 10.1021/acs.nanolett.7b03103}
  {\bibfield  {journal} {\bibinfo  {journal} {Nano letters}\ }\textbf {\bibinfo
  {volume} {17}},\ \bibinfo {pages} {6340} (\bibinfo {year}
  {2017})}\BibitemShut {NoStop}%
\bibitem [{\citenamefont {Huang}\ \emph {et~al.}(2020)\citenamefont {Huang},
  \citenamefont {Liao},\ and\ \citenamefont {Kuang}}]{Ulstrong-JC-2}%
  \BibitemOpen
  \bibfield  {author} {\bibinfo {author} {\bibfnamefont {J.-F.}\ \bibnamefont
  {Huang}}, \bibinfo {author} {\bibfnamefont {J.-Q.}\ \bibnamefont {Liao}}, \
  and\ \bibinfo {author} {\bibfnamefont {L.-M.}\ \bibnamefont {Kuang}},\ }\href
  {\doibase 10.1103/PhysRevA.101.043835} {\bibfield  {journal} {\bibinfo
  {journal} {Phys. Rev. A}\ }\textbf {\bibinfo {volume} {101}},\ \bibinfo
  {pages} {043835} (\bibinfo {year} {2020})}\BibitemShut {NoStop}%
\bibitem [{\citenamefont {Wang}\ \emph {et~al.}(2023)\citenamefont {Wang},
  \citenamefont {Ridolfo}, \citenamefont {Li}, \citenamefont {Savasta},
  \citenamefont {Nori}, \citenamefont {Nakamura},\ and\ \citenamefont
  {You}}]{WangYouJQ2023DeepStrong}%
  \BibitemOpen
  \bibfield  {author} {\bibinfo {author} {\bibfnamefont {S.-P.}\ \bibnamefont
  {Wang}}, \bibinfo {author} {\bibfnamefont {A.}~\bibnamefont {Ridolfo}},
  \bibinfo {author} {\bibfnamefont {T.}~\bibnamefont {Li}}, \bibinfo {author}
  {\bibfnamefont {S.}~\bibnamefont {Savasta}}, \bibinfo {author} {\bibfnamefont
  {F.}~\bibnamefont {Nori}}, \bibinfo {author} {\bibfnamefont {Y.}~\bibnamefont
  {Nakamura}}, \ and\ \bibinfo {author} {\bibfnamefont {J.~Q.}\ \bibnamefont
  {You}},\ }\href {\doibase 10.1038/s41467-023-40097-0} {\bibfield  {journal}
  {\bibinfo  {journal} {Nat. Commun.}\ }\textbf {\bibinfo {volume} {14}},\
  \bibinfo {pages} {4397} (\bibinfo {year} {2023})}\BibitemShut {NoStop}%
\bibitem [{\citenamefont {Liu}\ and\ \citenamefont
  {Huang}(2024)}]{DeepStrong-JC-Huang-2024}%
  \BibitemOpen
  \bibfield  {author} {\bibinfo {author} {\bibfnamefont {C.}~\bibnamefont
  {Liu}}\ and\ \bibinfo {author} {\bibfnamefont {J.-F.}\ \bibnamefont
  {Huang}},\ }\href {\doibase https://doi.org/10.1007/s11433-023-2243-7}
  {\bibfield  {journal} {\bibinfo  {journal} {Sci. China Phys. Mech. Astron.}\
  }\textbf {\bibinfo {volume} {67}},\ \bibinfo {pages} {210311} (\bibinfo
  {year} {2024})}\BibitemShut {NoStop}%
\bibitem [{\citenamefont {Braak}(2011)}]{Braak2011}%
  \BibitemOpen
  \bibfield  {author} {\bibinfo {author} {\bibfnamefont {D.}~\bibnamefont
  {Braak}},\ }\href {\doibase 10.1103/PhysRevLett.107.100401} {\bibfield
  {journal} {\bibinfo  {journal} {Phys. Rev. Lett.}\ }\textbf {\bibinfo
  {volume} {107}},\ \bibinfo {pages} {100401} (\bibinfo {year}
  {2011})}\BibitemShut {NoStop}%
\bibitem [{\citenamefont {Chen}\ \emph {et~al.}(2012)\citenamefont {Chen},
  \citenamefont {Wang}, \citenamefont {He}, \citenamefont {Liu},\ and\
  \citenamefont {Wang}}]{ChenQH2012}%
  \BibitemOpen
  \bibfield  {author} {\bibinfo {author} {\bibfnamefont {Q.-H.}\ \bibnamefont
  {Chen}}, \bibinfo {author} {\bibfnamefont {C.}~\bibnamefont {Wang}}, \bibinfo
  {author} {\bibfnamefont {S.}~\bibnamefont {He}}, \bibinfo {author}
  {\bibfnamefont {T.}~\bibnamefont {Liu}}, \ and\ \bibinfo {author}
  {\bibfnamefont {K.-L.}\ \bibnamefont {Wang}},\ }\href {\doibase
  10.1103/PhysRevA.86.023822} {\bibfield  {journal} {\bibinfo  {journal} {Phys.
  Rev. A}\ }\textbf {\bibinfo {volume} {86}},\ \bibinfo {pages} {023822}
  (\bibinfo {year} {2012})}\BibitemShut {NoStop}%
\bibitem [{\citenamefont {Batchelor}\ and\ \citenamefont
  {Zhou}(2015)}]{Batchelor2015}%
  \BibitemOpen
  \bibfield  {author} {\bibinfo {author} {\bibfnamefont {M.~T.}\ \bibnamefont
  {Batchelor}}\ and\ \bibinfo {author} {\bibfnamefont {H.-Q.}\ \bibnamefont
  {Zhou}},\ }\href {\doibase 10.1103/PhysRevA.91.053808} {\bibfield  {journal}
  {\bibinfo  {journal} {Phys. Rev. A}\ }\textbf {\bibinfo {volume} {91}},\
  \bibinfo {pages} {053808} (\bibinfo {year} {2015})}\BibitemShut {NoStop}%
\bibitem [{\citenamefont {Ashhab}(2013)}]{Ashhab2013}%
  \BibitemOpen
  \bibfield  {author} {\bibinfo {author} {\bibfnamefont {S.}~\bibnamefont
  {Ashhab}},\ }\href {\doibase 10.1103/PhysRevA.87.013826} {\bibfield
  {journal} {\bibinfo  {journal} {Phys. Rev. A}\ }\textbf {\bibinfo {volume}
  {87}},\ \bibinfo {pages} {013826} (\bibinfo {year} {2013})}\BibitemShut
  {NoStop}%
\bibitem [{\citenamefont {Ying}\ \emph {et~al.}(2015)\citenamefont {Ying},
  \citenamefont {Liu}, \citenamefont {Luo}, \citenamefont {Lin},\ and\
  \citenamefont {You}}]{Ying2015}%
  \BibitemOpen
  \bibfield  {author} {\bibinfo {author} {\bibfnamefont {Z.-J.}\ \bibnamefont
  {Ying}}, \bibinfo {author} {\bibfnamefont {M.}~\bibnamefont {Liu}}, \bibinfo
  {author} {\bibfnamefont {H.-G.}\ \bibnamefont {Luo}}, \bibinfo {author}
  {\bibfnamefont {H.-Q.}\ \bibnamefont {Lin}}, \ and\ \bibinfo {author}
  {\bibfnamefont {J.~Q.}\ \bibnamefont {You}},\ }\href {\doibase
  10.1103/PhysRevA.92.053823} {\bibfield  {journal} {\bibinfo  {journal} {Phys.
  Rev. A}\ }\textbf {\bibinfo {volume} {92}},\ \bibinfo {pages} {053823}
  (\bibinfo {year} {2015})}\BibitemShut {NoStop}%
\bibitem [{\citenamefont {Liu}\ \emph {et~al.}(2017)\citenamefont {Liu},
  \citenamefont {Chesi}, \citenamefont {Ying}, \citenamefont {Chen},
  \citenamefont {Luo},\ and\ \citenamefont {Lin}}]{LiuM2017PRL}%
  \BibitemOpen
  \bibfield  {author} {\bibinfo {author} {\bibfnamefont {M.}~\bibnamefont
  {Liu}}, \bibinfo {author} {\bibfnamefont {S.}~\bibnamefont {Chesi}}, \bibinfo
  {author} {\bibfnamefont {Z.-J.}\ \bibnamefont {Ying}}, \bibinfo {author}
  {\bibfnamefont {X.}~\bibnamefont {Chen}}, \bibinfo {author} {\bibfnamefont
  {H.-G.}\ \bibnamefont {Luo}}, \ and\ \bibinfo {author} {\bibfnamefont
  {H.-Q.}\ \bibnamefont {Lin}},\ }\href {\doibase
  10.1103/PhysRevLett.119.220601} {\bibfield  {journal} {\bibinfo  {journal}
  {Phys. Rev. Lett.}\ }\textbf {\bibinfo {volume} {119}},\ \bibinfo {pages}
  {220601} (\bibinfo {year} {2017})}\BibitemShut {NoStop}%
\bibitem [{\citenamefont {Liu}\ \emph {et~al.}(2021)\citenamefont {Liu},
  \citenamefont {Liu}, \citenamefont {Ying},\ and\ \citenamefont
  {Luo}}]{Liu2021AQT}%
  \BibitemOpen
  \bibfield  {author} {\bibinfo {author} {\bibfnamefont {J.}~\bibnamefont
  {Liu}}, \bibinfo {author} {\bibfnamefont {M.}~\bibnamefont {Liu}}, \bibinfo
  {author} {\bibfnamefont {Z.-J.}\ \bibnamefont {Ying}}, \ and\ \bibinfo
  {author} {\bibfnamefont {H.-G.}\ \bibnamefont {Luo}},\ }\href {\doibase
  https://doi.org/10.1002/qute.202000139} {\bibfield  {journal} {\bibinfo
  {journal} {Adv. Quantum Technol.}\ }\textbf {\bibinfo {volume} {4}},\
  \bibinfo {pages} {2000139} (\bibinfo {year} {2021})}\BibitemShut {NoStop}%
\bibitem [{\citenamefont {Hwang}\ \emph {et~al.}(2015)\citenamefont {Hwang},
  \citenamefont {Puebla},\ and\ \citenamefont {Plenio}}]{Hwang2015PRL}%
  \BibitemOpen
  \bibfield  {author} {\bibinfo {author} {\bibfnamefont {M.-J.}\ \bibnamefont
  {Hwang}}, \bibinfo {author} {\bibfnamefont {R.}~\bibnamefont {Puebla}}, \
  and\ \bibinfo {author} {\bibfnamefont {M.~B.}\ \bibnamefont {Plenio}},\
  }\href {\doibase 10.1103/PhysRevLett.115.180404} {\bibfield  {journal}
  {\bibinfo  {journal} {Phys. Rev. Lett.}\ }\textbf {\bibinfo {volume} {115}},\
  \bibinfo {pages} {180404} (\bibinfo {year} {2015})}\BibitemShut {NoStop}%
\bibitem [{\citenamefont {Hwang}\ and\ \citenamefont
  {Plenio}(2016)}]{Hwang2016PRL}%
  \BibitemOpen
  \bibfield  {author} {\bibinfo {author} {\bibfnamefont {M.-J.}\ \bibnamefont
  {Hwang}}\ and\ \bibinfo {author} {\bibfnamefont {M.~B.}\ \bibnamefont
  {Plenio}},\ }\href {\doibase 10.1103/PhysRevLett.117.123602} {\bibfield
  {journal} {\bibinfo  {journal} {Phys. Rev. Lett.}\ }\textbf {\bibinfo
  {volume} {117}},\ \bibinfo {pages} {123602} (\bibinfo {year}
  {2016})}\BibitemShut {NoStop}%
\bibitem [{\citenamefont {Irish}\ and\ \citenamefont
  {Larson}(2017)}]{Irish2017}%
  \BibitemOpen
  \bibfield  {author} {\bibinfo {author} {\bibfnamefont {E.~K.}\ \bibnamefont
  {Irish}}\ and\ \bibinfo {author} {\bibfnamefont {J.}~\bibnamefont {Larson}},\
  }\href {\doibase 10.1088/1751-8121/aa65dc} {\bibfield  {journal} {\bibinfo
  {journal} {J. Phys. A: Math. Theor.}\ }\textbf {\bibinfo {volume} {50}},\
  \bibinfo {pages} {174002} (\bibinfo {year} {2017})}\BibitemShut {NoStop}%
\bibitem [{\citenamefont {Ying}\ \emph
  {et~al.}(2020{\natexlab{a}})\citenamefont {Ying}, \citenamefont {Cong},\ and\
  \citenamefont {Sun}}]{Ying-2018-arxiv}%
  \BibitemOpen
  \bibfield  {author} {\bibinfo {author} {\bibfnamefont {Z.-J.}\ \bibnamefont
  {Ying}}, \bibinfo {author} {\bibfnamefont {L.}~\bibnamefont {Cong}}, \ and\
  \bibinfo {author} {\bibfnamefont {X.-M.}\ \bibnamefont {Sun}},\ }\href
  {\doibase 10.1088/1751-8121/ab9bd0} {\bibfield  {journal} {\bibinfo
  {journal} {J. Phys. A: Math. Theor.}\ }\textbf {\bibinfo {volume} {53}},\
  \bibinfo {pages} {345301} (\bibinfo {year} {2020}{\natexlab{a}})},\ \bibinfo
  {note} {arXiv:1804.08128}\BibitemShut {NoStop}%
\bibitem [{\citenamefont {Ying}(2021)}]{Ying2020-nonlinear-bias}%
  \BibitemOpen
  \bibfield  {author} {\bibinfo {author} {\bibfnamefont {Z.-J.}\ \bibnamefont
  {Ying}},\ }\href {\doibase 10.1103/PhysRevA.103.063701} {\bibfield  {journal}
  {\bibinfo  {journal} {Phys. Rev. A}\ }\textbf {\bibinfo {volume} {103}},\
  \bibinfo {pages} {063701} (\bibinfo {year} {2021})}\BibitemShut {NoStop}%
\bibitem [{\citenamefont {Ying}(2022{\natexlab{a}})}]{Ying-2021-AQT}%
  \BibitemOpen
  \bibfield  {author} {\bibinfo {author} {\bibfnamefont {Z.-J.}\ \bibnamefont
  {Ying}},\ }\href {\doibase 10.1002/qute.202100088} {\bibfield  {journal}
  {\bibinfo  {journal} {Adv. Quantum Technol.}\ }\textbf {\bibinfo {volume}
  {5}},\ \bibinfo {pages} {2100088} (\bibinfo {year}
  {2022}{\natexlab{a}})}\BibitemShut {NoStop}%
\bibitem [{\citenamefont {Ying}(2022{\natexlab{b}})}]{Ying-2021-AQT-Cover}%
  \BibitemOpen
  \bibfield  {author} {\bibinfo {author} {\bibfnamefont {Z.-J.}\ \bibnamefont
  {Ying}},\ }\href {\doibase 10.1002/qute.202270013} {\bibfield  {journal}
  {\bibinfo  {journal} {Adv. Quantum Technol.}\ }\textbf {\bibinfo {volume}
  {5}},\ \bibinfo {pages} {2270013} (\bibinfo {year} {2022}{\natexlab{b}})},\
  \bibinfo {note} {[Back Cover (Adv. Quantum Technol. 1/2022)]}\BibitemShut
  {NoStop}%
\bibitem [{\citenamefont {Ying}(2022{\natexlab{c}})}]{Ying-gapped-top}%
  \BibitemOpen
  \bibfield  {author} {\bibinfo {author} {\bibfnamefont {Z.-J.}\ \bibnamefont
  {Ying}},\ }\href {\doibase 10.1002/qute.202100165} {\bibfield  {journal}
  {\bibinfo  {journal} {Adv. Quantum Technol.}\ }\textbf {\bibinfo {volume}
  {5}},\ \bibinfo {pages} {2100165} (\bibinfo {year}
  {2022}{\natexlab{c}})}\BibitemShut {NoStop}%
\bibitem [{\citenamefont {Ying}(2023{\natexlab{a}})}]{Ying-Stark-top}%
  \BibitemOpen
  \bibfield  {author} {\bibinfo {author} {\bibfnamefont {Z.-J.}\ \bibnamefont
  {Ying}},\ }\href {\doibase 10.1002/qute.202200068} {\bibfield  {journal}
  {\bibinfo  {journal} {Adv. Quantum Technol.}\ }\textbf {\bibinfo {volume}
  {6}},\ \bibinfo {pages} {2200068} (\bibinfo {year}
  {2023}{\natexlab{a}})}\BibitemShut {NoStop}%
\bibitem [{\citenamefont {Ying}(2023{\natexlab{b}})}]{Ying-Stark-top-Cover}%
  \BibitemOpen
  \bibfield  {author} {\bibinfo {author} {\bibfnamefont {Z.-J.}\ \bibnamefont
  {Ying}},\ }\href {\doibase 10.1002/qute.202370011} {\bibfield  {journal}
  {\bibinfo  {journal} {Adv. Quantum Technol.}\ }\textbf {\bibinfo {volume}
  {6}},\ \bibinfo {pages} {2370011} (\bibinfo {year} {2023}{\natexlab{b}})},\
  \bibinfo {note} {[Front Cover (Adv. Quantum Technol. 1/2023)]}\BibitemShut
  {NoStop}%
\bibitem [{\citenamefont {Ying}(2023{\natexlab{c}})}]{Ying-Spin-Winding}%
  \BibitemOpen
  \bibfield  {author} {\bibinfo {author} {\bibfnamefont {Z.-J.}\ \bibnamefont
  {Ying}},\ }\href {\doibase 10.1002/qute.202200177} {\bibfield  {journal}
  {\bibinfo  {journal} {Adv. Quantum Technol.}\ }\textbf {\bibinfo {volume}
  {6}},\ \bibinfo {pages} {2200177} (\bibinfo {year}
  {2023}{\natexlab{c}})}\BibitemShut {NoStop}%
\bibitem [{\citenamefont {Ying}(2023{\natexlab{d}})}]{Ying-Spin-Winding-Cover}%
  \BibitemOpen
  \bibfield  {author} {\bibinfo {author} {\bibfnamefont {Z.-J.}\ \bibnamefont
  {Ying}},\ }\href {\doibase 10.1002/qute.202370071} {\bibfield  {journal}
  {\bibinfo  {journal} {Adv. Quantum Technol.}\ }\textbf {\bibinfo {volume}
  {6}},\ \bibinfo {pages} {2370071} (\bibinfo {year} {2023}{\natexlab{d}})},\
  \bibinfo {note} {[Front Cover: (Adv. Quantum Technol. 7/2023)]}\BibitemShut
  {NoStop}%
\bibitem [{\citenamefont {Ying}(2024{\natexlab{c}})}]{Ying-JCwinding}%
  \BibitemOpen
  \bibfield  {author} {\bibinfo {author} {\bibfnamefont {Z.-J.}\ \bibnamefont
  {Ying}},\ }\href {\doibase 10.1103/PhysRevA.109.053705} {\bibfield  {journal}
  {\bibinfo  {journal} {Phys. Rev. A}\ }\textbf {\bibinfo {volume} {109}},\
  \bibinfo {pages} {053705} (\bibinfo {year} {2024}{\natexlab{c}})}\BibitemShut
  {NoStop}%
\bibitem [{\citenamefont
  {Ying}(2024{\natexlab{d}})}]{Ying-Topo-JC-nonHermitian}%
  \BibitemOpen
  \bibfield  {author} {\bibinfo {author} {\bibfnamefont {Z.-J.}\ \bibnamefont
  {Ying}},\ }\href {\doibase 10.1002/qute.202400053} {\bibfield  {journal}
  {\bibinfo  {journal} {Adv. Quantum Technol.}\ }\textbf {\bibinfo {volume}
  {7}},\ \bibinfo {pages} {2400053} (\bibinfo {year}
  {2024}{\natexlab{d}})}\BibitemShut {NoStop}%
\bibitem [{\citenamefont
  {Ying}(2024{\natexlab{e}})}]{Ying-Topo-JC-nonHermitian-Cover}%
  \BibitemOpen
  \bibfield  {author} {\bibinfo {author} {\bibfnamefont {Z.-J.}\ \bibnamefont
  {Ying}},\ }\href {\doibase 10.1002/qute.202470017} {\bibfield  {journal}
  {\bibinfo  {journal} {Adv. Quantum Technol.}\ }\textbf {\bibinfo {volume}
  {7}},\ \bibinfo {pages} {2470017} (\bibinfo {year} {2024}{\natexlab{e}})},\
  \bibinfo {note} {[Front Cover: (Adv. Quantum Technol. 7/2024)]}\BibitemShut
  {NoStop}%
\bibitem [{\citenamefont {Ying}\ \emph {et~al.}(2024)\citenamefont {Ying},
  \citenamefont {Wang},\ and\ \citenamefont {Li}}]{Ying-gC-by-QFI-2024}%
  \BibitemOpen
  \bibfield  {author} {\bibinfo {author} {\bibfnamefont {Z.-J.}\ \bibnamefont
  {Ying}}, \bibinfo {author} {\bibfnamefont {W.-L.}\ \bibnamefont {Wang}}, \
  and\ \bibinfo {author} {\bibfnamefont {B.-J.}\ \bibnamefont {Li}},\ }\href
  {\doibase 10.1103/PhysRevA.110.033715} {\bibfield  {journal} {\bibinfo
  {journal} {Phys. Rev. A}\ }\textbf {\bibinfo {volume} {110}},\ \bibinfo
  {pages} {033715} (\bibinfo {year} {2024})}\BibitemShut {NoStop}%
\bibitem [{\citenamefont {Grimaudo}\ \emph
  {et~al.}(2023{\natexlab{a}})\citenamefont {Grimaudo}, \citenamefont
  {Magalh\~{a}es~de Castro}, \citenamefont {Messina}, \citenamefont {Solano},\
  and\ \citenamefont {Valenti}}]{Grimaudo2022q2QPT}%
  \BibitemOpen
  \bibfield  {author} {\bibinfo {author} {\bibfnamefont {R.}~\bibnamefont
  {Grimaudo}}, \bibinfo {author} {\bibfnamefont {A.~S.}\ \bibnamefont
  {Magalh\~{a}es~de Castro}}, \bibinfo {author} {\bibfnamefont
  {A.}~\bibnamefont {Messina}}, \bibinfo {author} {\bibfnamefont
  {E.}~\bibnamefont {Solano}}, \ and\ \bibinfo {author} {\bibfnamefont
  {D.}~\bibnamefont {Valenti}},\ }\href {\doibase
  10.1103/PhysRevLett.130.043602} {\bibfield  {journal} {\bibinfo  {journal}
  {Phys. Rev. Lett.}\ }\textbf {\bibinfo {volume} {130}},\ \bibinfo {pages}
  {043602} (\bibinfo {year} {2023}{\natexlab{a}})}\BibitemShut {NoStop}%
\bibitem [{\citenamefont {Grimaudo}\ \emph
  {et~al.}(2023{\natexlab{b}})\citenamefont {Grimaudo}, \citenamefont
  {Valenti}, \citenamefont {Sergi},\ and\ \citenamefont
  {Messina}}]{Grimaudo2023-Entropy}%
  \BibitemOpen
  \bibfield  {author} {\bibinfo {author} {\bibfnamefont {R.}~\bibnamefont
  {Grimaudo}}, \bibinfo {author} {\bibfnamefont {D.}~\bibnamefont {Valenti}},
  \bibinfo {author} {\bibfnamefont {A.}~\bibnamefont {Sergi}}, \ and\ \bibinfo
  {author} {\bibfnamefont {A.}~\bibnamefont {Messina}},\ }\href {\doibase
  10.3390/e25020187} {\bibfield  {journal} {\bibinfo  {journal} {Entropy}\
  }\textbf {\bibinfo {volume} {25}},\ \bibinfo {pages} {187} (\bibinfo {year}
  {2023}{\natexlab{b}})}\BibitemShut {NoStop}%
\bibitem [{\citenamefont {Grimaudo}\ \emph {et~al.}(2024)\citenamefont
  {Grimaudo}, \citenamefont {Falci}, \citenamefont {Messina}, \citenamefont
  {Paladino}, \citenamefont {Sergi}, \citenamefont {Solano},\ and\
  \citenamefont {Valenti}}]{Grimaudo2024PRR}%
  \BibitemOpen
  \bibfield  {author} {\bibinfo {author} {\bibfnamefont {R.}~\bibnamefont
  {Grimaudo}}, \bibinfo {author} {\bibfnamefont {G.}~\bibnamefont {Falci}},
  \bibinfo {author} {\bibfnamefont {A.}~\bibnamefont {Messina}}, \bibinfo
  {author} {\bibfnamefont {E.}~\bibnamefont {Paladino}}, \bibinfo {author}
  {\bibfnamefont {A.}~\bibnamefont {Sergi}}, \bibinfo {author} {\bibfnamefont
  {E.}~\bibnamefont {Solano}}, \ and\ \bibinfo {author} {\bibfnamefont
  {D.}~\bibnamefont {Valenti}},\ }\href {\doibase
  10.1103/PhysRevResearch.6.043298} {\bibfield  {journal} {\bibinfo  {journal}
  {Phys. Rev. Res.}\ }\textbf {\bibinfo {volume} {6}},\ \bibinfo {pages}
  {043298} (\bibinfo {year} {2024})}\BibitemShut {NoStop}%
\bibitem [{\citenamefont {Zhu}\ \emph {et~al.}(2024)\citenamefont {Zhu},
  \citenamefont {Hu}, \citenamefont {Wang}, \citenamefont {Qin}, \citenamefont
  {L\"u},\ and\ \citenamefont {Nori}}]{Zhu2024PRL}%
  \BibitemOpen
  \bibfield  {author} {\bibinfo {author} {\bibfnamefont {G.-L.}\ \bibnamefont
  {Zhu}}, \bibinfo {author} {\bibfnamefont {C.-S.}\ \bibnamefont {Hu}},
  \bibinfo {author} {\bibfnamefont {H.}~\bibnamefont {Wang}}, \bibinfo {author}
  {\bibfnamefont {W.}~\bibnamefont {Qin}}, \bibinfo {author} {\bibfnamefont
  {X.-Y.}\ \bibnamefont {L\"u}}, \ and\ \bibinfo {author} {\bibfnamefont
  {F.}~\bibnamefont {Nori}},\ }\href {\doibase 10.1103/PhysRevLett.132.193602}
  {\bibfield  {journal} {\bibinfo  {journal} {Phys. Rev. Lett.}\ }\textbf
  {\bibinfo {volume} {132}},\ \bibinfo {pages} {193602} (\bibinfo {year}
  {2024})}\BibitemShut {NoStop}%
\bibitem [{\citenamefont {Peng}\ \emph {et~al.}(2019)\citenamefont {Peng},
  \citenamefont {Rico}, \citenamefont {Zhong}, \citenamefont {Solano},\ and\
  \citenamefont {Egusquiza}}]{PengJie2019}%
  \BibitemOpen
  \bibfield  {author} {\bibinfo {author} {\bibfnamefont {J.}~\bibnamefont
  {Peng}}, \bibinfo {author} {\bibfnamefont {E.}~\bibnamefont {Rico}}, \bibinfo
  {author} {\bibfnamefont {J.}~\bibnamefont {Zhong}}, \bibinfo {author}
  {\bibfnamefont {E.}~\bibnamefont {Solano}}, \ and\ \bibinfo {author}
  {\bibfnamefont {I.~L.}\ \bibnamefont {Egusquiza}},\ }\href {\doibase
  10.1103/PhysRevA.100.063820} {\bibfield  {journal} {\bibinfo  {journal}
  {Phys. Rev. A}\ }\textbf {\bibinfo {volume} {100}},\ \bibinfo {pages}
  {063820} (\bibinfo {year} {2019})}\BibitemShut {NoStop}%
\bibitem [{\citenamefont {Peng}\ \emph {et~al.}(2021)\citenamefont {Peng},
  \citenamefont {Zheng}, \citenamefont {Yu}, \citenamefont {Tang},
  \citenamefont {Barrios}, \citenamefont {Zhong}, \citenamefont {Solano},
  \citenamefont {Albarr\'an-Arriagada},\ and\ \citenamefont
  {Lamata}}]{PengJ2021PRL}%
  \BibitemOpen
  \bibfield  {author} {\bibinfo {author} {\bibfnamefont {J.}~\bibnamefont
  {Peng}}, \bibinfo {author} {\bibfnamefont {J.}~\bibnamefont {Zheng}},
  \bibinfo {author} {\bibfnamefont {J.}~\bibnamefont {Yu}}, \bibinfo {author}
  {\bibfnamefont {P.}~\bibnamefont {Tang}}, \bibinfo {author} {\bibfnamefont
  {G.~A.}\ \bibnamefont {Barrios}}, \bibinfo {author} {\bibfnamefont
  {J.}~\bibnamefont {Zhong}}, \bibinfo {author} {\bibfnamefont
  {E.}~\bibnamefont {Solano}}, \bibinfo {author} {\bibfnamefont
  {F.}~\bibnamefont {Albarr\'an-Arriagada}}, \ and\ \bibinfo {author}
  {\bibfnamefont {L.}~\bibnamefont {Lamata}},\ }\href {\doibase
  10.1103/PhysRevLett.127.043604} {\bibfield  {journal} {\bibinfo  {journal}
  {Phys. Rev. Lett.}\ }\textbf {\bibinfo {volume} {127}},\ \bibinfo {pages}
  {043604} (\bibinfo {year} {2021})}\BibitemShut {NoStop}%
\bibitem [{\citenamefont {F.~Padilla}\ \emph {et~al.}(2022)\citenamefont
  {F.~Padilla}, \citenamefont {Pu}, \citenamefont {Cheng},\ and\ \citenamefont
  {Zhang}}]{Padilla2022}%
  \BibitemOpen
  \bibfield  {author} {\bibinfo {author} {\bibfnamefont {D.}~\bibnamefont
  {F.~Padilla}}, \bibinfo {author} {\bibfnamefont {H.}~\bibnamefont {Pu}},
  \bibinfo {author} {\bibfnamefont {G.-J.}\ \bibnamefont {Cheng}}, \ and\
  \bibinfo {author} {\bibfnamefont {Y.-Y.}\ \bibnamefont {Zhang}},\ }\href
  {\doibase 10.1103/PhysRevLett.129.183602} {\bibfield  {journal} {\bibinfo
  {journal} {Phys. Rev. Lett.}\ }\textbf {\bibinfo {volume} {129}},\ \bibinfo
  {pages} {183602} (\bibinfo {year} {2022})}\BibitemShut {NoStop}%
\bibitem [{\citenamefont {Cheng}\ \emph {et~al.}(2022)\citenamefont {Cheng},
  \citenamefont {Xu}, \citenamefont {Liu},\ and\ \citenamefont
  {Xianlong}}]{Gao2022Rabi-dimer}%
  \BibitemOpen
  \bibfield  {author} {\bibinfo {author} {\bibfnamefont {S.}~\bibnamefont
  {Cheng}}, \bibinfo {author} {\bibfnamefont {H.-G.}\ \bibnamefont {Xu}},
  \bibinfo {author} {\bibfnamefont {X.}~\bibnamefont {Liu}}, \ and\ \bibinfo
  {author} {\bibfnamefont {G.}~\bibnamefont {Xianlong}},\ }\href {\doibase
  https://doi.org/10.1016/j.physa.2022.127940} {\bibfield  {journal} {\bibinfo
  {journal} {Physica A}\ }\textbf {\bibinfo {volume} {604}},\ \bibinfo {pages}
  {127940} (\bibinfo {year} {2022})}\BibitemShut {NoStop}%
\bibitem [{\citenamefont {Cheng}\ \emph {et~al.}(2025)\citenamefont {Cheng},
  \citenamefont {Wang}, \citenamefont {Neto},\ and\ \citenamefont
  {Xianlong}}]{GaoXL2025SPT}%
  \BibitemOpen
  \bibfield  {author} {\bibinfo {author} {\bibfnamefont {S.}~\bibnamefont
  {Cheng}}, \bibinfo {author} {\bibfnamefont {S.-P.}\ \bibnamefont {Wang}},
  \bibinfo {author} {\bibfnamefont {G.~D.~M.}\ \bibnamefont {Neto}}, \ and\
  \bibinfo {author} {\bibfnamefont {G.}~\bibnamefont {Xianlong}},\ }\href
  {\doibase 10.1103/PhysRevA.111.053501} {\bibfield  {journal} {\bibinfo
  {journal} {Phys. Rev. A}\ }\textbf {\bibinfo {volume} {111}},\ \bibinfo
  {pages} {053501} (\bibinfo {year} {2025})}\BibitemShut {NoStop}%
\bibitem [{\citenamefont {Braak}(2019)}]{Braak2019Symmetry}%
  \BibitemOpen
  \bibfield  {author} {\bibinfo {author} {\bibfnamefont {D.}~\bibnamefont
  {Braak}},\ }\href {\doibase 10.3390/sym11101259} {\bibfield  {journal}
  {\bibinfo  {journal} {Symmetry}\ }\textbf {\bibinfo {volume} {11}},\ \bibinfo
  {pages} {1259} (\bibinfo {year} {2019})}\BibitemShut {NoStop}%
\bibitem [{\citenamefont {Mangazeev}\ \emph {et~al.}(2021)\citenamefont
  {Mangazeev}, \citenamefont {Batchelor},\ and\ \citenamefont
  {Bazhanov}}]{HiddenSymMangazeev2021}%
  \BibitemOpen
  \bibfield  {author} {\bibinfo {author} {\bibfnamefont {V.~V.}\ \bibnamefont
  {Mangazeev}}, \bibinfo {author} {\bibfnamefont {M.~T.}\ \bibnamefont
  {Batchelor}}, \ and\ \bibinfo {author} {\bibfnamefont {V.~V.}\ \bibnamefont
  {Bazhanov}},\ }\href {\doibase 10.1088/1751-8121/abe426} {\bibfield
  {journal} {\bibinfo  {journal} {J. Phys. A: Math. Theor.}\ }\textbf {\bibinfo
  {volume} {54}},\ \bibinfo {pages} {12LT01} (\bibinfo {year}
  {2021})}\BibitemShut {NoStop}%
\bibitem [{\citenamefont {Li}\ and\ \citenamefont
  {Batchelor}(2021)}]{HiddenSymLi2021}%
  \BibitemOpen
  \bibfield  {author} {\bibinfo {author} {\bibfnamefont {Z.-M.}\ \bibnamefont
  {Li}}\ and\ \bibinfo {author} {\bibfnamefont {M.~T.}\ \bibnamefont
  {Batchelor}},\ }\href {\doibase 10.1103/PhysRevA.103.023719} {\bibfield
  {journal} {\bibinfo  {journal} {Phys. Rev. A}\ }\textbf {\bibinfo {volume}
  {103}},\ \bibinfo {pages} {023719} (\bibinfo {year} {2021})}\BibitemShut
  {NoStop}%
\bibitem [{\citenamefont {Reyes-Bustos}\ \emph {et~al.}(2021)\citenamefont
  {Reyes-Bustos}, \citenamefont {Braak},\ and\ \citenamefont
  {Wakayama}}]{HiddenSymBustos2021}%
  \BibitemOpen
  \bibfield  {author} {\bibinfo {author} {\bibfnamefont {C.}~\bibnamefont
  {Reyes-Bustos}}, \bibinfo {author} {\bibfnamefont {D.}~\bibnamefont {Braak}},
  \ and\ \bibinfo {author} {\bibfnamefont {M.}~\bibnamefont {Wakayama}},\
  }\href {\doibase 10.1088/1751-8121/ac0508} {\bibfield  {journal} {\bibinfo
  {journal} {J. Phys. A: Math. Theor.}\ }\textbf {\bibinfo {volume} {54}},\
  \bibinfo {pages} {285202} (\bibinfo {year} {2021})}\BibitemShut {NoStop}%
\bibitem [{\citenamefont {T.~Irish}\ and\ \citenamefont
  {Armour}(2022)}]{Irish-class-quan-corresp}%
  \BibitemOpen
  \bibfield  {author} {\bibinfo {author} {\bibfnamefont {E.~K.}\ \bibnamefont
  {T.~Irish}}\ and\ \bibinfo {author} {\bibfnamefont {A.~D.}\ \bibnamefont
  {Armour}},\ }\href {\doibase 10.1103/PhysRevLett.129.183603} {\bibfield
  {journal} {\bibinfo  {journal} {Phys. Rev. Lett.}\ }\textbf {\bibinfo
  {volume} {129}},\ \bibinfo {pages} {183603} (\bibinfo {year}
  {2022})}\BibitemShut {NoStop}%
\bibitem [{\citenamefont {Felicetti}\ \emph {et~al.}(2015)\citenamefont
  {Felicetti}, \citenamefont {Pedernales}, \citenamefont {Egusquiza},
  \citenamefont {Romero}, \citenamefont {Lamata}, \citenamefont {Braak},\ and\
  \citenamefont {Solano}}]{Felicetti2015-TwoPhotonProcess}%
  \BibitemOpen
  \bibfield  {author} {\bibinfo {author} {\bibfnamefont {S.}~\bibnamefont
  {Felicetti}}, \bibinfo {author} {\bibfnamefont {J.~S.}\ \bibnamefont
  {Pedernales}}, \bibinfo {author} {\bibfnamefont {I.~L.}\ \bibnamefont
  {Egusquiza}}, \bibinfo {author} {\bibfnamefont {G.}~\bibnamefont {Romero}},
  \bibinfo {author} {\bibfnamefont {L.}~\bibnamefont {Lamata}}, \bibinfo
  {author} {\bibfnamefont {D.}~\bibnamefont {Braak}}, \ and\ \bibinfo {author}
  {\bibfnamefont {E.}~\bibnamefont {Solano}},\ }\href {\doibase
  10.1103/PhysRevA.92.033817} {\bibfield  {journal} {\bibinfo  {journal} {Phys.
  Rev. A}\ }\textbf {\bibinfo {volume} {92}},\ \bibinfo {pages} {033817}
  (\bibinfo {year} {2015})}\BibitemShut {NoStop}%
\bibitem [{\citenamefont {Garbe}\ \emph {et~al.}(2017)\citenamefont {Garbe},
  \citenamefont {Egusquiza}, \citenamefont {Solano}, \citenamefont {Ciuti},
  \citenamefont {Coudreau}, \citenamefont {Milman},\ and\ \citenamefont
  {Felicetti}}]{e-collpase-Garbe-2017}%
  \BibitemOpen
  \bibfield  {author} {\bibinfo {author} {\bibfnamefont {L.}~\bibnamefont
  {Garbe}}, \bibinfo {author} {\bibfnamefont {I.~L.}\ \bibnamefont
  {Egusquiza}}, \bibinfo {author} {\bibfnamefont {E.}~\bibnamefont {Solano}},
  \bibinfo {author} {\bibfnamefont {C.}~\bibnamefont {Ciuti}}, \bibinfo
  {author} {\bibfnamefont {T.}~\bibnamefont {Coudreau}}, \bibinfo {author}
  {\bibfnamefont {P.}~\bibnamefont {Milman}}, \ and\ \bibinfo {author}
  {\bibfnamefont {S.}~\bibnamefont {Felicetti}},\ }\href {\doibase
  10.1103/PhysRevA.95.053854} {\bibfield  {journal} {\bibinfo  {journal} {Phys.
  Rev. A}\ }\textbf {\bibinfo {volume} {95}},\ \bibinfo {pages} {053854}
  (\bibinfo {year} {2017})}\BibitemShut {NoStop}%
\bibitem [{\citenamefont {Duan}\ \emph {et~al.}(2016)\citenamefont {Duan},
  \citenamefont {Xie}, \citenamefont {Braak},\ and\ \citenamefont
  {Chen}}]{e-collpase-Duan-2016}%
  \BibitemOpen
  \bibfield  {author} {\bibinfo {author} {\bibfnamefont {L.}~\bibnamefont
  {Duan}}, \bibinfo {author} {\bibfnamefont {Y.-F.}\ \bibnamefont {Xie}},
  \bibinfo {author} {\bibfnamefont {D.}~\bibnamefont {Braak}}, \ and\ \bibinfo
  {author} {\bibfnamefont {Q.-H.}\ \bibnamefont {Chen}},\ }\href {\doibase
  10.1088/1751-8113/49/46/464002} {\bibfield  {journal} {\bibinfo  {journal}
  {J. Phys. A: Math. Theor.}\ }\textbf {\bibinfo {volume} {49}},\ \bibinfo
  {pages} {464002} (\bibinfo {year} {2016})}\BibitemShut {NoStop}%
\bibitem [{\citenamefont {Cong}\ \emph {et~al.}(2019)\citenamefont {Cong},
  \citenamefont {Sun}, \citenamefont {Liu}, \citenamefont {Ying},\ and\
  \citenamefont {Luo}}]{CongLei2019}%
  \BibitemOpen
  \bibfield  {author} {\bibinfo {author} {\bibfnamefont {L.}~\bibnamefont
  {Cong}}, \bibinfo {author} {\bibfnamefont {X.-M.}\ \bibnamefont {Sun}},
  \bibinfo {author} {\bibfnamefont {M.}~\bibnamefont {Liu}}, \bibinfo {author}
  {\bibfnamefont {Z.-J.}\ \bibnamefont {Ying}}, \ and\ \bibinfo {author}
  {\bibfnamefont {H.-G.}\ \bibnamefont {Luo}},\ }\href {\doibase
  10.1103/PhysRevA.99.013815} {\bibfield  {journal} {\bibinfo  {journal} {Phys.
  Rev. A}\ }\textbf {\bibinfo {volume} {99}},\ \bibinfo {pages} {013815}
  (\bibinfo {year} {2019})}\BibitemShut {NoStop}%
\bibitem [{\citenamefont {Rico}\ \emph {et~al.}(2020)\citenamefont {Rico},
  \citenamefont {Maldonado-Villamizar},\ and\ \citenamefont
  {Rodriguez-Lara}}]{Rico2020}%
  \BibitemOpen
  \bibfield  {author} {\bibinfo {author} {\bibfnamefont {R.~J.~A.}\
  \bibnamefont {Rico}}, \bibinfo {author} {\bibfnamefont {F.~H.}\ \bibnamefont
  {Maldonado-Villamizar}}, \ and\ \bibinfo {author} {\bibfnamefont {B.~M.}\
  \bibnamefont {Rodriguez-Lara}},\ }\href {\doibase
  10.1103/PhysRevA.101.063825} {\bibfield  {journal} {\bibinfo  {journal}
  {Phys. Rev. A}\ }\textbf {\bibinfo {volume} {101}},\ \bibinfo {pages}
  {063825} (\bibinfo {year} {2020})}\BibitemShut {NoStop}%
\bibitem [{\citenamefont {Wu}\ \emph {et~al.}(2024)\citenamefont {Wu},
  \citenamefont {Hu}, \citenamefont {Wang}, \citenamefont {Chen}, \citenamefont
  {Li}, \citenamefont {Zhao}, \citenamefont {L\"u},\ and\ \citenamefont
  {Peng}}]{PengXHPRL2024RabiNMR}%
  \BibitemOpen
  \bibfield  {author} {\bibinfo {author} {\bibfnamefont {Z.}~\bibnamefont
  {Wu}}, \bibinfo {author} {\bibfnamefont {C.}~\bibnamefont {Hu}}, \bibinfo
  {author} {\bibfnamefont {T.}~\bibnamefont {Wang}}, \bibinfo {author}
  {\bibfnamefont {Y.}~\bibnamefont {Chen}}, \bibinfo {author} {\bibfnamefont
  {Y.}~\bibnamefont {Li}}, \bibinfo {author} {\bibfnamefont {L.}~\bibnamefont
  {Zhao}}, \bibinfo {author} {\bibfnamefont {X.-Y.}\ \bibnamefont {L\"u}}, \
  and\ \bibinfo {author} {\bibfnamefont {X.}~\bibnamefont {Peng}},\ }\href
  {\doibase 10.1103/PhysRevLett.133.173602} {\bibfield  {journal} {\bibinfo
  {journal} {Phys. Rev. Lett.}\ }\textbf {\bibinfo {volume} {133}},\ \bibinfo
  {pages} {173602} (\bibinfo {year} {2024})}\BibitemShut {NoStop}%
\bibitem [{\citenamefont {Li}\ \emph {et~al.}(2021)\citenamefont {Li},
  \citenamefont {Ferri}, \citenamefont {Tilbrook},\ and\ \citenamefont
  {Batchelor}}]{Li2020conical}%
  \BibitemOpen
  \bibfield  {author} {\bibinfo {author} {\bibfnamefont {Z.-M.}\ \bibnamefont
  {Li}}, \bibinfo {author} {\bibfnamefont {D.}~\bibnamefont {Ferri}}, \bibinfo
  {author} {\bibfnamefont {D.}~\bibnamefont {Tilbrook}}, \ and\ \bibinfo
  {author} {\bibfnamefont {M.~T.}\ \bibnamefont {Batchelor}},\ }\href {\doibase
  10.1088/1751-8121/ac1fc1} {\bibfield  {journal} {\bibinfo  {journal} {J.
  Phys. A: Math. Theor.}\ }\textbf {\bibinfo {volume} {54}},\ \bibinfo {pages}
  {405201} (\bibinfo {year} {2021})}\BibitemShut {NoStop}%
\bibitem [{\citenamefont {Zhang}\ \emph {et~al.}(2024)\citenamefont {Zhang},
  \citenamefont {Zhou}, \citenamefont {Zuo}, \citenamefont {Zhang},
  \citenamefont {Chen}, \citenamefont {Jing},\ and\ \citenamefont
  {Kuang}}]{KuangLM2024AQT}%
  \BibitemOpen
  \bibfield  {author} {\bibinfo {author} {\bibfnamefont {J.}~\bibnamefont
  {Zhang}}, \bibinfo {author} {\bibfnamefont {Y.-L.}\ \bibnamefont {Zhou}},
  \bibinfo {author} {\bibfnamefont {Y.}~\bibnamefont {Zuo}}, \bibinfo {author}
  {\bibfnamefont {H.}~\bibnamefont {Zhang}}, \bibinfo {author} {\bibfnamefont
  {P.-X.}\ \bibnamefont {Chen}}, \bibinfo {author} {\bibfnamefont
  {H.}~\bibnamefont {Jing}}, \ and\ \bibinfo {author} {\bibfnamefont {L.-M.}\
  \bibnamefont {Kuang}},\ }\href {\doibase
  https://doi.org/10.1002/qute.202300350} {\bibfield  {journal} {\bibinfo
  {journal} {Adv. Quantum Technol.}\ }\textbf {\bibinfo {volume} {7}},\
  \bibinfo {pages} {2300350} (\bibinfo {year} {2024})}\BibitemShut {NoStop}%
\bibitem [{\citenamefont {Yan}\ \emph {et~al.}(2023)\citenamefont {Yan},
  \citenamefont {L\"u}, \citenamefont {Chen},\ and\ \citenamefont
  {Zheng}}]{Yan2023-AQT}%
  \BibitemOpen
  \bibfield  {author} {\bibinfo {author} {\bibfnamefont {Y.}~\bibnamefont
  {Yan}}, \bibinfo {author} {\bibfnamefont {Z.}~\bibnamefont {L\"u}}, \bibinfo
  {author} {\bibfnamefont {L.}~\bibnamefont {Chen}}, \ and\ \bibinfo {author}
  {\bibfnamefont {H.}~\bibnamefont {Zheng}},\ }\href {\doibase
  https://doi.org/10.1002/qute.202200191} {\bibfield  {journal} {\bibinfo
  {journal} {Advanced Quantum Technologies}\ }\textbf {\bibinfo {volume} {6}},\
  \bibinfo {pages} {2200191} (\bibinfo {year} {2023})}\BibitemShut {NoStop}%
\bibitem [{\citenamefont {L\"{u}}\ \emph {et~al.}(2017)\citenamefont {L\"{u}},
  \citenamefont {Zhao},\ and\ \citenamefont {Zheng}}]{ZhengHang2017}%
  \BibitemOpen
  \bibfield  {author} {\bibinfo {author} {\bibfnamefont {Z.}~\bibnamefont
  {L\"{u}}}, \bibinfo {author} {\bibfnamefont {C.}~\bibnamefont {Zhao}}, \ and\
  \bibinfo {author} {\bibfnamefont {H.}~\bibnamefont {Zheng}},\ }\href
  {\doibase 10.1088/1751-8121/aa5537} {\bibfield  {journal} {\bibinfo
  {journal} {J. Phys. A: Math. Theor.}\ }\textbf {\bibinfo {volume} {50}},\
  \bibinfo {pages} {074002} (\bibinfo {year} {2017})}\BibitemShut {NoStop}%
\bibitem [{\citenamefont {Le~Boit\'e}\ \emph {et~al.}(2016)\citenamefont
  {Le~Boit\'e}, \citenamefont {Hwang}, \citenamefont {Nha},\ and\ \citenamefont
  {Plenio}}]{Boite2016-Photon-Blockade}%
  \BibitemOpen
  \bibfield  {author} {\bibinfo {author} {\bibfnamefont {A.}~\bibnamefont
  {Le~Boit\'e}}, \bibinfo {author} {\bibfnamefont {M.-J.}\ \bibnamefont
  {Hwang}}, \bibinfo {author} {\bibfnamefont {H.}~\bibnamefont {Nha}}, \ and\
  \bibinfo {author} {\bibfnamefont {M.~B.}\ \bibnamefont {Plenio}},\ }\href
  {\doibase 10.1103/PhysRevA.94.033827} {\bibfield  {journal} {\bibinfo
  {journal} {Phys. Rev. A}\ }\textbf {\bibinfo {volume} {94}},\ \bibinfo
  {pages} {033827} (\bibinfo {year} {2016})}\BibitemShut {NoStop}%
\bibitem [{\citenamefont {Ridolfo}\ \emph {et~al.}(2012)\citenamefont
  {Ridolfo}, \citenamefont {Leib}, \citenamefont {Savasta},\ and\ \citenamefont
  {Hartmann}}]{Ridolfo2012-Photon-Blockade}%
  \BibitemOpen
  \bibfield  {author} {\bibinfo {author} {\bibfnamefont {A.}~\bibnamefont
  {Ridolfo}}, \bibinfo {author} {\bibfnamefont {M.}~\bibnamefont {Leib}},
  \bibinfo {author} {\bibfnamefont {S.}~\bibnamefont {Savasta}}, \ and\
  \bibinfo {author} {\bibfnamefont {M.~J.}\ \bibnamefont {Hartmann}},\ }\href
  {\doibase 10.1103/PhysRevLett.109.193602} {\bibfield  {journal} {\bibinfo
  {journal} {Phys. Rev. Lett.}\ }\textbf {\bibinfo {volume} {109}},\ \bibinfo
  {pages} {193602} (\bibinfo {year} {2012})}\BibitemShut {NoStop}%
\bibitem [{\citenamefont {Zou}\ \emph {et~al.}(2020)\citenamefont {Zou},
  \citenamefont {Zhang}, \citenamefont {Xu}, \citenamefont {Huang},\ and\
  \citenamefont {Liao}}]{LiaoJQ2020BlockadeJC}%
  \BibitemOpen
  \bibfield  {author} {\bibinfo {author} {\bibfnamefont {F.}~\bibnamefont
  {Zou}}, \bibinfo {author} {\bibfnamefont {X.-Y.}\ \bibnamefont {Zhang}},
  \bibinfo {author} {\bibfnamefont {X.-W.}\ \bibnamefont {Xu}}, \bibinfo
  {author} {\bibfnamefont {J.-F.}\ \bibnamefont {Huang}}, \ and\ \bibinfo
  {author} {\bibfnamefont {J.-Q.}\ \bibnamefont {Liao}},\ }\href {\doibase
  10.1103/PhysRevA.102.053710} {\bibfield  {journal} {\bibinfo  {journal}
  {Phys. Rev. A}\ }\textbf {\bibinfo {volume} {102}},\ \bibinfo {pages}
  {053710} (\bibinfo {year} {2020})}\BibitemShut {NoStop}%
\bibitem [{\citenamefont {Ma}\ \emph {et~al.}(2026)\citenamefont {Ma},
  \citenamefont {Tang}, \citenamefont {Zuo}, \citenamefont {Huang},
  \citenamefont {Miranowicz}, \citenamefont {Nori},\ and\ \citenamefont
  {Jing}}]{Ma2026PRL-Strong-2PB}%
  \BibitemOpen
  \bibfield  {author} {\bibinfo {author} {\bibfnamefont {T.-T.}\ \bibnamefont
  {Ma}}, \bibinfo {author} {\bibfnamefont {J.}~\bibnamefont {Tang}}, \bibinfo
  {author} {\bibfnamefont {Y.-L.}\ \bibnamefont {Zuo}}, \bibinfo {author}
  {\bibfnamefont {R.}~\bibnamefont {Huang}}, \bibinfo {author} {\bibfnamefont
  {A.}~\bibnamefont {Miranowicz}}, \bibinfo {author} {\bibfnamefont
  {F.}~\bibnamefont {Nori}}, \ and\ \bibinfo {author} {\bibfnamefont
  {H.}~\bibnamefont {Jing}},\ }\href {\doibase 10.1103/nfz3-txyt} {\bibfield
  {journal} {\bibinfo  {journal} {Phys. Rev. Lett.}\ }\textbf {\bibinfo
  {volume} {136}},\ \bibinfo {pages} {033601} (\bibinfo {year}
  {2026})}\BibitemShut {NoStop}%
\bibitem [{\citenamefont {Dong}\ \emph {et~al.}(2026)\citenamefont {Dong},
  \citenamefont {Shah}, \citenamefont {Kirton}, \citenamefont {Alaeian},\ and\
  \citenamefont {Felicetti}}]{Felicetti2026PRXQuantumBlockade}%
  \BibitemOpen
  \bibfield  {author} {\bibinfo {author} {\bibfnamefont {L.}~\bibnamefont
  {Dong}}, \bibinfo {author} {\bibfnamefont {A.~J.}\ \bibnamefont {Shah}},
  \bibinfo {author} {\bibfnamefont {P.}~\bibnamefont {Kirton}}, \bibinfo
  {author} {\bibfnamefont {H.}~\bibnamefont {Alaeian}}, \ and\ \bibinfo
  {author} {\bibfnamefont {S.}~\bibnamefont {Felicetti}},\ }\href {\doibase
  10.1103/l477-z8jr} {\bibfield  {journal} {\bibinfo  {journal} {PRX Quantum}\
  }\textbf {\bibinfo {volume} {7}},\ \bibinfo {pages} {020349} (\bibinfo {year}
  {2026})}\BibitemShut {NoStop}%
\bibitem [{\citenamefont {Carmichael}(1985)}]{Carmichael1985}%
  \BibitemOpen
  \bibfield  {author} {\bibinfo {author} {\bibfnamefont {H.~J.}\ \bibnamefont
  {Carmichael}},\ }\href {\doibase 10.1103/PhysRevLett.55.2790} {\bibfield
  {journal} {\bibinfo  {journal} {Phys. Rev. Lett.}\ }\textbf {\bibinfo
  {volume} {55}},\ \bibinfo {pages} {2790} (\bibinfo {year}
  {1985})}\BibitemShut {NoStop}%
\bibitem [{\citenamefont {Birnbaum}\ \emph {et~al.}(2005)\citenamefont
  {Birnbaum}, \citenamefont {Boca}, \citenamefont {Miller}, \citenamefont
  {Boozer}, \citenamefont {Northup},\ and\ \citenamefont
  {Kimble}}]{Birnbaum2005}%
  \BibitemOpen
  \bibfield  {author} {\bibinfo {author} {\bibfnamefont {K.~M.}\ \bibnamefont
  {Birnbaum}}, \bibinfo {author} {\bibfnamefont {A.}~\bibnamefont {Boca}},
  \bibinfo {author} {\bibfnamefont {R.}~\bibnamefont {Miller}}, \bibinfo
  {author} {\bibfnamefont {A.~D.}\ \bibnamefont {Boozer}}, \bibinfo {author}
  {\bibfnamefont {T.~E.}\ \bibnamefont {Northup}}, \ and\ \bibinfo {author}
  {\bibfnamefont {H.~J.}\ \bibnamefont {Kimble}},\ }\href {\doibase
  10.1038/nature03804} {\bibfield  {journal} {\bibinfo  {journal} {Nature}\
  }\textbf {\bibinfo {volume} {436}},\ \bibinfo {pages} {87} (\bibinfo {year}
  {2005})}\BibitemShut {NoStop}%
\bibitem [{\citenamefont {Shamailov}\ \emph {et~al.}(2010)\citenamefont
  {Shamailov}, \citenamefont {Parkins}, \citenamefont {Collett},\ and\
  \citenamefont {Carmichael}}]{Shamailov2010}%
  \BibitemOpen
  \bibfield  {author} {\bibinfo {author} {\bibfnamefont {S.~S.}\ \bibnamefont
  {Shamailov}}, \bibinfo {author} {\bibfnamefont {A.~S.}\ \bibnamefont
  {Parkins}}, \bibinfo {author} {\bibfnamefont {M.~J.}\ \bibnamefont
  {Collett}}, \ and\ \bibinfo {author} {\bibfnamefont {H.~J.}\ \bibnamefont
  {Carmichael}},\ }\href {\doibase 10.1016/j.optcom.2009.10.057} {\bibfield
  {journal} {\bibinfo  {journal} {Optics Communications}\ }\textbf {\bibinfo
  {volume} {283}},\ \bibinfo {pages} {766} (\bibinfo {year}
  {2010})}\BibitemShut {NoStop}%
\bibitem [{\citenamefont {Liew}\ and\ \citenamefont {Savona}(2010)}]{Liew2010}%
  \BibitemOpen
  \bibfield  {author} {\bibinfo {author} {\bibfnamefont {T.~C.~H.}\
  \bibnamefont {Liew}}\ and\ \bibinfo {author} {\bibfnamefont {V.}~\bibnamefont
  {Savona}},\ }\href {\doibase 10.1103/PhysRevLett.104.183601} {\bibfield
  {journal} {\bibinfo  {journal} {Phys. Rev. Lett.}\ }\textbf {\bibinfo
  {volume} {104}},\ \bibinfo {pages} {183601} (\bibinfo {year}
  {2010})}\BibitemShut {NoStop}%
\bibitem [{\citenamefont {Hamsen}\ \emph {et~al.}(2017)\citenamefont {Hamsen},
  \citenamefont {Tolazzi}, \citenamefont {Wilk},\ and\ \citenamefont
  {Rempe}}]{Hamsen2017}%
  \BibitemOpen
  \bibfield  {author} {\bibinfo {author} {\bibfnamefont {C.}~\bibnamefont
  {Hamsen}}, \bibinfo {author} {\bibfnamefont {K.~N.}\ \bibnamefont {Tolazzi}},
  \bibinfo {author} {\bibfnamefont {T.}~\bibnamefont {Wilk}}, \ and\ \bibinfo
  {author} {\bibfnamefont {G.}~\bibnamefont {Rempe}},\ }\href {\doibase
  10.1103/PhysRevLett.118.133604} {\bibfield  {journal} {\bibinfo  {journal}
  {Phys. Rev. Lett.}\ }\textbf {\bibinfo {volume} {118}},\ \bibinfo {pages}
  {133604} (\bibinfo {year} {2017})}\BibitemShut {NoStop}%
\bibitem [{\citenamefont {Garziano}\ \emph {et~al.}(2017)\citenamefont
  {Garziano}, \citenamefont {Ridolfo}, \citenamefont {De~Liberato},\ and\
  \citenamefont {Savasta}}]{Garziano2017}%
  \BibitemOpen
  \bibfield  {author} {\bibinfo {author} {\bibfnamefont {L.}~\bibnamefont
  {Garziano}}, \bibinfo {author} {\bibfnamefont {A.}~\bibnamefont {Ridolfo}},
  \bibinfo {author} {\bibfnamefont {S.}~\bibnamefont {De~Liberato}}, \ and\
  \bibinfo {author} {\bibfnamefont {S.}~\bibnamefont {Savasta}},\ }\href
  {\doibase 10.1021/acsphotonics.7b00635} {\bibfield  {journal} {\bibinfo
  {journal} {ACS Photonics}\ }\textbf {\bibinfo {volume} {4}},\ \bibinfo
  {pages} {2345} (\bibinfo {year} {2017})}\BibitemShut {NoStop}%
\bibitem [{\citenamefont {Flayac}\ and\ \citenamefont
  {Savona}(2017)}]{Flayac2017}%
  \BibitemOpen
  \bibfield  {author} {\bibinfo {author} {\bibfnamefont {H.}~\bibnamefont
  {Flayac}}\ and\ \bibinfo {author} {\bibfnamefont {V.}~\bibnamefont
  {Savona}},\ }\href {\doibase 10.1103/PhysRevA.96.053810} {\bibfield
  {journal} {\bibinfo  {journal} {Phys. Rev. A}\ }\textbf {\bibinfo {volume}
  {96}},\ \bibinfo {pages} {053810} (\bibinfo {year} {2017})}\BibitemShut
  {NoStop}%
\bibitem [{\citenamefont {Han}\ and\ \citenamefont
  {Ying}(2026)}]{Han2026WeakTPB}%
  \BibitemOpen
  \bibfield  {author} {\bibinfo {author} {\bibfnamefont {H.-H.}\ \bibnamefont
  {Han}}\ and\ \bibinfo {author} {\bibfnamefont {Z.-J.}\ \bibnamefont {Ying}},\
  }\href {\doibase https://doi.org/10.48550/arXiv.2607.08882} {\bibfield
  {journal} {\bibinfo  {journal} {arXiv:2607.08882}\ } (\bibinfo {year}
  {2026}),\ https://doi.org/10.48550/arXiv.2607.08882}\BibitemShut {NoStop}%
\bibitem [{\citenamefont {Qiao}\ \emph {et~al.}(2026)\citenamefont {Qiao},
  \citenamefont {Chen},\ and\ \citenamefont {Ying}}]{QiaoFeng2016AsymPolaron}%
  \BibitemOpen
  \bibfield  {author} {\bibinfo {author} {\bibfnamefont {F.}~\bibnamefont
  {Qiao}}, \bibinfo {author} {\bibfnamefont {Q.-Y.}\ \bibnamefont {Chen}}, \
  and\ \bibinfo {author} {\bibfnamefont {Z.-J.}\ \bibnamefont {Ying}},\ }\href
  {\doibase 10.1103/nc6b-691t} {\bibfield  {journal} {\bibinfo  {journal}
  {Phys. Rev. A}\ }\textbf {\bibinfo {volume} {114}},\ \bibinfo {pages}
  {022457} (\bibinfo {year} {2026})}\BibitemShut {NoStop}%
\bibitem [{\citenamefont {Rabi}(1937)}]{rabi1936}%
  \BibitemOpen
  \bibfield  {author} {\bibinfo {author} {\bibfnamefont {I.~I.}\ \bibnamefont
  {Rabi}},\ }\href {\doibase 10.1103/PhysRev.51.652} {\bibfield  {journal}
  {\bibinfo  {journal} {Phys. Rev.}\ }\textbf {\bibinfo {volume} {51}},\
  \bibinfo {pages} {652} (\bibinfo {year} {1937})}\BibitemShut {NoStop}%
\bibitem [{\citenamefont {Braak}\ \emph {et~al.}(2016)\citenamefont {Braak},
  \citenamefont {Chen}, \citenamefont {Batchelor},\ and\ \citenamefont
  {Solano}}]{Rabi-Braak}%
  \BibitemOpen
  \bibfield  {author} {\bibinfo {author} {\bibfnamefont {D.}~\bibnamefont
  {Braak}}, \bibinfo {author} {\bibfnamefont {Q.-H.}\ \bibnamefont {Chen}},
  \bibinfo {author} {\bibfnamefont {M.~T.}\ \bibnamefont {Batchelor}}, \ and\
  \bibinfo {author} {\bibfnamefont {E.}~\bibnamefont {Solano}},\ }\href
  {\doibase 10.1088/1751-8113/49/30/300301} {\bibfield  {journal} {\bibinfo
  {journal} {J. Phys. A: Math. Theor.}\ }\textbf {\bibinfo {volume} {49}},\
  \bibinfo {pages} {300301} (\bibinfo {year} {2016})}\BibitemShut {NoStop}%
\bibitem [{\citenamefont {Xie}\ \emph {et~al.}(2014)\citenamefont {Xie},
  \citenamefont {Cui}, \citenamefont {Cao}, \citenamefont {Amico},\ and\
  \citenamefont {Fan}}]{PRX-Xie-Anistropy}%
  \BibitemOpen
  \bibfield  {author} {\bibinfo {author} {\bibfnamefont {Q.-T.}\ \bibnamefont
  {Xie}}, \bibinfo {author} {\bibfnamefont {S.}~\bibnamefont {Cui}}, \bibinfo
  {author} {\bibfnamefont {J.-P.}\ \bibnamefont {Cao}}, \bibinfo {author}
  {\bibfnamefont {L.}~\bibnamefont {Amico}}, \ and\ \bibinfo {author}
  {\bibfnamefont {H.}~\bibnamefont {Fan}},\ }\href {\doibase
  10.1103/PhysRevX.4.021046} {\bibfield  {journal} {\bibinfo  {journal} {Phys.
  Rev. X}\ }\textbf {\bibinfo {volume} {4}},\ \bibinfo {pages} {021046}
  (\bibinfo {year} {2014})}\BibitemShut {NoStop}%
\bibitem [{\citenamefont {Solano}(2011)}]{Solano2011}%
  \BibitemOpen
  \bibfield  {author} {\bibinfo {author} {\bibfnamefont {E.}~\bibnamefont
  {Solano}},\ }\href {https://physics.aps.org/articles/v4/68} {\bibfield
  {journal} {\bibinfo  {journal} {Physics}\ }\textbf {\bibinfo {volume} {4}},\
  \bibinfo {pages} {68} (\bibinfo {year} {2011})}\BibitemShut {NoStop}%
\bibitem [{\citenamefont {Le~Boit\'e}(2020)}]{Boite2020}%
  \BibitemOpen
  \bibfield  {author} {\bibinfo {author} {\bibfnamefont {A.}~\bibnamefont
  {Le~Boit\'e}},\ }\href {\doibase 10.1002/qute.201900140} {\bibfield
  {journal} {\bibinfo  {journal} {Adv. Quantum Technol.}\ }\textbf {\bibinfo
  {volume} {3}},\ \bibinfo {pages} {1900140} (\bibinfo {year}
  {2020})}\BibitemShut {NoStop}%
\bibitem [{\citenamefont {Bera}\ \emph {et~al.}(2014)\citenamefont {Bera},
  \citenamefont {Florens}, \citenamefont {Baranger}, \citenamefont {Roch},
  \citenamefont {Nazir},\ and\ \citenamefont {Chin}}]{Bera2014Polaron}%
  \BibitemOpen
  \bibfield  {author} {\bibinfo {author} {\bibfnamefont {S.}~\bibnamefont
  {Bera}}, \bibinfo {author} {\bibfnamefont {S.}~\bibnamefont {Florens}},
  \bibinfo {author} {\bibfnamefont {H.~U.}\ \bibnamefont {Baranger}}, \bibinfo
  {author} {\bibfnamefont {N.}~\bibnamefont {Roch}}, \bibinfo {author}
  {\bibfnamefont {A.}~\bibnamefont {Nazir}}, \ and\ \bibinfo {author}
  {\bibfnamefont {A.~W.}\ \bibnamefont {Chin}},\ }\href {\doibase
  10.1103/PhysRevB.89.121108} {\bibfield  {journal} {\bibinfo  {journal} {Phys.
  Rev. B}\ }\textbf {\bibinfo {volume} {89}},\ \bibinfo {pages} {121108}
  (\bibinfo {year} {2014})}\BibitemShut {NoStop}%
\bibitem [{\citenamefont {Cong}\ \emph {et~al.}(2017)\citenamefont {Cong},
  \citenamefont {Sun}, \citenamefont {Liu}, \citenamefont {Ying},\ and\
  \citenamefont {Luo}}]{CongLei2017}%
  \BibitemOpen
  \bibfield  {author} {\bibinfo {author} {\bibfnamefont {L.}~\bibnamefont
  {Cong}}, \bibinfo {author} {\bibfnamefont {X.-M.}\ \bibnamefont {Sun}},
  \bibinfo {author} {\bibfnamefont {M.}~\bibnamefont {Liu}}, \bibinfo {author}
  {\bibfnamefont {Z.-J.}\ \bibnamefont {Ying}}, \ and\ \bibinfo {author}
  {\bibfnamefont {H.-G.}\ \bibnamefont {Luo}},\ }\href {\doibase
  10.1103/PhysRevA.95.063803} {\bibfield  {journal} {\bibinfo  {journal} {Phys.
  Rev. A}\ }\textbf {\bibinfo {volume} {95}},\ \bibinfo {pages} {063803}
  (\bibinfo {year} {2017})}\BibitemShut {NoStop}%
\bibitem [{\citenamefont {Wolf}\ \emph {et~al.}(2012)\citenamefont {Wolf},
  \citenamefont {Kollar},\ and\ \citenamefont {Braak}}]{Wolf2012}%
  \BibitemOpen
  \bibfield  {author} {\bibinfo {author} {\bibfnamefont {F.~A.}\ \bibnamefont
  {Wolf}}, \bibinfo {author} {\bibfnamefont {M.}~\bibnamefont {Kollar}}, \ and\
  \bibinfo {author} {\bibfnamefont {D.}~\bibnamefont {Braak}},\ }\href
  {\doibase 10.1103/PhysRevA.85.053817} {\bibfield  {journal} {\bibinfo
  {journal} {Phys. Rev. A}\ }\textbf {\bibinfo {volume} {85}},\ \bibinfo
  {pages} {053817} (\bibinfo {year} {2012})}\BibitemShut {NoStop}%
\bibitem [{\citenamefont {Felicetti}\ and\ \citenamefont
  {Le~Boit\'e}(2020)}]{FelicettiPRL2020}%
  \BibitemOpen
  \bibfield  {author} {\bibinfo {author} {\bibfnamefont {S.}~\bibnamefont
  {Felicetti}}\ and\ \bibinfo {author} {\bibfnamefont {A.}~\bibnamefont
  {Le~Boit\'e}},\ }\href {\doibase 10.1103/PhysRevLett.124.040404} {\bibfield
  {journal} {\bibinfo  {journal} {Phys. Rev. Lett.}\ }\textbf {\bibinfo
  {volume} {124}},\ \bibinfo {pages} {040404} (\bibinfo {year}
  {2020})}\BibitemShut {NoStop}%
\bibitem [{\citenamefont {Felicetti}\ \emph
  {et~al.}(2018{\natexlab{a}})\citenamefont {Felicetti}, \citenamefont
  {Rossatto}, \citenamefont {Rico}, \citenamefont {Solano},\ and\ \citenamefont
  {Forn-D\'{\i}az}}]{Felicetti2018-mixed-TPP-SPP}%
  \BibitemOpen
  \bibfield  {author} {\bibinfo {author} {\bibfnamefont {S.}~\bibnamefont
  {Felicetti}}, \bibinfo {author} {\bibfnamefont {D.~Z.}\ \bibnamefont
  {Rossatto}}, \bibinfo {author} {\bibfnamefont {E.}~\bibnamefont {Rico}},
  \bibinfo {author} {\bibfnamefont {E.}~\bibnamefont {Solano}}, \ and\ \bibinfo
  {author} {\bibfnamefont {P.}~\bibnamefont {Forn-D\'{\i}az}},\ }\href
  {\doibase 10.1103/PhysRevA.97.013851} {\bibfield  {journal} {\bibinfo
  {journal} {Phys. Rev. A}\ }\textbf {\bibinfo {volume} {97}},\ \bibinfo
  {pages} {013851} (\bibinfo {year} {2018}{\natexlab{a}})}\BibitemShut
  {NoStop}%
\bibitem [{\citenamefont {Felicetti}\ \emph
  {et~al.}(2018{\natexlab{b}})\citenamefont {Felicetti}, \citenamefont
  {Hwang},\ and\ \citenamefont {Le~Boit\'e}}]{Simone2018}%
  \BibitemOpen
  \bibfield  {author} {\bibinfo {author} {\bibfnamefont {S.}~\bibnamefont
  {Felicetti}}, \bibinfo {author} {\bibfnamefont {M.-J.}\ \bibnamefont
  {Hwang}}, \ and\ \bibinfo {author} {\bibfnamefont {A.}~\bibnamefont
  {Le~Boit\'e}},\ }\href {\doibase 10.1103/PhysRevA.98.053859} {\bibfield
  {journal} {\bibinfo  {journal} {Phys. Rev. A}\ }\textbf {\bibinfo {volume}
  {98}},\ \bibinfo {pages} {053859} (\bibinfo {year}
  {2018}{\natexlab{b}})}\BibitemShut {NoStop}%
\bibitem [{\citenamefont {Alushi}\ \emph {et~al.}(2023)\citenamefont {Alushi},
  \citenamefont {Ramos}, \citenamefont {Garc\'{\i}a-Ripoll}, \citenamefont
  {Di~Candia},\ and\ \citenamefont {Felicetti}}]{Alushi2023PRX}%
  \BibitemOpen
  \bibfield  {author} {\bibinfo {author} {\bibfnamefont {U.}~\bibnamefont
  {Alushi}}, \bibinfo {author} {\bibfnamefont {T.}~\bibnamefont {Ramos}},
  \bibinfo {author} {\bibfnamefont {J.~J.}\ \bibnamefont {Garc\'{\i}a-Ripoll}},
  \bibinfo {author} {\bibfnamefont {R.}~\bibnamefont {Di~Candia}}, \ and\
  \bibinfo {author} {\bibfnamefont {S.}~\bibnamefont {Felicetti}},\ }\href
  {\doibase 10.1103/PRXQuantum.4.030326} {\bibfield  {journal} {\bibinfo
  {journal} {PRX Quantum}\ }\textbf {\bibinfo {volume} {4}},\ \bibinfo {pages}
  {030326} (\bibinfo {year} {2023})}\BibitemShut {NoStop}%
\bibitem [{\citenamefont {Irish}\ and\ \citenamefont
  {Gea-Banacloche}(2014)}]{Irish2014}%
  \BibitemOpen
  \bibfield  {author} {\bibinfo {author} {\bibfnamefont {E.~K.}\ \bibnamefont
  {Irish}}\ and\ \bibinfo {author} {\bibfnamefont {J.}~\bibnamefont
  {Gea-Banacloche}},\ }\href {\doibase 10.1103/PhysRevB.89.085421} {\bibfield
  {journal} {\bibinfo  {journal} {Phys. Rev. B}\ }\textbf {\bibinfo {volume}
  {89}},\ \bibinfo {pages} {085421} (\bibinfo {year} {2014})}\BibitemShut
  {NoStop}%
\bibitem [{\citenamefont {Xie}\ \emph {et~al.}(2017)\citenamefont {Xie},
  \citenamefont {Zhong}, \citenamefont {Batchelor},\ and\ \citenamefont
  {Lee}}]{XieQ-2017JPA}%
  \BibitemOpen
  \bibfield  {author} {\bibinfo {author} {\bibfnamefont {Q.}~\bibnamefont
  {Xie}}, \bibinfo {author} {\bibfnamefont {H.}~\bibnamefont {Zhong}}, \bibinfo
  {author} {\bibfnamefont {M.~T.}\ \bibnamefont {Batchelor}}, \ and\ \bibinfo
  {author} {\bibfnamefont {C.}~\bibnamefont {Lee}},\ }\href {\doibase
  10.1088/1751-8121/aa5a65} {\bibfield  {journal} {\bibinfo  {journal} {J.
  Phys. A: Math. Theor.}\ }\textbf {\bibinfo {volume} {50}},\ \bibinfo {pages}
  {113001} (\bibinfo {year} {2017})}\BibitemShut {NoStop}%
\bibitem [{\citenamefont {Ma}(2020)}]{Ma2020Nonlinear}%
  \BibitemOpen
  \bibfield  {author} {\bibinfo {author} {\bibfnamefont {K.~K.~W.}\
  \bibnamefont {Ma}},\ }\href {\doibase 10.1103/PhysRevA.102.053709} {\bibfield
   {journal} {\bibinfo  {journal} {Phys. Rev. A}\ }\textbf {\bibinfo {volume}
  {102}},\ \bibinfo {pages} {053709} (\bibinfo {year} {2020})}\BibitemShut
  {NoStop}%
\bibitem [{\citenamefont {Zhang}(2016)}]{ZhangYY2016}%
  \BibitemOpen
  \bibfield  {author} {\bibinfo {author} {\bibfnamefont {Y.-Y.}\ \bibnamefont
  {Zhang}},\ }\href {\doibase 10.1103/PhysRevA.94.063824} {\bibfield  {journal}
  {\bibinfo  {journal} {Phys. Rev. A}\ }\textbf {\bibinfo {volume} {94}},\
  \bibinfo {pages} {063824} (\bibinfo {year} {2016})}\BibitemShut {NoStop}%
\bibitem [{\citenamefont {Shen}\ \emph {et~al.}(2017)\citenamefont {Shen},
  \citenamefont {Yang}, \citenamefont {Wu},\ and\ \citenamefont
  {Zheng}}]{Zheng2017}%
  \BibitemOpen
  \bibfield  {author} {\bibinfo {author} {\bibfnamefont {L.-T.}\ \bibnamefont
  {Shen}}, \bibinfo {author} {\bibfnamefont {Z.-B.}\ \bibnamefont {Yang}},
  \bibinfo {author} {\bibfnamefont {H.-Z.}\ \bibnamefont {Wu}}, \ and\ \bibinfo
  {author} {\bibfnamefont {S.-B.}\ \bibnamefont {Zheng}},\ }\href {\doibase
  10.1103/PhysRevA.95.013819} {\bibfield  {journal} {\bibinfo  {journal} {Phys.
  Rev. A}\ }\textbf {\bibinfo {volume} {95}},\ \bibinfo {pages} {013819}
  (\bibinfo {year} {2017})}\BibitemShut {NoStop}%
\bibitem [{\citenamefont {Chen}\ \emph {et~al.}(2021)\citenamefont {Chen},
  \citenamefont {Wu}, \citenamefont {Jiang}, \citenamefont {L\"{u}},
  \citenamefont {Peng},\ and\ \citenamefont {Du}}]{Chen-2021-NC}%
  \BibitemOpen
  \bibfield  {author} {\bibinfo {author} {\bibfnamefont {X.}~\bibnamefont
  {Chen}}, \bibinfo {author} {\bibfnamefont {Z.}~\bibnamefont {Wu}}, \bibinfo
  {author} {\bibfnamefont {M.}~\bibnamefont {Jiang}}, \bibinfo {author}
  {\bibfnamefont {X.-Y.}\ \bibnamefont {L\"{u}}}, \bibinfo {author}
  {\bibfnamefont {X.}~\bibnamefont {Peng}}, \ and\ \bibinfo {author}
  {\bibfnamefont {J.}~\bibnamefont {Du}},\ }\href {\doibase
  https://doi.org/10.1038/s41467-021-26573-5} {\bibfield  {journal} {\bibinfo
  {journal} {Nat. Commun.}\ }\textbf {\bibinfo {volume} {12}},\ \bibinfo
  {pages} {6281} (\bibinfo {year} {2021})}\BibitemShut {NoStop}%
\bibitem [{\citenamefont {Liu}\ \emph {et~al.}(2015)\citenamefont {Liu},
  \citenamefont {Ying}, \citenamefont {An},\ and\ \citenamefont
  {Luo}}]{Liu2015}%
  \BibitemOpen
  \bibfield  {author} {\bibinfo {author} {\bibfnamefont {M.}~\bibnamefont
  {Liu}}, \bibinfo {author} {\bibfnamefont {Z.-J.}\ \bibnamefont {Ying}},
  \bibinfo {author} {\bibfnamefont {J.-H.}\ \bibnamefont {An}}, \ and\ \bibinfo
  {author} {\bibfnamefont {H.-G.}\ \bibnamefont {Luo}},\ }\href {\doibase
  10.1088/1367-2630/17/4/043001} {\bibfield  {journal} {\bibinfo  {journal}
  {New Journal of Physics}\ }\textbf {\bibinfo {volume} {17}},\ \bibinfo
  {pages} {043001} (\bibinfo {year} {2015})}\BibitemShut {NoStop}%
\bibitem [{\citenamefont {Zhang}\ \emph {et~al.}(2011)\citenamefont {Zhang},
  \citenamefont {Chen}, \citenamefont {Yu}, \citenamefont {Liang},
  \citenamefont {Liang},\ and\ \citenamefont {Jia}}]{ChenGang2011-GVM}%
  \BibitemOpen
  \bibfield  {author} {\bibinfo {author} {\bibfnamefont {Y.}~\bibnamefont
  {Zhang}}, \bibinfo {author} {\bibfnamefont {G.}~\bibnamefont {Chen}},
  \bibinfo {author} {\bibfnamefont {L.}~\bibnamefont {Yu}}, \bibinfo {author}
  {\bibfnamefont {Q.}~\bibnamefont {Liang}}, \bibinfo {author} {\bibfnamefont
  {J.-Q.}\ \bibnamefont {Liang}}, \ and\ \bibinfo {author} {\bibfnamefont
  {S.}~\bibnamefont {Jia}},\ }\href {\doibase 10.1103/PhysRevA.83.065802}
  {\bibfield  {journal} {\bibinfo  {journal} {Phys. Rev. A}\ }\textbf {\bibinfo
  {volume} {83}},\ \bibinfo {pages} {065802} (\bibinfo {year}
  {2011})}\BibitemShut {NoStop}%
\bibitem [{\citenamefont {Yu}\ \emph {et~al.}(2012)\citenamefont {Yu},
  \citenamefont {Zhu}, \citenamefont {Liang}, \citenamefont {Chen},\ and\
  \citenamefont {Jia}}]{ChenGang2012}%
  \BibitemOpen
  \bibfield  {author} {\bibinfo {author} {\bibfnamefont {L.}~\bibnamefont
  {Yu}}, \bibinfo {author} {\bibfnamefont {S.}~\bibnamefont {Zhu}}, \bibinfo
  {author} {\bibfnamefont {Q.}~\bibnamefont {Liang}}, \bibinfo {author}
  {\bibfnamefont {G.}~\bibnamefont {Chen}}, \ and\ \bibinfo {author}
  {\bibfnamefont {S.}~\bibnamefont {Jia}},\ }\href {\doibase
  10.1103/PhysRevA.86.015803} {\bibfield  {journal} {\bibinfo  {journal} {Phys.
  Rev. A}\ }\textbf {\bibinfo {volume} {86}},\ \bibinfo {pages} {015803}
  (\bibinfo {year} {2012})}\BibitemShut {NoStop}%
\bibitem [{\citenamefont {Liu}\ \emph {et~al.}(2013)\citenamefont {Liu},
  \citenamefont {Feng}, \citenamefont {Yang}, \citenamefont {Zou},
  \citenamefont {Li}, \citenamefont {Fan},\ and\ \citenamefont
  {Wang}}]{FengMang2013}%
  \BibitemOpen
  \bibfield  {author} {\bibinfo {author} {\bibfnamefont {T.}~\bibnamefont
  {Liu}}, \bibinfo {author} {\bibfnamefont {M.}~\bibnamefont {Feng}}, \bibinfo
  {author} {\bibfnamefont {W.~L.}\ \bibnamefont {Yang}}, \bibinfo {author}
  {\bibfnamefont {J.~H.}\ \bibnamefont {Zou}}, \bibinfo {author} {\bibfnamefont
  {L.}~\bibnamefont {Li}}, \bibinfo {author} {\bibfnamefont {Y.~X.}\
  \bibnamefont {Fan}}, \ and\ \bibinfo {author} {\bibfnamefont {K.~L.}\
  \bibnamefont {Wang}},\ }\href {\doibase 10.1103/PhysRevA.88.013820}
  {\bibfield  {journal} {\bibinfo  {journal} {Phys. Rev. A}\ }\textbf {\bibinfo
  {volume} {88}},\ \bibinfo {pages} {013820} (\bibinfo {year}
  {2013})}\BibitemShut {NoStop}%
\bibitem [{\citenamefont {Casanova}\ \emph {et~al.}(2018)\citenamefont
  {Casanova}, \citenamefont {Puebla}, \citenamefont {Moya-Cessa},\ and\
  \citenamefont {Plenio}}]{Casanova2018npj}%
  \BibitemOpen
  \bibfield  {author} {\bibinfo {author} {\bibfnamefont {J.}~\bibnamefont
  {Casanova}}, \bibinfo {author} {\bibfnamefont {R.}~\bibnamefont {Puebla}},
  \bibinfo {author} {\bibfnamefont {H.}~\bibnamefont {Moya-Cessa}}, \ and\
  \bibinfo {author} {\bibfnamefont {M.~B.}\ \bibnamefont {Plenio}},\ }\href
  {\doibase 10.1038/s41534-018-0096-9} {\bibfield  {journal} {\bibinfo
  {journal} {npj Quantum Information}\ }\textbf {\bibinfo {volume} {4}},\
  \bibinfo {pages} {47} (\bibinfo {year} {2018})}\BibitemShut {NoStop}%
\bibitem [{\citenamefont {Larson}\ and\ \citenamefont
  {Mavrogordatos}(2021)}]{JC-Larson2021}%
  \BibitemOpen
  \bibfield  {author} {\bibinfo {author} {\bibfnamefont {J.}~\bibnamefont
  {Larson}}\ and\ \bibinfo {author} {\bibfnamefont {T.}~\bibnamefont
  {Mavrogordatos}},\ }\href@noop {} {\emph {\bibinfo {title} {The
  Jaynes-Cummings Model and Its Descendants}}}\ (\bibinfo  {publisher} {IOP,
  London},\ \bibinfo {year} {2021})\BibitemShut {NoStop}%
\bibitem [{\citenamefont {Eckle}\ and\ \citenamefont
  {Johannesson}(2017)}]{Eckle-2017JPA}%
  \BibitemOpen
  \bibfield  {author} {\bibinfo {author} {\bibfnamefont {H.-P.}\ \bibnamefont
  {Eckle}}\ and\ \bibinfo {author} {\bibfnamefont {H.}~\bibnamefont
  {Johannesson}},\ }\href {\doibase 10.1088/1751-8121/aa785a} {\bibfield
  {journal} {\bibinfo  {journal} {J. Phys. A: Math. Theor.}\ }\textbf {\bibinfo
  {volume} {50}},\ \bibinfo {pages} {294004} (\bibinfo {year}
  {2017})}\BibitemShut {NoStop}%
\bibitem [{\citenamefont {Eckle}\ and\ \citenamefont
  {Johannesson}(2023)}]{Eckle-2017JPA-b}%
  \BibitemOpen
  \bibfield  {author} {\bibinfo {author} {\bibfnamefont {H.-P.}\ \bibnamefont
  {Eckle}}\ and\ \bibinfo {author} {\bibfnamefont {H.}~\bibnamefont
  {Johannesson}},\ }\href {\doibase 10.1088/1751-8121/acea07} {\bibfield
  {journal} {\bibinfo  {journal} {J. Phys. A: Math. Theor.}\ }\textbf {\bibinfo
  {volume} {56}},\ \bibinfo {pages} {345302} (\bibinfo {year}
  {2023})}\BibitemShut {NoStop}%
\bibitem [{\citenamefont {Grimsmo}\ and\ \citenamefont
  {Parkins}(2013)}]{Stark-Grimsmo2013}%
  \BibitemOpen
  \bibfield  {author} {\bibinfo {author} {\bibfnamefont {A.~L.}\ \bibnamefont
  {Grimsmo}}\ and\ \bibinfo {author} {\bibfnamefont {S.}~\bibnamefont
  {Parkins}},\ }\href {\doibase 10.1103/PhysRevA.87.033814} {\bibfield
  {journal} {\bibinfo  {journal} {Phys. Rev. A}\ }\textbf {\bibinfo {volume}
  {87}},\ \bibinfo {pages} {033814} (\bibinfo {year} {2013})}\BibitemShut
  {NoStop}%
\bibitem [{\citenamefont {Grimsmo}\ and\ \citenamefont
  {Parkins}(2014)}]{Stark-Grimsmo2014}%
  \BibitemOpen
  \bibfield  {author} {\bibinfo {author} {\bibfnamefont {A.~L.}\ \bibnamefont
  {Grimsmo}}\ and\ \bibinfo {author} {\bibfnamefont {S.}~\bibnamefont
  {Parkins}},\ }\href {\doibase 10.1103/PhysRevA.89.033802} {\bibfield
  {journal} {\bibinfo  {journal} {Phys. Rev. A}\ }\textbf {\bibinfo {volume}
  {89}},\ \bibinfo {pages} {033802} (\bibinfo {year} {2014})}\BibitemShut
  {NoStop}%
\bibitem [{\citenamefont {L\"u}\ \emph {et~al.}(2018)\citenamefont {L\"u},
  \citenamefont {Zheng}, \citenamefont {Zhu},\ and\ \citenamefont
  {Wu}}]{Lu-2018-1}%
  \BibitemOpen
  \bibfield  {author} {\bibinfo {author} {\bibfnamefont {X.-Y.}\ \bibnamefont
  {L\"u}}, \bibinfo {author} {\bibfnamefont {L.-L.}\ \bibnamefont {Zheng}},
  \bibinfo {author} {\bibfnamefont {G.-L.}\ \bibnamefont {Zhu}}, \ and\
  \bibinfo {author} {\bibfnamefont {Y.}~\bibnamefont {Wu}},\ }\href {\doibase
  10.1103/PhysRevApplied.9.064006} {\bibfield  {journal} {\bibinfo  {journal}
  {Phys. Rev. Appl.}\ }\textbf {\bibinfo {volume} {9}},\ \bibinfo {pages}
  {064006} (\bibinfo {year} {2018})}\BibitemShut {NoStop}%
\bibitem [{\citenamefont {Xie}\ \emph {et~al.}(2019)\citenamefont {Xie},
  \citenamefont {Duan},\ and\ \citenamefont {Chen}}]{Xie2019-Stark}%
  \BibitemOpen
  \bibfield  {author} {\bibinfo {author} {\bibfnamefont {Y.-F.}\ \bibnamefont
  {Xie}}, \bibinfo {author} {\bibfnamefont {L.}~\bibnamefont {Duan}}, \ and\
  \bibinfo {author} {\bibfnamefont {Q.-H.}\ \bibnamefont {Chen}},\ }\href
  {\doibase 10.1088/1751-8121/ab1cf6} {\bibfield  {journal} {\bibinfo
  {journal} {J. Phys. A: Math. Theor.}\ }\textbf {\bibinfo {volume} {52}},\
  \bibinfo {pages} {245304} (\bibinfo {year} {2019})}\BibitemShut {NoStop}%
\bibitem [{\citenamefont {Cong}\ \emph {et~al.}(2020)\citenamefont {Cong},
  \citenamefont {Felicetti}, \citenamefont {Casanova}, \citenamefont {Lamata},
  \citenamefont {Solano},\ and\ \citenamefont {Arrazola}}]{Stark-Cong2020}%
  \BibitemOpen
  \bibfield  {author} {\bibinfo {author} {\bibfnamefont {L.}~\bibnamefont
  {Cong}}, \bibinfo {author} {\bibfnamefont {S.}~\bibnamefont {Felicetti}},
  \bibinfo {author} {\bibfnamefont {J.}~\bibnamefont {Casanova}}, \bibinfo
  {author} {\bibfnamefont {L.}~\bibnamefont {Lamata}}, \bibinfo {author}
  {\bibfnamefont {E.}~\bibnamefont {Solano}}, \ and\ \bibinfo {author}
  {\bibfnamefont {I.}~\bibnamefont {Arrazola}},\ }\href {\doibase
  10.1103/PhysRevA.101.032350} {\bibfield  {journal} {\bibinfo  {journal}
  {Phys. Rev. A}\ }\textbf {\bibinfo {volume} {101}},\ \bibinfo {pages}
  {032350} (\bibinfo {year} {2020})}\BibitemShut {NoStop}%
\bibitem [{\citenamefont {Shi}\ \emph {et~al.}(2022)\citenamefont {Shi},
  \citenamefont {Cong},\ and\ \citenamefont {Eckle}}]{Cong2022Peter}%
  \BibitemOpen
  \bibfield  {author} {\bibinfo {author} {\bibfnamefont {Y.-Q.}\ \bibnamefont
  {Shi}}, \bibinfo {author} {\bibfnamefont {L.}~\bibnamefont {Cong}}, \ and\
  \bibinfo {author} {\bibfnamefont {H.-P.}\ \bibnamefont {Eckle}},\ }\href
  {\doibase 10.1103/PhysRevA.105.062450} {\bibfield  {journal} {\bibinfo
  {journal} {Phys. Rev. A}\ }\textbf {\bibinfo {volume} {105}},\ \bibinfo
  {pages} {062450} (\bibinfo {year} {2022})}\BibitemShut {NoStop}%
\bibitem [{\citenamefont {Xu}\ \emph {et~al.}(2024)\citenamefont {Xu},
  \citenamefont {Montenegro}, \citenamefont {Xianlong}, \citenamefont {Jin},\
  and\ \citenamefont {Neto}}]{Gao2022Rabi-aniso}%
  \BibitemOpen
  \bibfield  {author} {\bibinfo {author} {\bibfnamefont {H.-G.}\ \bibnamefont
  {Xu}}, \bibinfo {author} {\bibfnamefont {V.}~\bibnamefont {Montenegro}},
  \bibinfo {author} {\bibfnamefont {G.}~\bibnamefont {Xianlong}}, \bibinfo
  {author} {\bibfnamefont {J.}~\bibnamefont {Jin}}, \ and\ \bibinfo {author}
  {\bibfnamefont {G.~D. d.~M.}\ \bibnamefont {Neto}},\ }\href {\doibase
  10.1103/PhysRevResearch.6.013001} {\bibfield  {journal} {\bibinfo  {journal}
  {Phys. Rev. Res.}\ }\textbf {\bibinfo {volume} {6}},\ \bibinfo {pages}
  {013001} (\bibinfo {year} {2024})}\BibitemShut {NoStop}%
\bibitem [{\citenamefont {Feng}\ and\ \citenamefont
  {Gong}(2021)}]{FengLJ2021TPB}%
  \BibitemOpen
  \bibfield  {author} {\bibinfo {author} {\bibfnamefont {L.-J.}\ \bibnamefont
  {Feng}}\ and\ \bibinfo {author} {\bibfnamefont {S.-Q.}\ \bibnamefont
  {Gong}},\ }\href {\doibase 10.1103/PhysRevA.103.043509} {\bibfield  {journal}
  {\bibinfo  {journal} {Phys. Rev. A}\ }\textbf {\bibinfo {volume} {103}},\
  \bibinfo {pages} {043509} (\bibinfo {year} {2021})}\BibitemShut {NoStop}%
\bibitem [{\citenamefont {Kowalewska-Kud\l{}aszyk}\ \emph
  {et~al.}(2019)\citenamefont {Kowalewska-Kud\l{}aszyk}, \citenamefont {Abo},
  \citenamefont {Chimczak}, \citenamefont {Pe\ifmmode~\check{r}\else
  \v{r}\fi{}ina}, \citenamefont {Nori},\ and\ \citenamefont
  {Miranowicz}}]{Kudlaszyk2019PRATPB}%
  \BibitemOpen
  \bibfield  {author} {\bibinfo {author} {\bibfnamefont {A.}~\bibnamefont
  {Kowalewska-Kud\l{}aszyk}}, \bibinfo {author} {\bibfnamefont {S.~I.}\
  \bibnamefont {Abo}}, \bibinfo {author} {\bibfnamefont {G.}~\bibnamefont
  {Chimczak}}, \bibinfo {author} {\bibfnamefont {J.}~\bibnamefont
  {Pe\ifmmode~\check{r}\else \v{r}\fi{}ina}}, \bibinfo {author} {\bibfnamefont
  {F.}~\bibnamefont {Nori}}, \ and\ \bibinfo {author} {\bibfnamefont
  {A.}~\bibnamefont {Miranowicz}},\ }\href {\doibase
  10.1103/PhysRevA.100.053857} {\bibfield  {journal} {\bibinfo  {journal}
  {Phys. Rev. A}\ }\textbf {\bibinfo {volume} {100}},\ \bibinfo {pages}
  {053857} (\bibinfo {year} {2019})}\BibitemShut {NoStop}%
\bibitem [{\citenamefont {Bin}\ \emph {et~al.}(2018)\citenamefont {Bin},
  \citenamefont {L\"u}, \citenamefont {Bin},\ and\ \citenamefont
  {Wu}}]{QianBin2018praTPB}%
  \BibitemOpen
  \bibfield  {author} {\bibinfo {author} {\bibfnamefont {Q.}~\bibnamefont
  {Bin}}, \bibinfo {author} {\bibfnamefont {X.-Y.}\ \bibnamefont {L\"u}},
  \bibinfo {author} {\bibfnamefont {S.-W.}\ \bibnamefont {Bin}}, \ and\
  \bibinfo {author} {\bibfnamefont {Y.}~\bibnamefont {Wu}},\ }\href {\doibase
  10.1103/PhysRevA.98.043858} {\bibfield  {journal} {\bibinfo  {journal} {Phys.
  Rev. A}\ }\textbf {\bibinfo {volume} {98}},\ \bibinfo {pages} {043858}
  (\bibinfo {year} {2018})}\BibitemShut {NoStop}%
\bibitem [{\citenamefont {Miranowicz}\ \emph {et~al.}(2013)\citenamefont
  {Miranowicz}, \citenamefont {Paprzycka}, \citenamefont {Liu}, \citenamefont
  {Bajer},\ and\ \citenamefont {Nori}}]{Miranowicz2013praTPB}%
  \BibitemOpen
  \bibfield  {author} {\bibinfo {author} {\bibfnamefont {A.}~\bibnamefont
  {Miranowicz}}, \bibinfo {author} {\bibfnamefont {M.}~\bibnamefont
  {Paprzycka}}, \bibinfo {author} {\bibfnamefont {Y.-x.}\ \bibnamefont {Liu}},
  \bibinfo {author} {\bibfnamefont {J.}~\bibnamefont {Bajer}}, \ and\ \bibinfo
  {author} {\bibfnamefont {F.}~\bibnamefont {Nori}},\ }\href {\doibase
  10.1103/PhysRevA.87.023809} {\bibfield  {journal} {\bibinfo  {journal} {Phys.
  Rev. A}\ }\textbf {\bibinfo {volume} {87}},\ \bibinfo {pages} {023809}
  (\bibinfo {year} {2013})}\BibitemShut {NoStop}%
\bibitem [{\citenamefont {Ying}\ and\ \citenamefont
  {Han}(2026)}]{Ying2026StarkTPB}%
  \BibitemOpen
  \bibfield  {author} {\bibinfo {author} {\bibfnamefont {Z.-J.}\ \bibnamefont
  {Ying}}\ and\ \bibinfo {author} {\bibfnamefont {H.-H.}\ \bibnamefont {Han}},\
  }\href {\doibase 10.48550/ARXIV.2608.10942} {\bibfield  {journal} {\bibinfo
  {journal} {arXiv:2608.10942}\ } (\bibinfo {year} {2026}),\
  10.48550/ARXIV.2608.10942}\BibitemShut {NoStop}%
\bibitem [{\citenamefont {Jaynes}\ and\ \citenamefont
  {Cummings}(1963)}]{JC-model}%
  \BibitemOpen
  \bibfield  {author} {\bibinfo {author} {\bibfnamefont {E.~T.}\ \bibnamefont
  {Jaynes}}\ and\ \bibinfo {author} {\bibfnamefont {F.~W.}\ \bibnamefont
  {Cummings}},\ }\href@noop {} {\bibfield  {journal} {\bibinfo  {journal}
  {Proceedings of the IEEE}\ }\textbf {\bibinfo {volume} {51}},\ \bibinfo
  {pages} {89} (\bibinfo {year} {1963})}\BibitemShut {NoStop}%
\bibitem [{\citenamefont {Twyeffort~Irish}(2007)}]{Irish2007GRWA}%
  \BibitemOpen
  \bibfield  {author} {\bibinfo {author} {\bibfnamefont {E.~K.}\ \bibnamefont
  {Twyeffort~Irish}},\ }\href {\doibase 10.1103/PhysRevLett.99.173601}
  {\bibfield  {journal} {\bibinfo  {journal} {Phys. Rev. Lett.}\ }\textbf
  {\bibinfo {volume} {99}},\ \bibinfo {pages} {173601} (\bibinfo {year}
  {2007})}\BibitemShut {NoStop}%
\bibitem [{\citenamefont {Twyeffort~Irish}\ \emph {et~al.}(2005)\citenamefont
  {Twyeffort~Irish}, \citenamefont {Gea-Banacloche}, \citenamefont {Martin},\
  and\ \citenamefont {Schwab}}]{Irish-2005-AA}%
  \BibitemOpen
  \bibfield  {author} {\bibinfo {author} {\bibfnamefont {E.~K.}\ \bibnamefont
  {Twyeffort~Irish}}, \bibinfo {author} {\bibfnamefont {J.}~\bibnamefont
  {Gea-Banacloche}}, \bibinfo {author} {\bibfnamefont {I.}~\bibnamefont
  {Martin}}, \ and\ \bibinfo {author} {\bibfnamefont {K.~C.}\ \bibnamefont
  {Schwab}},\ }\href {\doibase 10.1103/PhysRevB.72.195410} {\bibfield
  {journal} {\bibinfo  {journal} {Phys. Rev. B}\ }\textbf {\bibinfo {volume}
  {72}},\ \bibinfo {pages} {195410} (\bibinfo {year} {2005})}\BibitemShut
  {NoStop}%
\bibitem [{\citenamefont {Hwang}\ and\ \citenamefont {Choi}(2010)}]{Hwang2010}%
  \BibitemOpen
  \bibfield  {author} {\bibinfo {author} {\bibfnamefont {M.-J.}\ \bibnamefont
  {Hwang}}\ and\ \bibinfo {author} {\bibfnamefont {M.-S.}\ \bibnamefont
  {Choi}},\ }\href {\doibase 10.1103/PhysRevA.82.025802} {\bibfield  {journal}
  {\bibinfo  {journal} {Phys. Rev. A}\ }\textbf {\bibinfo {volume} {82}},\
  \bibinfo {pages} {025802} (\bibinfo {year} {2010})}\BibitemShut {NoStop}%
\bibitem [{\citenamefont {Ying}\ \emph {et~al.}(2016)\citenamefont {Ying},
  \citenamefont {Gentile}, \citenamefont {Ortix},\ and\ \citenamefont
  {Cuoco}}]{Ying2016Ellipse}%
  \BibitemOpen
  \bibfield  {author} {\bibinfo {author} {\bibfnamefont {Z.-J.}\ \bibnamefont
  {Ying}}, \bibinfo {author} {\bibfnamefont {P.}~\bibnamefont {Gentile}},
  \bibinfo {author} {\bibfnamefont {C.}~\bibnamefont {Ortix}}, \ and\ \bibinfo
  {author} {\bibfnamefont {M.}~\bibnamefont {Cuoco}},\ }\href {\doibase
  10.1103/PhysRevB.94.081406} {\bibfield  {journal} {\bibinfo  {journal} {Phys.
  Rev. B}\ }\textbf {\bibinfo {volume} {94}},\ \bibinfo {pages} {081406(R)}
  (\bibinfo {year} {2016})}\BibitemShut {NoStop}%
\bibitem [{\citenamefont {Ying}\ \emph {et~al.}(2017)\citenamefont {Ying},
  \citenamefont {Cuoco}, \citenamefont {Ortix},\ and\ \citenamefont
  {Gentile}}]{Ying2017curvedSC}%
  \BibitemOpen
  \bibfield  {author} {\bibinfo {author} {\bibfnamefont {Z.-J.}\ \bibnamefont
  {Ying}}, \bibinfo {author} {\bibfnamefont {M.}~\bibnamefont {Cuoco}},
  \bibinfo {author} {\bibfnamefont {C.}~\bibnamefont {Ortix}}, \ and\ \bibinfo
  {author} {\bibfnamefont {P.}~\bibnamefont {Gentile}},\ }\href {\doibase
  10.1103/PhysRevB.96.100506} {\bibfield  {journal} {\bibinfo  {journal} {Phys.
  Rev. B}\ }\textbf {\bibinfo {volume} {96}},\ \bibinfo {pages} {100506(R)}
  (\bibinfo {year} {2017})}\BibitemShut {NoStop}%
\bibitem [{\citenamefont {Ying}\ \emph
  {et~al.}(2020{\natexlab{b}})\citenamefont {Ying}, \citenamefont {Gentile},
  \citenamefont {Baltan\'as}, \citenamefont {Frustaglia}, \citenamefont
  {Ortix},\ and\ \citenamefont {Cuoco}}]{Ying2020PRR}%
  \BibitemOpen
  \bibfield  {author} {\bibinfo {author} {\bibfnamefont {Z.-J.}\ \bibnamefont
  {Ying}}, \bibinfo {author} {\bibfnamefont {P.}~\bibnamefont {Gentile}},
  \bibinfo {author} {\bibfnamefont {J.~P.}\ \bibnamefont {Baltan\'as}},
  \bibinfo {author} {\bibfnamefont {D.}~\bibnamefont {Frustaglia}}, \bibinfo
  {author} {\bibfnamefont {C.}~\bibnamefont {Ortix}}, \ and\ \bibinfo {author}
  {\bibfnamefont {M.}~\bibnamefont {Cuoco}},\ }\href {\doibase
  10.1103/PhysRevResearch.2.023167} {\bibfield  {journal} {\bibinfo  {journal}
  {Phys. Rev. Res.}\ }\textbf {\bibinfo {volume} {2}},\ \bibinfo {pages}
  {023167} (\bibinfo {year} {2020}{\natexlab{b}})}\BibitemShut {NoStop}%
\bibitem [{\citenamefont {Gentile}\ \emph {et~al.}(2022)\citenamefont
  {Gentile}, \citenamefont {Cuoco}, \citenamefont {Volkov}, \citenamefont
  {Ying}, \citenamefont {VeraMarun}, \citenamefont {Makarov},\ and\
  \citenamefont {Ortix}}]{Gentile2022NatElec}%
  \BibitemOpen
  \bibfield  {author} {\bibinfo {author} {\bibfnamefont {P.}~\bibnamefont
  {Gentile}}, \bibinfo {author} {\bibfnamefont {M.}~\bibnamefont {Cuoco}},
  \bibinfo {author} {\bibfnamefont {O.~M.}\ \bibnamefont {Volkov}}, \bibinfo
  {author} {\bibfnamefont {Z.-J.}\ \bibnamefont {Ying}}, \bibinfo {author}
  {\bibfnamefont {I.~J.}\ \bibnamefont {VeraMarun}}, \bibinfo {author}
  {\bibfnamefont {D.}~\bibnamefont {Makarov}}, \ and\ \bibinfo {author}
  {\bibfnamefont {C.}~\bibnamefont {Ortix}},\ }\href {\doibase
  10.1038/s41928-022-00820-z} {\bibfield  {journal} {\bibinfo  {journal}
  {Nature Electronics}\ }\textbf {\bibinfo {volume} {5}},\ \bibinfo {pages}
  {551} (\bibinfo {year} {2022})}\BibitemShut {NoStop}%
\bibitem [{\citenamefont {Nagasawa}\ \emph {et~al.}(2013)\citenamefont
  {Nagasawa}, \citenamefont {Frustaglia}, \citenamefont {Saarikoski},
  \citenamefont {Richter},\ and\ \citenamefont {Nitta}}]{Nagasawa2013Rings}%
  \BibitemOpen
  \bibfield  {author} {\bibinfo {author} {\bibfnamefont {F.}~\bibnamefont
  {Nagasawa}}, \bibinfo {author} {\bibfnamefont {D.}~\bibnamefont
  {Frustaglia}}, \bibinfo {author} {\bibfnamefont {H.}~\bibnamefont
  {Saarikoski}}, \bibinfo {author} {\bibfnamefont {K.}~\bibnamefont {Richter}},
  \ and\ \bibinfo {author} {\bibfnamefont {J.}~\bibnamefont {Nitta}},\ }\href
  {\doibase 10.1038/ncomms3526} {\bibfield  {journal} {\bibinfo  {journal}
  {Nat. Commun.}\ }\textbf {\bibinfo {volume} {4}},\ \bibinfo {pages} {2526}
  (\bibinfo {year} {2013})}\BibitemShut {NoStop}%
\bibitem [{\citenamefont {Mooij}\ \emph {et~al.}(1999)\citenamefont {Mooij},
  \citenamefont {Orlando}, \citenamefont {Levitov}, \citenamefont {Tian},
  \citenamefont {van~der Wal},\ and\ \citenamefont
  {Lloyd}}]{flux-qubit-Mooij-1999}%
  \BibitemOpen
  \bibfield  {author} {\bibinfo {author} {\bibfnamefont {J.~E.}\ \bibnamefont
  {Mooij}}, \bibinfo {author} {\bibfnamefont {T.~P.}\ \bibnamefont {Orlando}},
  \bibinfo {author} {\bibfnamefont {L.}~\bibnamefont {Levitov}}, \bibinfo
  {author} {\bibfnamefont {L.}~\bibnamefont {Tian}}, \bibinfo {author}
  {\bibfnamefont {C.~H.}\ \bibnamefont {van~der Wal}}, \ and\ \bibinfo {author}
  {\bibfnamefont {S.}~\bibnamefont {Lloyd}},\ }\href {\doibase
  10.1126/science.285.5430.1036} {\bibfield  {journal} {\bibinfo  {journal}
  {Science}\ }\textbf {\bibinfo {volume} {285}},\ \bibinfo {pages} {1036}
  (\bibinfo {year} {1999})}\BibitemShut {NoStop}%
\bibitem [{\citenamefont {Bertet}\ \emph {et~al.}()\citenamefont {Bertet},
  \citenamefont {Chiorescu}, \citenamefont {Harmans},\ and\ \citenamefont
  {Mooij}}]{Bertet2005mixedModel}%
  \BibitemOpen
  \bibfield  {author} {\bibinfo {author} {\bibfnamefont {P.}~\bibnamefont
  {Bertet}}, \bibinfo {author} {\bibfnamefont {I.}~\bibnamefont {Chiorescu}},
  \bibinfo {author} {\bibfnamefont {C.~J. P.~M.}\ \bibnamefont {Harmans}}, \
  and\ \bibinfo {author} {\bibfnamefont {J.~E.}\ \bibnamefont {Mooij}},\ }\href
  {https://doi.org/10.48550/arXiv.cond-mat/0507290} {\bibinfo  {journal}
  {arXiv:cond-mat/0507290}\ }\BibitemShut {NoStop}%
\bibitem [{\citenamefont {You}\ \emph {et~al.}(2005)\citenamefont {You},
  \citenamefont {Nakamura},\ and\ \citenamefont {Nori}}]{you024532}%
  \BibitemOpen
\bibfield  {journal} {  }\bibfield  {author} {\bibinfo {author} {\bibfnamefont
  {J.~Q.}\ \bibnamefont {You}}, \bibinfo {author} {\bibfnamefont
  {Y.}~\bibnamefont {Nakamura}}, \ and\ \bibinfo {author} {\bibfnamefont
  {F.}~\bibnamefont {Nori}},\ }\href {\doibase 10.1103/PhysRevB.71.024532}
  {\bibfield  {journal} {\bibinfo  {journal} {Phys. Rev. B}\ }\textbf {\bibinfo
  {volume} {71}},\ \bibinfo {pages} {024532} (\bibinfo {year}
  {2005})}\BibitemShut {NoStop}%
\bibitem [{\citenamefont {Sachdev}(2011)}]{Sachdev-QPT}%
  \BibitemOpen
  \bibfield  {author} {\bibinfo {author} {\bibfnamefont {S.}~\bibnamefont
  {Sachdev}},\ }\href@noop {} {\emph {\bibinfo {title} {Quantum Phase
  Transitions}}},\ \bibinfo {edition} {2nd}\ ed.\ (\bibinfo  {publisher}
  {Cambridge University Press},\ \bibinfo {address} {Cambridge},\ \bibinfo
  {year} {2011})\BibitemShut {NoStop}%
\bibitem [{\citenamefont {Emary}\ and\ \citenamefont
  {Brandes}(2003)}]{EmaryPRE2003DickeChaos}%
  \BibitemOpen
  \bibfield  {author} {\bibinfo {author} {\bibfnamefont {C.}~\bibnamefont
  {Emary}}\ and\ \bibinfo {author} {\bibfnamefont {T.}~\bibnamefont
  {Brandes}},\ }\href {\doibase 10.1103/PhysRevE.67.066203} {\bibfield
  {journal} {\bibinfo  {journal} {Phys. Rev. E}\ }\textbf {\bibinfo {volume}
  {67}},\ \bibinfo {pages} {066203} (\bibinfo {year} {2003})}\BibitemShut
  {NoStop}%
\bibitem [{\citenamefont {Aharonov}\ and\ \citenamefont
  {Anandan}(1987)}]{AAphase1987}%
  \BibitemOpen
  \bibfield  {author} {\bibinfo {author} {\bibfnamefont {Y.}~\bibnamefont
  {Aharonov}}\ and\ \bibinfo {author} {\bibfnamefont {J.}~\bibnamefont
  {Anandan}},\ }\href {\doibase 10.1103/PhysRevLett.58.1593} {\bibfield
  {journal} {\bibinfo  {journal} {Phys. Rev. Lett.}\ }\textbf {\bibinfo
  {volume} {58}},\ \bibinfo {pages} {1593} (\bibinfo {year}
  {1987})}\BibitemShut {NoStop}%
\bibitem [{\citenamefont {Ying}(2026)}]{Ying2026GeomNonhermitian}%
  \BibitemOpen
  \bibfield  {author} {\bibinfo {author} {\bibfnamefont {Z.-J.}\ \bibnamefont
  {Ying}},\ }\href@noop {} {\bibfield  {journal} {\bibinfo  {journal} {arXiv:
  to appear}\ } (\bibinfo {year} {2026})}\BibitemShut {NoStop}%
\end{thebibliography}%

\end{document}